\documentclass[a4paper,11pt]{article}
\usepackage{jheppub} 
\usepackage{lineno}

\usepackage{newtxtext}
\usepackage{newtxmath}
\usepackage{mathrsfs}

\usepackage{graphicx}
\usepackage{subfigure}
\usepackage{amsmath, bm}
\usepackage{float}
\usepackage{multirow}
\usepackage{slashed}
\usepackage{xcolor}
\usepackage{mathtools,braket}
\usepackage{color,physics,tensor}
\usepackage{indentfirst}
\usepackage{soul}

\newcommand{\be}{\begin{eqnarray}}
\newcommand{\ee}{\end{eqnarray}}
\newcommand{\ba}{\begin{array}}
\newcommand{\ea}{\end{array}}
\newcommand{\bea}{\begin{eqnarray}}
\newcommand{\eea
}{\end{eqnarray}}
\newcommand{\bi}{\begin{itemize}}
\newcommand{\ei}{\end{itemize}}

\newcommand{\nc}{\newcommand}
\nc{\non}{\nonumber}
\nc{\hc}{\hbox {H.c.}}
\nc{\noi}{\noindent}
\nc{\barx}{\bar{x}}
\nc{\pbarn}{\;\hbox {pb}}
\nc{\fbarn}{\;\hbox {fb}}
\nc{\cone}{{\scriptsize \textcolor{orange}{\circled{1}}}}
\nc{\ctwo}{{\scriptsize \textcolor{orange}{\circled{2}}}}
\nc{\cthr}{{\scriptsize \textcolor{orange}{\circled{3}}}}

\nc{\hsp}{\hspace{0.5cm}}
\nc{\lsp}{\hspace{1cm}}
\nc{\Lsp}{\hspace{2cm}}
\nc{\LLsp}{\lsp\lsp}
\nc{\lra}{\longrightarrow}
\nc{\p}{\prime}
\nc{\sgn}{\text{sgn}}
\nc{\ph}{\varphi}
\nc{\eff}{\mathrm{eff}}
\nc{\sqM}{\sqrt{M}}
\nc{\NL}{\mathrm{NL}}
\nc{\moc}{\mathcal{M}} 
\nc{\rd}[1]{\textcolor{red}{#1}}
\nc{\bl}[1]{\textcolor{blue}{#1}}

\nc{\pbar}{\bar{\psi}}
\nc{\beq}{\begin{equation}}  \nc{\eeq}{\end{equation}}
\nc{\baa}{\begin{array}}     \nc{\eaa}{\end{array}}
\nc{\bit}{\begin{itemize}}   \nc{\eit}{\end{itemize}}
\nc{\ben}{\begin{enumerate}} \nc{\een}{\end{enumerate}}
\nc{\bce}{\begin{center}}    \nc{\ece}{\end{center}}
\nc{\bpm}{\begin{pmatrix}}   \nc{\epm}{\end{pmatrix}}
\nc{\bvt}{\begin{verbatim}}  \nc{\evt}{\end{verbatim}}
\nc{\vp}[1]{\mathbf{p}_{#1}}
\nc{\vk}{{\bm k}_\perp}
\nc{\vkp}{{\bm k}'_\perp}
\nc{\vq}{{\bm q}_\perp}
\nc{\uu}{{\uparrow\uparrow}}
\nc{\ud}{{\uparrow\downarrow}}
\nc{\du}{{\downarrow\uparrow}}

\def\lsim{\mathrel{\raise.3ex\hbox{$<$\kern-.75em\lower1ex\hbox{$\sim$}}}}
\def\gsim{\mathrel{\raise.3ex\hbox{$>$\kern-.75em\lower1ex\hbox{$\sim$}}}}

\def\udots{\mathinner{\mkern1mu\raise1pt\vbox{\kern7pt\hbox{.}}\mkern2mu\raise4pt\hbox{.}\mkern2mu\raise7pt\hbox{.}\mkern1mu}}

\def\ep{\epsilon}

\def\la{\langle}
\def\ra{\rangle}

\def\vp{{\bf p}}

\def\lam{\lambda}
\def\al{\alpha}

\def\be{\begin{equation}}
\def\ee{\end{equation}}
\def\bea{\begin{eqnarray}}
\def\eea{\end{eqnarray}}

\def\ep{\epsilon}

\def\la{\langle}
\def\ra{\rangle}

\def\vp{{\bf p}}

\def\lam{\lambda}
\def\al{\alpha}

\def\be{\begin{equation}}
\def\ee{\end{equation}}
\def\bea{\begin{eqnarray}}
\def\eea{\end{eqnarray}}

\def\orcid#1{\kern .08em\href{https://orcid.org/#1}{\includegraphics[keepaspectratio,width=0.7em]{ORCID_iD.png}}}

\arxivnumber{2512.21642} 

\title{\boldmath Kaon T-even transverse-momentum-dependent distributions and form factors\\ 
in a self-consistent light-front quark model}

\author[a,b]{Yongwoo Choi,}
\author[c]{Ho-Meoyng Choi,}
\author[d,e]{Ahmad Jafar Arifi,}
\author[f]{and Chueng-Ryong Ji}
\affiliation[a]{School of AIT, Hoseo University, Asan 31499, Republic of Korea}
\affiliation[b]{Quantum AI Innovative Technology Institute (Q-AIT), Hoseo University, Asan 31499, Republic of Korea}
\affiliation[c]{Department of Physics Education, Teachers College, Kyungpook National University, Daegu 41566, Republic of Korea}
\affiliation[d]{Research Center for Nuclear Physics (RCNP), The University of Osaka, Ibaraki 567-0047, Japan}
\affiliation[e]{Advanced Science Research Center, Japan Atomic Energy Agency (JAEA), Ibaraki 319-1195, Japan}
\affiliation[f]{Department of Physics and Astronomy, North Carolina State University, Raleigh, NC 27695-8202, USA}

\emailAdd{ywchoi@hoseo.edu}
\emailAdd{homyoung@knu.ac.kr}
\emailAdd{aj.arifi01@gmail.com}
\emailAdd{crji@ncsu.edu}

\abstract{
We present a self-consistent full Bakamjian--Thomas (fBT) light-front quark model (LFQM) for the kaon and apply it to the electromagnetic (EM) and scalar form factors, as well as to the full set of unpolarized $T$-even transverse-momentum-dependent distributions (TMDs) and their collinear parton distribution functions (PDFs). 
The defining feature of the fBT-LFQM is the uniform implementation of the invariant mass $M_0$ in both the hadronic matrix elements and the associated Lorentz prefactors, yielding current-component--independent results for the observables considered here. 
For the EM form factor, we explicitly demonstrate that the fBT-LFQM produces a unique result from the $\gamma^+$, $\gamma^\perp$, and $\gamma^-$ projections, which couldn't be achieved in the previous partially BT-based LFQM (pBT-LFQM) unless an effective LF zero-mode contribution prone to the $\gamma^-$ extraction is separately taken into account.
Within the manifestly covariant field-theoretic Bethe--Salpeter (BS) framework, we explicitly demonstrate that the scalar form factor receives a nonvanishing LF zero-mode and illustrate the accountability of 
the corresponding effective LF zero-mode contribution through the BT-consistent replacement $M\to M_0(x,\bm k_\perp)$ in the 
LFQM.
At the model scale, the $x$-integrated transverse profile of the twist-2 TMD $f_1^q(x,\bm k_\perp)$ exhibits an exact Gaussian transverse-momentum dependence, 
while the higher-twist TMDs $f^{\perp q}$, $e^q$, and $f_4^q$ display systematic twist and flavor dependence. The fBT-LFQM satisfies the forward-limit sum rule for $f_4^q(x)$ and yields self-consistent extractions of the twist-3 PDF $e^q(x)$.
We further perform NNLO DGLAP evolution of the valence PDFs for the pion and kaon, comparing the evolved distributions and Mellin moments with lattice QCD, phenomenological extractions, and other theoretical approaches. Overall, the fBT-LFQM provides a unified BT-consistent
framework for meson form factors, TMDs, and PDFs with the zero-mode-sensitive projections consistently treated within the BT-based valence LF formalism.
}

\begin{document}
\maketitle
\flushbottom

\section{Introduction}

Exploring the quark–gluon substructure of light mesons is a longstanding challenge in hadron physics, offering crucial tests of our understanding 
of QCD in the nonperturbative regime. 
The kaon, like the pion, is a pseudo-Nambu--Goldstone boson arising from spontaneous chiral-symmetry breaking, and both share a simple valence quark–antiquark composition. 
However, unlike the pion, the kaon contains a strange quark, introducing explicit SU(3) flavor-symmetry breaking due to the large mass difference between 
the strange and light quarks. 
This difference introduces a heavier mass scale, modifies its internal momentum distributions, and makes the kaon a 
unique laboratory for studying how flavor-symmetry breaking influences hadronic structure. 

While pion and kaon observables—such as decay constants, 
distribution amplitudes (DAs)~\cite{Kock21,Bali:2019dqc,Cui:2020tdf,Hua22,CJ07}, 
electromagnetic form factors (EMFFs)~\cite{CJ06,Brom07,Nam08,CJ08,FLS09}, 
parton distribution functions (PDFs)~\cite{Nguyen:2011jy, Nam:2012vm, Hutauruk:2016sug,Watanabe:2017pvl, Bednar:2018mtf, Lan:2019rba,Wu:2022iiu,Hutauruk:2025wkn}
and generalized parton distributions (GPDs)~\cite{CJK01,CJK02,Pasq05,CHEN20,Vary21,Raya:2022,SH24,Luan24}—have been extensively studied, 
their transverse-momentum–dependent distributions (TMDs)~\cite{Lorce:2015,Lorce:2016, Puhan:2023ekt,Zhu23,Liu:2025mbl} remain comparatively less developed. 
The PDF and TMD studies cited above primarily refer to model calculations. Phenomenological extractions from data are of course also available: for the collinear pion PDFs, from the classic parametrizations of Refs.~\cite{Gluck:1991ey,Gluck:1997ww} to the recent global QCD analyses of the JAM Collaboration~\cite{Cao:2021aci,Barry:2021osv}, and for the unpolarized quark TMDs of the pion, from Refs.~\cite{Vladimirov:2019bfa,Cerutti:2022lmb,Barry:2023qqh}, including a recent valence--sea separation~\cite{Rossi:2025plb}. We compare with some of these extractions in Sec.~\ref{sec:Num}.
In particular, kaon studies are generally less advanced than those for the pion, with TMDs being especially underexplored. 
Extending such analyses to the kaon provides valuable insight into the spatial and momentum distributions of its quark constituents 
and deepens our understanding of nonperturbative QCD dynamics.

TMDs~\cite{Metz09,Diehl:2016epja,BC08} extend the concept of collinear PDFs by 
incorporating the intrinsic transverse momentum $\boldsymbol{k}_\perp$ of partons, thereby offering a three-dimensional description of hadrons in momentum space~\cite{Collins:1981uk,Collins:1984kg,Mulders:1995dh,Bacchetta:2006tn}. 
They encode information not only on the longitudinal momentum fraction $x$ of partons but also on correlations between $\boldsymbol{k}_\perp$ 
and the spin of the hadron or the parton itself. 
Comprehensive reviews of the TMD formalism and of its phenomenology can be found in Refs.~\cite{Boussarie:2023izj,Angeles-Martinez:2015sea}. Besides the light-cone quark model of Ref.~\cite{BC08}, quark and gluon TMDs have also been computed in the spectator framework~\cite{Bacchetta:2008af,Bacchetta:2020vty,Bacchetta:2024fci}.
Experimental \bl{progresses} in semi-inclusive deep inelastic scattering (SIDIS)~\cite{Bacchetta:2017jhep}, 
Drell--Yan (DY) processes~\cite{Metz09,Yuan10}, and high-energy hadron collisions 
at the LHC \bl{have} significantly advanced our knowledge of TMDs. 
These data have made possible a series of increasingly refined global extractions of the unpolarized quark TMDs~\cite{Scimemi:2017etj,Bertone:2019nxa,Scimemi:2019cmh,Bacchetta:2019sam,Bacchetta:2022awv,Bury:2022czx,Moos:2023yfa,Moos:2025sal}, which by now reach N$^3$LL--NNLO perturbative accuracy~\cite{Bacchetta:2022awv}, include the flavor dependence of the $k_\perp$ distributions~\cite{Bacchetta:2024qre}, and have been performed with neural-network parametrizations of the nonperturbative input~\cite{Bacchetta:2025ara}.
The upcoming Electron-Ion Collider (EIC)~\cite{AbdulKhalek:2021gbh,Arrington:2021biu} and other planned facilities, such as the EicC~\cite{Anderle:2021wcy}, COMPASS++/AMBER~\cite{Adams:2018pwt}, and the fixed-target LHCb/LHCSpin program~\cite{Aidala:2019pit}, promise to deliver precision data over broad kinematic regimes, enabling deeper exploration of the 
partonic structure of mesons and baryons.

Despite these developments, experimental data on the kaon’s parton structure remain scarce. 
At present, the only direct constraint comes from the valence-quark ratio $u^\pi/u^K$ measured by 
the NA3 Collaboration~\cite{Saclay-CERN-CollegedeFrance-EcolePoly-Orsay:1980fhh}.
Recent proposals suggest exploiting $J/\psi$ production in kaon-induced processes to extract additional information~\cite{Bourrely:2023yzi,Chang:2024rbs}. 
Lattice QCD has also begun providing calculations of meson PDFs~\cite{Detmold:2003tm, Sufian:2019bol, Fan:2021bcr, Lin:2020ssv, Lin:2025hka, Alexandrou:2021mmi, Salas-Chavira:2021wui,ExtendedTwistedMass:2024kjf,NieMiera:2025inn}, 
DAs, and other observables from first-principles, offering valuable theoretical inputs. These efforts, together with future experimental programs, 
are expected to significantly enhance our understanding of the kaon’s structure.

Among the various theoretical approaches used to study hadron structure,
the light-front quark model (LFQM)~\cite{LB80,BPP}, formulated in light-front dynamics (LFD), offers distinct advantages,
enabling a direct partonic interpretation of various parton distribution observables, including DAs, PDFs, GPDs, TMDs, and their unified representation through generalized transverse-momentum–dependent distributions, all of which can be expressed in terms of light–front wave functions (LFWFs).
In the LF formalism, TMDs can be related to the EMFF through their representation in terms of hadronic matrix elements of bilinear quark–field correlators.
As shown in Refs.~\cite{Lorce:2015,Lorce:2016}, unpolarized TMDs — at both leading and higher 
twist — can be formulated in the LFQM.  
In this approach, four T-even unpolarized TMDs, \{$f_1^q(x,{\bm k}_\perp)$, $f^{\perp q}(x,{\bm k}_\perp)$, 
$f_4^q(x,{\bm k}_\perp)$, $e^q(x,{\bm k}_\perp)$\}
form a complete set of $T$-even distributions accessible through the vector and scalar current 
projections for unpolarized targets, 
with a Lorentz–invariance relation among them 
holding in the absence of explicit gauge degrees of freedom.
Of these, \{$f_1^q(x,{\bm k}_\perp)$, $f^{\perp q}(x,{\bm k}_\perp)$, $f_4^q(x,{\bm k}_\perp)$\} are directly linked to the forward matrix elements 
of the electromagnetic vector current $\langle h| \bar{\psi}(0) \gamma^\mu \psi(0) |h \rangle$,
with $\mu = +$, $\perp$, and $-$ corresponding to twist-2, twist-3, and twist-4 TMDs, respectively.
Since these TMDs are related to the EMFF through specific current components, an extraction of the EMFF that is independent of the chosen current component is essential for obtaining physically consistent TMDs.
The remaining one, $e^q(x,{\bm k}_\perp)$, is the twist-3 TMD associated with the matrix element of the scalar current $\bar{\psi} \psi$.  
Importantly, higher-twist TMDs such as $f^{\perp q}$ and $e^q$ play a 
non-negligible role in DY observables~\cite{Metz09,Lorce:2016}, 
making their study important for understanding quark dynamics beyond the leading-twist picture.
\footnote{We note that the leading-twist $T$-odd Boer--Mulders function $h_1^{\perp q}(x,{\bm k}_\perp)$~\cite{Boer:1997nt}, which drives the
$\cos 2\phi$ asymmetry in DY processes~\cite{Lam:1978pu,Boer:1999mm},
requires the phase generated by the gauge link and hence goes beyond the lowest Fock-state constituent quark model without explicit gluon degrees of
freedom. Its description lies beyond the present $T$-even analysis and is
left for future work.}

Furthermore, Refs.~\cite{Lorce:2015,Lorce:2016} noted the difficulties in computing the twist-4 quark TMD $f^q_4(x, {\bm k}_\perp)$, which corresponds
to the minus component of the vector current, and in satisfying the sum rule for its twist-4 PDF $f^q_4(x)$. 
While the twist-4 TMD is hard to access experimentally, its correct extraction is very important to check the self-consistency of the model.
This issue is closely tied to the so-called LF zero-mode problem~\cite{LFzeroMB,LFzeroCJ,BCJ01,LFzero05} 
associated with the use of the ``bad" (minus) component of the current in LF calculations.
In the recent pion TMD analysis~\cite{Choi:2024ptc}, two of us explicitly identified the
effective LF zero-mode correction to $f_4^q$ and demonstrated that a consistent
treatment of this correction leads to current-component–independent results in the LFQM.
While the zero-mode structure of the electromagnetic current has been extensively
studied in the vector channel, the corresponding effective LF zero-mode correction for
the twist-3 TMD $e^q(x,{\bm k}_\perp)$ has not yet been identified, and
it has remained unclear whether the scalar form factor receives a nonvanishing
explicit LF zero-mode contribution. 

In this work, we provide the first explicit demonstration to the best of our knowledge within the
solvable manifestly covariant Bethe--Salpeter (BS) model employed here that the scalar form factor
indeed receives a nonvanishing explicit LF zero-mode contribution.
As shown in Appendix~\ref{app:scalarZM}, this explicity zero mode originates from
the endpoint behavior of the LF energy in the shrinking nonvalence region
and can be mapped onto an effective valence operator within the BS framework itself.
We illustrate in Appendix~\ref{app:scalarZM} that the treatment of the corresponding zero-mode-sensitive projections in the on-shell valence LFQM
requires a self-consistent implementation based on the Bakamjian--Thomas (BT)
construction~\cite{BT53,Poly10,CCKP} as discussed in the BT-based
LFQM~\cite{CJ14,CJ15,CJ17,Jafar1,Jafar2,Choi21,Choi:2024ptc,Choi:2025rto}.

In the BT-based LFQM,
the interaction effects between the quark and antiquark constituents are incorporated in the mass operator $M = M_0 + V_{q\bar{q}}$~\cite{CJ97,CJ99a,CJLR15,ACJO22,Nisha,Bhoo24}, that is the Casimir operator commuting with all other Poincaré operators. 
As the constituent quark and antiquark are taken to be on their respective
mass shells, their free two-body LF energy is defined by
\be
P^-_{\rm BT}\equiv p_q^-+p_{\bar q}^-
=
\frac{M_0^2+{\bm P}_\perp^2}{P^+}
=
\frac{m_q^2+{\bm p}_{q\perp}^2}{p_q^+}
+
\frac{m_{\bar q}^2+{\bm p}_{\bar q\perp}^2}{p_{\bar q}^+}.
\ee
Here, $M_0$ is the invariant mass of the free on-shell $q\bar q$ pair, and
$P^-_{\rm BT}$ denotes its kinematic LF energy. It should be distinguished
from the physical meson LF energy,
\[
P^-_{\rm phys}=\frac{M^2+{\bm P}_\perp^2}{P^+},
\]
where $M$ is the eigenvalue of the interacting mass operator. When current
matrix elements are evaluated in this on-shell valence representation, the
BT prescription requires the replacement $M\to M_0$ to be implemented
uniformly at the integrand level, not only in the matrix element but also in
all associated Lorentz structures. A partial implementation mixes the
on-shell constituent kinematics with Lorentz prefactors constructed from the
physical meson mass, leading to an inconsistent extraction of
zero-mode-sensitive observables and potentially violating rotational
covariance. This integrand-level prescription is specific to the on-shell
BT-based LFQM and should be distinguished from the off-shell field-theoretic
formulation \textit{\`a la} the BS model discussed in
Appendix~\ref{app:scalarZM}.

Depending on how this prescription is implemented, one may distinguish between
the partially BT-based (pBT) LFQM and the full BT-based (fBT) LFQM.
For instance, consider a matrix element written as
\be
\langle P'|\mathcal O^\mu|P\rangle={\cal P}^\mu{\cal F},
\ee
where ${\cal P}^\mu$ denotes the kinematic Lorentz structure and ${\cal F}$ is the
corresponding physical observable. For a given component $\mu$, the pBT-LFQM extracts ${\cal F}$ as
\begin{align}
{\cal F}_{\rm pBT}^{(\mu)}
=
\frac{1}{{\cal P}^\mu}
\bra{P'}\mathcal O^\mu\ket{P}_{\rm BT}.
\end{align}
In this case, the BT construction is implemented only in the matrix element, while
the associated Lorentz structure is not treated consistently at the integrand level.
For the zero-mode-sensitive projections,
this partial implementation leads to an inconsistency
associated with the effective LF zero-mode correction missing from the pBT extraction.

In contrast, the fBT prescription~\cite{CJ14,CJ15,CJ17,Jafar1,Jafar2,Choi21,Choi:2024ptc,Choi:2025rto} evaluates
\begin{align}\label{eq:3}
{\cal F}_{\rm fBT}^{(\mu)} = \bra{P'}\frac{\mathcal{O}^\mu }{{\cal{P}}^\mu}\ket{P}_{\rm BT},
\end{align}
so that the BT construction is implemented consistently at the integrand level,
including both the matrix element and the Lorentz structure.
The difference
$\Delta{\cal F}^{(\mu)} =
{\cal F}_{\rm fBT}^{(\mu)} - {\cal F}_{\rm pBT}^{(\mu)}$
signals the effective LF zero-mode correction missing 
from the pBT extraction for the given $\mu$ component.
This observable-level identification is consistent with the explicit LF zero-mode contribution explicitly derived in 
the covariant BS analysis, as discussed in Appendix~\ref{app:scalarZM}.

The main goal of this work is to investigate how explicit SU(3) flavor-symmetry
breaking is reflected in the unpolarized $T$-even partonic observables of the kaon
within the self-consistent fBT-LFQM framework. Building on our recent pion
analysis~\cite{Choi:2024ptc}, we extend the formalism to the kaon and include both
the vector-current sector associated with
$\{f_1^q,\,f^{\perp q},\,f_4^q\}$ and the scalar-current sector associated with the
twist-3 distribution $e^q$.

A central new aspect of the present work is the analysis of zero-mode
sensitivity in the scalar-current sector. The BS-model demonstration quoted
above motivates the BT-consistent extraction of the twist-3 TMD
$e^q(x,{\bm k}_\perp)$ within the on-shell valence LFQM. Thus, the scalar
form factor and $e^q$ provide a complementary testing ground, beyond the
electromagnetic sector, for the BT-consistent treatment of zero-mode-sensitive observables.
The comparative analysis of both kaon and pion further provides a useful setting
for examining flavor asymmetry, current-component independence, and zero-mode
structure in a system with unequal constituent masses. We also evolve the
twist-2 PDFs to higher scales and compare their Mellin moments and distributions
with other approaches in order to trace how the SU(3)-breaking pattern at the
hadronic scale appears after QCD evolution.

The paper is organized as follows. Section~\ref{sec:LFQM} reviews our LFQM and its BT-based implementation.
Section~\ref{sec:EMFF} presents the EM and scalar form factors of pseudoscalar mesons, contrasts the fBT-LFQM with the pBT-LFQM, and discusses
zero-mode sensitivity and its treatment in the $(-)$ component of the vector current and in the scalar channel.
Section~\ref{sec:TMD} defines the TMDs and PDFs at various twists, relates them to forward matrix elements in the vector and scalar channels, and compares with existing formulations in the literature.
Section~\ref{sec:Num} presents numerical results for the unpolarized T-even TMDs and PDFs at the model scale $\mu_0$, 
and performs NNLO DGLAP evolution of the twist-2 PDFs for comparison with other approaches.
Section~\ref{sec:summary} summarizes our findings.
Appendix~\ref{app:scalarZM} provides a covariant BS-model demonstration of the
nonvanishing explicit LF zero-mode contribution to the scalar form factor and clarifies the distinction between the explicit BS-model LF zero-mode 
contribution and the effective LF zero-mode correction signaled by the fBT--pBT difference within the BT-based LFQM, without implying an operator-level mapping between the two.
Appendix~\ref{app:forward} derives the relation between the TMD correlator and the forward matrix elements in the vector and scalar channels.

\section{Model description}
\label{sec:LFQM}
Our fBT-LFQM for $q\bar{q}$ bound-state mesons with total momentum $P$ describes the meson as a Fock state 
of non-interacting constituent quark $q$ and antiquark $\bar{q}$, with interaction effects incorporated through 
the mass operator $M = M_0 + V_{q\bar{q}}$~\cite{CJ97,CJ99a,CJLR15,ACJO22,Nisha,Bhoo24}, that is the Casimir operator commuting with all other Poincaré operators. 
Thus, this formulation preserves the Poincaré algebra at the level of the two-particle bound-state system
within the BT construction. 
The interaction dynamics are encoded in the LFWF which is the eigenfunction of the mass operator. 

The four-momentum $P$ of a meson in LF coordinates is expressed as $P=(P^+, P^-,  \bm{P}_\perp)$, where
$P^{+}=P^0 + P^3$ represents the longitudinal LF momentum, $P^-= P^0-P^3$ corresponds to the LF energy, and
$\bm{P}_\perp=(P^1, P^2)$ denotes the transverse momentum components. 
We adopt the Minkowski metric convention, in which the inner product of two four-vectors is defined as
$a\cdot b=\frac{1}{2} (a^+b^- + a^-b^+) - \bm{a}_\perp\cdot \bm{b}_\perp$.

The meson state $\ket{P}$ with momentum $P$ and total angular momentum $(J, J_z)$ is constructed as
\begin{align}\label{eq:KP}
\ket{P} =& \int \left[ \mathrm{d}^3\bm{p}_1 \right] \left[ \mathrm{d}^3\bm{p}_2 \right]  2(2\pi)^3 \delta^3 
\left(\bm{P}-\bm{p}_1 -\bm{p}_2 \right)\nonumber\\
        &\times \sum_{\lambda_1,\lambda_2} \Psi_{\lambda_1 \lambda_2}^{JJ_z}(x, {\bm k}_\perp)
        \ket{q(p_1,\lambda_1) \bar{q}(p_2,\lambda_2) },
\end{align}
where $p_{1(2)}$  are the on-shell four-momenta and $\lambda_{1(2)}$ the helicities of the constituent quark and antiquark,
respectively. The LF three-momentum is ${\bm p}=(p^{+},{\bm p}_{\perp})$  and the integration
measure is $\left[ d^3{\bm p} \right]=\frac{{\rm d}p^+ {\rm d}^2\bm{p}_{\perp}}{2(2\pi)^3}$.
The LF relative momentum variables $(x, \bm{k}_\perp)$ are defined by $x_i=p^+_i/P^+$ and 
${\bm k}_{i\perp}={\bm p}_{i\perp}-x_i{\bm P}_\perp$,
with $\sum_i x_i=1$ and $\sum_i {\bm k}_{i\perp}=0$. 
In this work, we set $x_1\equiv x$ and $\bm{k}_{1\perp}\equiv  {\bm k}_{\perp}$. Thus, the LFWF depends on the internal variables
$(x, {\bm k}_\perp)$, while the single-particle basis states are labeled by the on-shell LF momenta ${\bm p}_1$ and ${\bm p}_2$.

We normalize the meson state $\ket{P}$ as~\cite{Choi:2025rto} 
\bea\label{eq:statenorm}
\Braket{P'|P}=2 (2\pi)^3 P^+  \delta^{(3)}({\bm P'}-{\bm P}),
\eea
so that the LFWF of a pseudoscalar meson $(J=0, J_z=0)$ satisfies the normalization condition
\be\label{eq:Psinorm}
\int^1_0 dx\int \frac{d^2{\bm k}_\perp}{2(2\pi)^3} \sum_{\lambda_1\lambda_2}
\Psi^{\dagger 00}_{\lam_1{\lam_2}}(x,{\bm k}_{\perp})\Psi^{00}_{\lam_1{\lam_2}}(x,{\bm k}_{\perp})=1.
\ee
The momentum-space LFWF of a pseudoscalar meson can be written as
\begin{align}\label{eq:6}
\Psi^{00}_{\lam_1{\lam_2}}(x,{\bm k}_{\perp})
=\phi(x,{\bm k}_{\perp}){\cal R}^{00}_{\lam_1{\lam_2}}(x,{\bm k}_{\perp}),
\end{align}
where $\phi(x, {\bm k}_{\perp})$ is the radial wave function and 
${\cal R}^{00}_{\lam_1{\lam_2}}$ denotes the spin-orbit wave function obtained via the 
interaction-independent Melosh transformation~\cite{Melosh}.

The covariant form of ${\cal R}^{00}_{\lam_1{\lam_2}}$ is given by~\cite{SLF2,SLF3}
\begin{equation}\label{eq:2.5}
\mathcal{R}^{00}_{\lambda_1{\lambda_2}}
= \frac{\bar{u}_{\lambda_1}(p_1)\gamma_5 v_{\lambda_2}(p_2)}{\sqrt{2} \sqrt{M_0^2 - (m_1 -m_2)^2}},
\end{equation}
where $m_{1(2)}$ is the mass of quark (antiquark), and 
\begin{equation}\label{eq:M0}
M^2_{0} = \frac{ \bm{k}^{2}_\perp + m^2_1}{x} + \frac{ \bm{k}^{2}_\perp + m^2_2}{1-x},
\end{equation}
is the invariant mass squared of the $q\bar{q}$ system.

By explicitly evaluating the above covariant expression with Dirac spinors~\cite{LB80,SLF2,SLF3}, one obtains the matrix
representation of $\mathcal{R}^{00}_{\lambda_1{\lambda_2}}$ in terms of $(x,{\bm k}_\perp)$ as
\begin{eqnarray}\label{eq:RMat}
\mathcal{R}^{00}_{\lambda_1{\lambda_2}}
= \frac{1}{\sqrt{2}\sqrt{\mathcal{A}^2+\bm{k}_\perp^2}}
\begin{pmatrix}
-k^L & \mathcal{A} \\ 
-\mathcal{A} & -k^R
\end{pmatrix},
\end{eqnarray}
where $k^{R(L)} = k_x \pm i k_y$ and ${\cal A}= (1-x)m_1 + x m_2$.
We note that ${\cal R}^{00}_{\lambda_1{\lambda_2}}$ satisfies the normalization condition
$\sum_{\lambda_1,\lambda_2} \mathcal{R}^{00\dagger}_{\lambda_1\lambda_2}
\mathcal{R}^{00}_{\lambda_1\lambda_2}=1$.

The expression in Eq.~(2.6) follows from the on-mass-shell constituent
kinematics and the free two-body LF relation
$P^-_{\rm BT}=p_1^-+p_2^-$.
This leads to the relation between $x$ and the $z$--component $k_z$ 
of the quark’s three-momentum~\cite{SLF2,SLF3}:
\bea\label{eq:k_z}
    k_z = \left( x - \frac{1}{2} \right) M_0 + \frac{(m^2_2 -m^2_1)}{2M_0},
\eea 
from which the Jacobian factor for the transformation $(x,{\bm k}_\perp) \to (k_z,{\bm k}_\perp)$ is
\begin{equation}
\frac{\partial k_z}{\partial x} = \frac{M_0}{4x(1-x)} \left[ 1 - \frac{ (m_1^2 - m_2^2)^2}{M_0^4} \right].
\end{equation}

The quark--antiquark interaction is incorporated through the mass operator~\cite{BT53,Poly10,CCKP}   
$M=M_0 + V_{q{\bar q}}$, whose eigenvalues yield the meson mass spectrum.
In our BT-based LFQM, the radial wave function $\phi(x,{\bm k}_\perp)$ is treated as a variational trial function, optimized 
for a QCD-motivated effective Hamiltonian, $H_{q\bar{q}}\ket{\Psi}=(M_0 + V_{q\bar{q}})\ket{\Psi}= M_{q{\bar q}}\ket{\Psi}$, 
where $M_{q{\bar q}}$ denotes the $q{\bar q}$ meson mass eigenvalue.
Comprehensive meson spectroscopy analyses in this approach are reported in Refs.~\cite{CJ97,CJ99a,CJLR15,ACJO22,Nisha,Bhoo24}.

For the $1S$ meson state, we adopt a Gaussian form:
\begin{align}\label{eq:8}
\phi(x,{\bm k}_{\perp})=
\frac{4\pi^{3/4}}{\beta^{3/2}} 
\sqrt{\frac{\partial k_z}{\partial x}} \,
\exp\!\left(-\frac{\bm{k}^2}{2\beta^2}\right),
\end{align}
where $\bm{k}^2 = \bm{k}_\perp^2 + k_z^2$, and $\beta$ is a variational parameter fixed from mass 
spectrum fits~\cite{CJ97,CJ99a,CJLR15,ACJO22,Nisha,Bhoo24}.
The state normalization in Eq.~\eqref{eq:statenorm}, together with the unitarity of the spin--orbit wave function, ensures that $\phi$ satisfies
\begin{align}\label{eq:9}
\int^1_0 dx \int \frac{d^2{\bm k}_\perp}{2(2\pi)^3}~
|\phi(x,{\bm k}_{\perp})|^2 = 1.
\end{align}
In our numerical analysis for the pseudoscalar mesons, we employ the model parameters determined within the linear confining potential model~\cite{CJ97}.  
The model parameters are given by  
\begin{equation}\label{eq:params}
(m_q,\, m_s,\, \beta_{q\bar{q}},\, \beta_{q\bar{s}}) = (0.22,\, 0.45,\, 0.3659,\, 0.3886)\,{\rm GeV},
\end{equation}
where $q=u=d$ is assumed under SU(2) flavor symmetry.  
The parameters $(m_q, \beta_{q\bar{q}})$ correspond to the pion ($\pi^+ = u\bar{d}$), while $(m_q, m_s, \beta_{q\bar{s}})$ are used for the kaons ($K^+=u\bar{s}$ and $K^0=d\bar{s}$).  
These parameter sets provide reasonable descriptions of the ground-state pseudoscalar mesons.  
For the pion, the model predicts the charge radius and decay constant as $\langle r^2_{\pi^+} \rangle^{1/2} = 0.654$~fm and $f_\pi = 130$~MeV, 
which are consistent with the experimental values of $(0.659 \pm 0.004)$~fm and $131$~MeV, respectively~\cite{PDG2024}.  
For the kaon, the predictions are $\langle r^2_{K^+} \rangle^{1/2} = 0.595$~fm and $f_K = 161$~MeV, in reasonable agreement with the experimental data of $(0.560 \pm 0.031)$~fm and $(155.7 \pm 0.3)$~MeV, respectively~\cite{PDG2024}.

\section{Electromagnetic and scalar form factors}
\label{sec:EMFF}
\subsection{Comparison of fBT-LFQM and pBT-LFQM}
Before discussing EMFFs, we outline the formalism for weak transitions between two pseudoscalar mesons, 
specifically the process in which an initial meson with momentum $P$ and mass $M$ transitions to a final meson with momentum $P'$ and mass $M'$.
The four-momentum transfer is defined as $q = P' - P$. 
The corresponding weak-current matrix element  admits the covariant decomposition
\bea\label{eq:Jweak}
\la\mathcal{J}^\mu_W\ra &\equiv& \langle P'|\bar{q}(0)\gamma^\mu(1-\gamma_5) q(0)|P\rangle
\nonumber\\
&=& \left( {\bar P}^\mu  - q^\mu \frac{{\bar P}\cdot q}{q^2} \right) F(q^2) 
+ q^\mu \frac{{\bar P}\cdot q}{q^2} H(q^2),\quad 
\eea
where ${\bar P}=P + P'$ and ${\bar P}\cdot q=M^{\prime 2} - M^2$.
The Lorentz structure multiplying $F(q^2)$ is transverse to $q^\mu$ and therefore 
satisfies the transversality condition for conserved currents,
while the term multiplying $H(q^2)$ 
is longitudinal
and vanishes for conserved currents (e.g.\ EM current) and contributes only when the current is not conserved.

In the EM transition for a pseudoscalar meson, current conservation ($q_\mu \langle J^\mu_{\rm em}\rangle=0$) alone is sufficient 
to eliminate the longitudinal structure: contracting Eq.~\eqref{eq:Jweak} with $q_\mu$ gives
$q_\mu\langle \mathcal{J}^\mu_{\rm em}\rangle = (\bar P\!\cdot q)\,H(q^2)$.
Hence, for a conserved current one must have $H(q^2)=0$ whenever
$\bar P\!\cdot q\neq0$, while for the physical elastic process ($M=M'$) the
longitudinal term vanishes identically and $H(q^2)$ drops out of the decomposition.
In particular, within the fBT-LFQM the Lorentz structure is expressed in terms of the invariant masses $M_0$ and $M_0'$ 
constructed from internal momenta. As a result, even  for an elastic EM process ($M=M'$), 
the quantity entering the Lorentz prefactor remains nonvanishing point-by-point in the integrand, since it
effectively behaves as $M_0^{\prime 2} - M_0^2 \neq 0$.
After integration, however, current conservation enforces transversality of the full matrix element,
and only the gauge-invariant term survives:
\begin{eqnarray}\label{eq:formfactor}
\la\mathcal{J}^\mu_\mathrm{em}\ra
&\equiv&
\langle P'|\bar{q}(0)\gamma^\mu q(0)| P\rangle
=
\left( {\bar P}^\mu  - q^\mu \frac{{\bar P}\cdot q}{q^2} \right)
F_{\rm em}(Q^2),
\end{eqnarray}
where $Q^2=-q^2$.
In contrast to the pBT implementation, where the Lorentz prefactor is taken as 
${\bar P}^\mu$, the fBT formulation employs the explicitly transverse structure
${\bar P}^\mu  - q^\mu \frac{{\bar P}\cdot q}{q^2}$,
which is orthogonal to $q^\mu$ by construction. 
Its use at the integrand level should be understood as a BT-consistent
extraction prescription rather than as a direct consequence of current
conservation in the physical equal-mass elastic kinematics. With this
prescription, the integrated matrix element is conserved and the extracted
form factor is current-component independent.

For the analysis of the twist–3 TMD $e^q(x,\bm{k}_\perp)$ associated with the scalar current~\cite{Lorce:2015,Lorce:2016}, 
we also study the scalar form factor of a pseudoscalar meson. We use the 
convention~\cite{JAFFE1992527,Anatoli,Gasser:1990ap},\footnote{The direct convention $\langle{\cal J}_S\rangle=f_S(Q^2)$ is also used
in the literature~\cite{Aoki09,ETM22,Wang22,Puhan25}. However, this convention is
not used in the extraction of $e^q(x,\bm{k}_\perp)$ considered here.}
\begin{equation}
\label{eq:massfactFS}
\langle{\cal J_S}\rangle \equiv \langle P'|\,\bar q(0)\,\mathbf{1}\,q(0)\,|P\rangle\;\equiv\;2M\,F_S(Q^2).
\end{equation}

The calculations of the hadronic matrix elements for the EM vector and scalar currents,
$\langle P'|\bar{q}(0)\,\Gamma\,q(0)|P\rangle$ with $\Gamma\in\{\gamma^\mu,\mathbf{1}\}$,
are evaluated in both the fBT-LFQM  and the pBT-LFQM. 
The essential difference between the two schemes lies in the
treatment of the Lorentz prefactors. In the pBT-LFQM one retains the physical masses
(with $M=M'$ for elastic EM kinematics) in the prefactors, whereas in the fBT-LFQM the BT construction
is applied uniformly to both the matrix element and the Lorentz prefactor to ensure consistency.
This distinction reflects the different implementations of the BT construction in the LFQM framework.

In the pBT-LFQM, the form factor is extracted as
\be\label{eq:conForm}
F^{[\Gamma]}_{\rm pBT}(Q^2) 
= \frac{1}{{\cal P}^{[\Gamma]}_{\rm pBT}}\langle P'|\bar{q}(0)\Gamma q(0)| P\rangle_{\rm BT},
\ee
with
\bea\label{eq:conFormP}
{\cal P}^{[\gamma^\mu]}_{\rm pBT} &=& {\bar P}^\mu,\qquad
{\cal P}^{[{\bf 1}]}_{\rm pBT} = 2 M.
\eea
By contrast, in the fBT-LFQM we extract
\be\label{eq:BTForm}
F^{[\Gamma]}_{\rm fBT}(Q^2) 
= \langle P'| \frac{\bar{q}(0)\Gamma q(0)}{{\cal P}^{[\Gamma]}_{\rm fBT}} |P\rangle_{\rm BT},
\ee
with the covariant prefactor
\bea\label{eq:BTFormP}
{\cal P}^{[\gamma^\mu]}_{\rm fBT} &=& {\bar P}^\mu  - q^\mu \frac{{\bar P}\cdot q}{q^2},\qquad
{\cal P}^{[{\bf 1}]}_{\rm fBT} = 2 M_0,
\eea
where $M_0$ is evaluated at the integrand level,
and the BT replacements $M\to M_0(x,\bm{k}_\perp)$ and $M'\to M_0(x,\bm{k}_\perp')$ are made
inside the $(x,\bm{k}_\perp)$ integration for both the matrix element and the prefactor.
This uniform implementation ensures consistency between the two sides of the matrix-element decomposition 
and effectively incorporates, at the level of the observable, the effective LF zero-mode correction missing from the pBT extraction, 
thereby yielding current-component-independent observables.

As shown below, this treatment of the Lorentz prefactor 
for the minus component of the vector current is consistently applied to 
the scalar channel. This provides a nontrivial consistency test of the
BT-based formulation and its treatment of effective zero-mode corrections.

\subsection{Form Factors in the BT-based LFQM Framework}
To compute the EMFF and scalar form factor, we adopt the Drell--Yan--West ($q^+=0$) frame
with $\bm{P}_\perp =0$, in which $q^2=-\bm{q}^2_\perp\equiv -Q^2$. 
In this frame, the four-momenta of the initial and final meson states are 
\begin{eqnarray}\label{eq:DYW}
P &=& \left(P^+, \frac{M^2}{P^+}, {\bm 0}_\perp \right),\quad
P' =\left(P^+,  \frac{M'^2 + \bm{q}^2_\perp}{P^+}, \bm{q}_\perp \right), \quad
q = \left (0, \frac{M^{\prime 2} - M^2 + \bm{q}^2_\perp}{P^+}, \bm{q}_\perp \right).
\end{eqnarray}

The LF on-shell momenta $p_{1(2)}$ of the incoming quark (antiquark) and $p'_{1(2)}$ of the outgoing quark (antiquark)
for the transition $\ket{P(q\bar q)}\to \ket{P' (q' \bar q')}$ are given by
\be\label{eq:15}
p^+_i = p^{\prime +}_i =x_i P^+, \quad
{\bm p}_{i\perp} = x_i {\bm P}_\perp + {\bm k}_{i\perp}, \quad
{\bm p}'_{i\perp} = x_i {\bm P}'_\perp + {\bm k}'_{i\perp},
\ee
where $x_1=x$ and ${\bm k}_{1\perp}= {\bm k}_\perp$.
Since the spectator antiquark ($i=2$) satisfies $p^+_{2}=p'^+_{2}$ and ${\bm p}_{2\perp}={\bm p}'_{2\perp}$,
it follows that ${\bm k}'_\perp = {\bm k}_\perp + (1-x) {\bm q}_\perp$.

In both BT-based LFQMs, which employ a noninteracting $q{\bar q}$ representation,
the one-loop contribution to the matrix elements of the vector and scalar currents
is obtained by convoluting the initial and final state LFWFs~\cite{Choi:2024ptc}:
\begin{align}\label{eq:16}
\langle P'|\bar{q}(0)\Gamma q(0)| P\rangle_{\rm BT} =
\int [d^3{{\bm p}_1}]\
\phi'(x,{\bm k}^\prime_\perp)  \phi(x,{\bm k}_\perp) S^{[\Gamma]},
\end{align}
where $[d^3{\bm p}_1]\equiv\frac{P^+ \dd x\dd^2{\bm k}_\perp}{2(2\pi)^3}$ and
$S^{[\Gamma]}$ denotes the  spin factors for the vector and scalar currents, given by
\begin{align}\label{eq:17}
S^{[\Gamma]} &=
 \mathcal{R}^\dagger_{\lambda'_1\lambda_2} 
 \bigg[
\frac{\bar u_{\lambda'_1}(p'_1)}{\sqrt{p^{\prime +}_1}} \Gamma \frac{ u_{\lambda_1}(p_1)}{\sqrt{p_1^+}}
    \bigg]
\mathcal{R}_{\lambda_1\lambda_2},
\end{align}
with an implicit sum over quark helicities. 
The explicit Dirac matrix elements of the vector and scalar currents for helicity spinors are listed in Table~\ref{tab:dirac}. 

\begin{table*}[t]
\centering
\renewcommand{\arraystretch}{1.5}
\setlength{\tabcolsep}{8pt}
\begin{tabular}{ccc} \hline\hline
  & \multicolumn{2}{c}{Helicity ($\lambda\to\lambda'$)} \\
 Matrix elements    & $\uparrow~\to~\uparrow$ & $\uparrow~\to~\downarrow$ \\
         & $\downarrow~\to~\downarrow$ & $\downarrow~\to~\uparrow$ \\
\hline
$\frac{{\bar u}_{\lambda'}(p'_1)}{\sqrt{{p'_1}^+}}~\gamma^+~\frac{u_\lambda(p_1)}{\sqrt{{p_1}^+}}$ & 2 & 0  \\
$\frac{{\bar u}_{\lambda'}(p'_1)}{\sqrt{{p'_1}^+}}~\gamma^-~ \frac{u_\lambda(p_1)}{\sqrt{{p_1}^+}}$  & $\frac{2}{{p^{\prime +}_1}{p^+_1}}(\bm{p}'_{1\perp}\cdot \bm{p}_{1\perp} \pm i \bm{p}'_{1\perp}\times \bm{p}_{1\perp} + m_1^{2})$ 
& $\mp\frac{2 m_1}{{p^{\prime +}_1}{p^+_1}}[({p'}_{1}^{R(L)}-p_{1}^{R(L)})]$   \\
$\frac{{\bar u}_{\lambda'}(p'_1)}{\sqrt{{p'_1}^+}}~{\bm\gamma}^i_{\perp}~ \frac{u_\lambda(p_1)}{\sqrt{{p_1}^+}}$   
& $\frac{ \bm{p'}^i_{1\perp} \mp i\epsilon^{ij}\bm{p'}^j_{1\perp}}{p^{\prime +}_1} + \frac{ \bm{p}^i_{1\perp} \pm i\epsilon^{ij}\bm{p}^j_{1\perp}}{p^+_1}$
&   $\mp m_{1}\left(\frac{p^{\prime +}_1 - p^+_1}{{p^{\prime +}_1}{p^+_1}}\right)(\delta^{ix}\pm i\delta^{iy})$ \\
$\frac{{\bar u}_{\lambda'}(p'_1)}{\sqrt{{p'_1}^+}}~\bm{1}~\frac{u_\lambda(p_1)}{\sqrt{{p_1}^+}}$   
& $\dfrac{m_1 ({p^{\prime +}_1}+{p^+_1})}{{p^{\prime +}_1}{p^+_1}}$ 
&   $\mp\dfrac{({p^+_1} {p^{\prime}_1}^{R(L)} - {p^{\prime +}_1} {p^{R(L)}_1})}{{p^{\prime +}_1}{p^+_1} }$  \\
\hline\hline
\end{tabular}
\caption{Dirac matrix elements for the helicity spinors.
Note that $p^{L(R)} q^{R(L)} = \bm{p}_\perp\cdot\bm{q}_\perp \pm i \bm{p}_\perp\times\bm{q}_\perp$. \label{tab:dirac} }
\end{table*}

\begin{table*}[t]
\centering
\renewcommand{\arraystretch}{2.5}
\setlength{\tabcolsep}{8pt}
\begin{tabular}{cccc} \hline\hline 
$\Gamma=(\gamma^\mu, {\bf 1})$ & $S^{[\Gamma]}$ 
& $\mathcal{P}^{[\Gamma]}_{\rm pBT}$
& $\mathcal{P}^{[\Gamma]}_{\rm fBT}$ \\
\hline
$\gamma^+$ & $\dfrac{2 (\mathcal{A}^2 + \bm{k}_\perp\cdot \bm{k}'_\perp)}
{\sqrt{\mathcal{A}^2 + \bm{k}^2_\perp}\sqrt{\mathcal{A}^2 + \bm{k}'^2_\perp}}$  
& $2 P^+$ 
& $2 P^+$ \\
${\bm\gamma}^j_\perp$ & $\dfrac{(\mathcal{A}^2 + \bm{k}_\perp\cdot \bm{k}'_\perp) 
(\bm{q}^j_\perp + 2 \bm{k}^j_\perp)}
{x P^+ \sqrt{\mathcal{A}^2 + \bm{k}^2_\perp}\sqrt{\mathcal{A}^2 + \bm{k}'^2_\perp}}$  
& $\bm{q}^j_\perp$
& $\bm{q}^j_\perp\left(1 + \dfrac{M^2_-}{\bm{q}^2_\perp}\right)$ \\
$\gamma^-$   
& $\dfrac{2\mathcal{A}m_1 (1-x) \bm{q}^2_\perp + 2 \mathcal{B}}
{(x P^+)^2\sqrt{\mathcal{A}^2 + \bm{k}^2_\perp}\sqrt{\mathcal{A}^2 + \bm{k}'^2_\perp}}$ 
& $\dfrac{2 M^2 + {\bm q}^2_\perp}{P^+}$
& $\dfrac{ 2 M^{\prime 2}_0 \bm{q}^2_\perp + \bm{q}^4_\perp + (M^2_-)^2}{\bm{q}^2_\perp P^+}$ \\
${\bf 1}$ 
& $\dfrac{2 m_1 (\mathcal{A}^2 + \bm{k}_\perp\cdot \bm{k}'_\perp)+{\mathcal{A} (1-x) \bm{q}^{2}_\perp}}
{x P^+ \sqrt{\mathcal{A}^2 + \bm{k}^2_\perp}\sqrt{\mathcal{A}^2 + \bm{k}'^2_\perp}}$ 
& $2M$ 
& $2M_0$ \\
\hline\hline
\end{tabular}
\caption{Spin factors $S^{[\Gamma]}$ and Lorentz prefactors 
 $\mathcal{P}^{[\Gamma]}_{\rm pBT}$ and $\mathcal{P}^{[\Gamma]}_{\rm fBT}$
for the vector ($\Gamma=\gamma^\mu$) and scalar ($\Gamma=\mathbf{1}$) currents, where  
$\mathcal{B} 
= ({\cal A}^2 + {\bm k}_\perp\cdot{\bm k}'_\perp)(\bm{k}_\perp^2+ {\bm{k}_\perp \cdot \bm{q}_\perp}  + m_1^2) + (1-x) |\bm{k}_\perp\times\bm{q}_\perp|^2$, and $M^2_{-}=M_0^{\prime 2} - M_0^2$. \label{tab:helicity} }
\end{table*}

Table~\ref{tab:helicity} summarizes the spin factors $S^{[\Gamma]}$ and the 
Lorentz prefactors $\mathcal{P}^{[\Gamma]}_{\rm pBT}$ and $\mathcal{P}^{[\Gamma]}_{\rm fBT}$
for all components of the vector current 
$(\Gamma=\gamma^\mu)$ and for the scalar current $(\Gamma=\mathbf{1})$. 
In both BT-based LFQMs, the EM current matrix element is computed within the same
BT framework using on-shell quark kinematics. 
As a result, the spin factor $S^{[\Gamma]}$ is identical in the two formulations.
For $S^{[\gamma^\mu]}$, the $\gamma^+$ and ${\bm\gamma}^\perp$ components receive contributions 
only from helicity–nonflip terms, whereas the $\gamma^-$ component contains both a helicity–nonflip term proportional 
to $\mathcal{B}$ and a helicity–flip term proportional to $\mathcal{A}$.

The EMFF $(\Gamma=\gamma^\mu)$ and scalar FF $(\Gamma={\bf 1})$ in our fBT-LFQM are defined by
\bea\label{eq:19}
F^{[\Gamma]}_{\rm fBT} (Q^2)
&=& \int [d^3{\bm p}_1]
\phi'(x, {\bm k}^\prime_\perp)  \phi(x, {\bm k}_\perp)  
\frac{S^{[\Gamma]}}{{\cal P}^{[\Gamma]}_{\rm fBT}},
\eea
where all occurrences of the meson masses $M^{(\prime)}$ in ${\cal P}^{[\Gamma]}_{\rm fBT}$ of Table~\ref{tab:helicity}
are replaced by the corresponding invariant masses $M^{(\prime)}_0$, 
with $M'_0=M_0(x,{\bm k}'_\perp)$. 
For the EMFF calculation, we find that all three current components yield identical results, confirming the 
self-consistency of the fBT-LFQM: 
$F^{[\gamma^+]}_{\rm fBT} =F^{[{\bm\gamma}^\perp]}_{\rm fBT} =F^{[\gamma^-]}_{\rm fBT}$.
These three extractions, however, become equal in different ways.
For $\Gamma={\bm\gamma}^\perp$, the equality with the $\gamma^+$ result
already holds at the integrand level. Upon contraction with
${\bm q}_\perp$, the Lorentz prefactor gives
\[
{\cal P}^{[{\bm\gamma}^\perp]}_{\rm fBT}\cdot{\bm q}_\perp
=
\frac{{\bm q}_\perp^2
+2{\bm k}_\perp\cdot{\bm q}_\perp}{x},
\]
which cancels identically against the corresponding factor in
$S^{[{\bm\gamma}^\perp]}\cdot{\bm q}_\perp$, leaving precisely the
$\gamma^+$ integrand,
\[
\frac{{\cal A}^2+{\bm k}_\perp\cdot{\bm k}'_\perp}
{P^+\sqrt{{\cal A}^2+{\bm k}_\perp^2}
\sqrt{{\cal A}^2+{\bm k}'_\perp{}^2}}.
\]

For the zero-mode-sensitive $\gamma^-$ projection, by contrast, the
integrands of $F^{[\gamma^+]}_{\rm fBT}$ and
$F^{[\gamma^-]}_{\rm fBT}$ are different pointwise in
$(x,{\bm k}_\perp)$. Their equality is recovered only after the
$(x,{\bm k}_\perp)$ integration, and holds for all values of $Q^2$.
This nontrivial equality is realized through the consistent BT
implementation in both the hadronic matrix element and the associated
Lorentz structure, and is not reproduced in the pBT extraction, where
the BT construction is applied only to the matrix element.

By contrast, in the pBT-LFQM the EMFF is obtained as
\be\label{eq:19con}
F^{[\Gamma]}_{\rm pBT} (Q^2) 
= \frac{1}{{\cal P}^{[\Gamma]}_{\rm pBT}}\int [d^3{\bm p}_1]
\phi'(x, {\bm k}^\prime_\perp)  \phi(x, {\bm k}_\perp) S^{[\Gamma]},
\ee
where the physical mass $M$ in $\mathcal{P}^{[\gamma^-]}_{\rm pBT}$ and 
$\mathcal{P}^{[{\bf 1}]}_{\rm pBT}$ is a constant and
does not depend on the internal kinematics. In this case, the $\gamma^+$ and $\gamma^\perp$ 
components reproduce the current-component independent fBT result, 
$F_{\rm pBT}^{[\gamma^+]}=F_{\rm pBT}^{[{\bm\gamma}^\perp]}=F_{\rm fBT}^{[\gamma^\mu]}$,
whereas the $\gamma^-$ component 
requires an effective LF zero-mode correction within the pBT kinematical treatment.
Consequently, without this effective correction,
one obtains $F_{\rm pBT}^{[\gamma^-]} \neq F_{\rm fBT}^{[\gamma^-]}$.

Finally, Eqs.~\eqref{eq:19} and~\eqref{eq:19con} give the quark-sector contribution to the EMFF
obtained in fBT- and pBT-LFQMs, respectively.
In this contribution, we identify
$m_1=m_q$ and $m_2=m_{\bar q}$, and write it as $F^{[\gamma^\mu]}_{\rm fBT (pBT)}(Q^2, m_q, m_{\bar q})$.  
The corresponding antiquark-sector contribution is obtained by interchanging 
$m_q \leftrightarrow m_{\bar q}$, namely
$F^{[\gamma^\mu]}_{\rm fBT(pBT)}(Q^2,m_{\bar q},m_q)$.
Thus, the full EMFF is
\bea
F_{\rm em} (Q^2) &=& e_q F^{[\gamma^\mu]}_{\rm fBT(pBT)}(Q^2, m_q, m_{\bar q}) 
+ e_{\bar q} F^{[\gamma^\mu]}_{\rm fBT(pBT)}(Q^2, m_{\bar q}, m_q).
\eea

\begin{table*}[t]
\centering
\renewcommand{\arraystretch}{2.5}
\setlength{\tabcolsep}{8pt}
\begin{tabular}{lccc} \hline\hline 
$\Gamma$ & $\big\langle P\big| \bar{q}(0)\,\Gamma\, q(0)\big|P\big\rangle_{\rm BT}$
& $\mathcal{P}^{[\Gamma]}_{\rm pBT}(0)$
& $\mathcal{P}^{[\Gamma]}_{\rm fBT}(0)$ \\
\hline
$\gamma^+$ & $\int [\dd^3 {\bm p}_1] |\phi(x,{\bm k}_\perp)|^2 \cdot 2$ & $2P^+$ & $2P^+$\\
${\bm\gamma}^j_\perp$ 
&  $\int [\dd^3 {\bm p}_1] |\phi(x,{\bm k}_\perp)|^2 \cdot \left(\frac{2 {\bm k}^j_\perp}{x P^+}\right)$ 
&0 & $\dfrac{2 \bm{k}^j_{\perp}}{x}$ \\
$\gamma^-$ & $\int [\dd^3 {\bm p}_1] |\phi(x,{\bm k}_\perp)|^2 \cdot \frac{2 ({\bm k}^2_\perp + m^2_1)}{(x P^+)^2}$ 
& $\dfrac{2M^2}{P^+}$ & $\dfrac{2\left(M^2_0 + \dfrac{ 2 |{\bm k}_\perp|^2\cos^2\theta }{x^2} \right)}{P^+}$\\
${\bf 1}$ 
& $\int [\dd^3 {\bm p}_1] |\phi(x,{\bm k}_\perp)|^2 \cdot \left(\frac{2 m_1}{x P^+} \right)$ & $2M$ & $2M_0$ \\
\hline\hline
\end{tabular}
\caption{Forward limit ($Q^2\to 0$) expressions for $\langle P|\bar q(0)\Gamma q(0)|P\rangle_{\rm BT}$ and the corresponding 
Lorentz prefactors $\mathcal{P}^{[\Gamma]}_{\rm fBT}(0)$ and $\mathcal{P}^{[\Gamma]}_{\rm pBT}(0)$ in the 
fBT- and pBT-LFQMs.\label{tab:forward} }
\end{table*}
The pBT- and fBT-LFQM results for the matrix elements
$\langle P|\,\bar q(0)\Gamma\,q(0)\,|P\rangle_{\rm BT}$ and for the associated Lorentz prefactors
$\mathcal{P}^{[\Gamma]}_{\rm pBT}(0)$ and $\mathcal{P}^{[\Gamma]}_{\rm fBT}(0)$ in the forward limit ($Q^2\!\to\!0$) are summarized in
Table~\ref{tab:forward}. For the transverse component $\Gamma={\bm\gamma}^\perp$, the entries in
Table~\ref{tab:forward} should be understood in the contracted forward-limit sense:
one first contracts the transverse current with $\bm q_\perp$ and then takes
$Q^2\to 0$. At the integrand level, this yields terms proportional to
$\bm k_\perp\!\cdot\!\bm q_\perp$, and the expressions listed in
Table~\ref{tab:forward} are obtained by factoring out the common $\bm q_\perp$
dependence, leaving the effective $\bm k_\perp^j$ structure.

We note that the angle $\theta$ appearing in $\mathcal{P}^{[\gamma^-]}_{\rm fBT}(0)$
originates from the relative orientation between ${\bm k}_\perp$ and ${\bm q}_\perp$,
through
${\bm k}_\perp \cdot {\bm q}_\perp
= |{\bm k}_\perp|\,|{\bm q}_\perp|\cos\theta$.
This $\theta$ dependence arises from the term $M_-^2 = M_0'^2 - M_0^2$ in
$\mathcal{P}^{[\gamma^-]}_{\rm fBT}$ and reflects the geometric structure of the LF projection at the integrand level. 
The angle $\theta$ is therefore an internal integration variable spanning
$0 \le \theta < 2\pi$, rather than a fixed transverse direction or an observable anisotropy of the unpolarized system.

Two remarks are in order for the vector projections
$\Gamma=\gamma^\perp$ and $\Gamma=\gamma^-$.

\paragraph{Transverse component $\Gamma={\bm\gamma}^\perp$.}
For the transverse current (${\bm\gamma}^\perp\equiv\{\gamma^1,\gamma^2\}$; $j=1,2$), the extraction of
$F^{(\perp)}(0)$ from Table~\ref{tab:helicity} and Eq.~\eqref{eq:16} requires a careful forward–limit
procedure. In the strict $Q^2\to 0$ limit, both the matrix element
$\langle P|\bar q\,{\bm\gamma}^j_\perp q|P\rangle_{\rm BT}$ and the corresponding Lorentz prefactor
$\mathcal P^{[{\bm\gamma}^j_\perp]}_{\rm pBT/fBT}(0)$ vanish,
leading to an indeterminate $0/0$ structure.
To obtain a finite result, we first contract the current ${\bm\gamma}^j_\perp$ with
${\bm q}_\perp^j$ and then take the forward limit:
\begin{align}
F^{[{\bm\gamma}^\perp]}_{\rm pBT}(0)
&= \lim_{Q^2\to0}
\frac{\langle P'|\,\bar q(0)\,({\bm\gamma}^\perp\cdot\bm q_\perp)\,q(0)\,|P\rangle_{\rm BT}}
{\mathcal P^{[\gamma^\perp]}_{\rm pBT}(Q^2)\!\cdot\!\bm q_\perp},
\\[2pt]
F^{[{\bm\gamma}^\perp]}_{\rm fBT}(0)
&= \lim_{Q^2\to0}
\Big\langle P'\Big|\,
\frac{\bar q(0)\,({\bm\gamma}^\perp\cdot\bm q_\perp)\,q(0)}
{\mathcal P^{[\gamma^\perp]}_{\rm fBT}(Q^2)\!\cdot\!\bm q_\perp}
\,\Big|P\Big\rangle_{\rm BT}.
\end{align}
One finds that
$F^{[{\bm\gamma}^\perp]}_{\rm pBT}(Q^2)=F^{[{\bm\gamma}^\perp]}_{\rm fBT}(Q^2)$
for all $Q^2$, including the normalization point
$F^{[{\bm\gamma}^\perp]}(0)=1$,
demonstrating that the transverse component requires no effective LF zero-mode correction in either formulation.

At the integrand level in the forward limit, the fBT-LFQM gives
\begin{align}
\lim_{Q^2\to 0} S^{[{\bm\gamma}^\perp\;\cdot\; {\bm q}_\perp]}
&= \lim_{Q^2\to 0}\,\frac{2\,\bm k_\perp\!\cdot\!\bm q_\perp}{x\,P^+},\\
\lim_{Q^2\to 0}\big(\mathcal P^{[{\bm\gamma}^\perp]}_{\rm fBT}(0)\!\cdot\!\bm q_\perp\big)
&= \lim_{Q^2\to 0}\,\frac{2\,\bm k_\perp\!\cdot\!\bm q_\perp}{x},
\end{align}
whereas in the pBT-LFQM
\begin{equation}
\lim_{Q^2\to 0}\big(\mathcal P^{[{\bm\gamma}^\perp]}_{\rm pBT}(0)\!\cdot\!\bm q_\perp\big)
= \lim_{Q^2\to 0}\,\bm q_\perp^{2}.
\end{equation}
Thus, in the fBT-LFQM the transverse-component normalization in the forward limit
coincides with that obtained from the plus component,
\begin{align}
    \lim_{Q^2\to 0}
\Big\langle P\Big|\,\frac{\bar q(0)\,({\bm\gamma}^\perp\!\cdot\!\bm q_\perp)\,q(0)}
{\mathcal P^{[{\bm\gamma}^\perp]}_{\rm fBT}(0)\!\cdot\!\bm q_\perp}\,\Big|P\Big\rangle_{\rm BT}
= \int [d^3{\bm p}_1]\;|\phi(x,\bm k_\perp)|^2 =1.
\end{align}
Accordingly, the forward-limit expressions for the ${\bm\gamma}^\perp$ matrix element
and prefactors listed in Table~\ref{tab:forward} are obtained by first contracting
with $\bm q_\perp$, taking $Q^2\to0$, and then factoring out the common
$\bm q_\perp$ dependence, leaving the effective ${\bm k}_\perp^j$ structure.

\paragraph{Minus component $\Gamma=\gamma^-$.}
In the pBT-LFQM, the Lorentz prefactor in the forward limit reduces to
\begin{equation}\label{eq:MBT}
\mathcal P^{[\gamma^-]}_{\rm pBT}(0)=\frac{2M^2}{P^+}.
\end{equation}
In the fBT-LFQM, the covariant prefactor is implemented pointwise under the
$(x,\bm k_\perp)$ integral with the BT replacement $M\to M_0$. 
For the minus component this
yields, at the integrand level,
\begin{equation}\label{CalMBT}
\mathcal P^{[\gamma^-]}_{\rm fBT}(0)
=\frac{2}{P^+}\!\left(M^2_0 + \dfrac{ 2 |{\bm k}_\perp|^2\cos^2\theta }{x^2} \right) \equiv \frac{2 {\cal M}^2_{\rm BT}}{P^+},
\end{equation}
where we define
\be\label{CalMBT2}
{\cal M}^2_{\rm BT} \equiv M^2_0 + \dfrac{ 2 |{\bm k}_\perp|^2\cos^2\theta }{x^2}.
\ee
The $\cos^2\theta$ term reflects the projection of ${\bm k}_\perp$ onto the direction of ${\bm q}_\perp$ 
and originates from the LF structure of the $\gamma^-$ projector at the integrand level.
Although $\mathcal P^{[\gamma^-]}_{\rm fBT}$ retains this angular dependence locally, the
extracted form factor remains component–independent:
\begin{equation}
F^{[\gamma^-]}_{\rm fBT}(0)
=
\lim_{Q^2\to0}
\Big\langle P'\Big|
\frac{\bar q(0)\gamma^-q(0)}
{\mathcal P^{[\gamma^-]}_{\rm fBT}(Q^2)}
\Big|P\Big\rangle_{\rm BT}
=1.
\end{equation}
By contrast, in the pBT-LFQM the use of the physical mass $M$ in the Lorentz prefactor, together
with the omission of the effective LF zero-mode correction, generally 
leads to a violation of the normalization,
\begin{equation}
F^{[\gamma^-]}_{\rm pBT}(0)
=
\lim_{Q^2\to0}
\frac{\langle P'|\,\bar q(0)\gamma^-q(0)\,|P\rangle_{\rm BT}}
{\mathcal P^{[\gamma^-]}_{\rm pBT}(Q^2)}
\neq 1.
\end{equation}
Restoring the correct charge normalization requires the inclusion of the
effective LF zero-mode correction in the pBT-LFQM.

For the $\gamma^-$ projection, the difference between the fBT and pBT
results signals the effective LF zero-mode correction missing from the pBT extraction. Including this correction in the pBT result
restores the charge normalization and reproduces the common fBT result
obtained from all current components.

\begin{figure}[t]
\centering
\includegraphics[width=.45\textwidth]{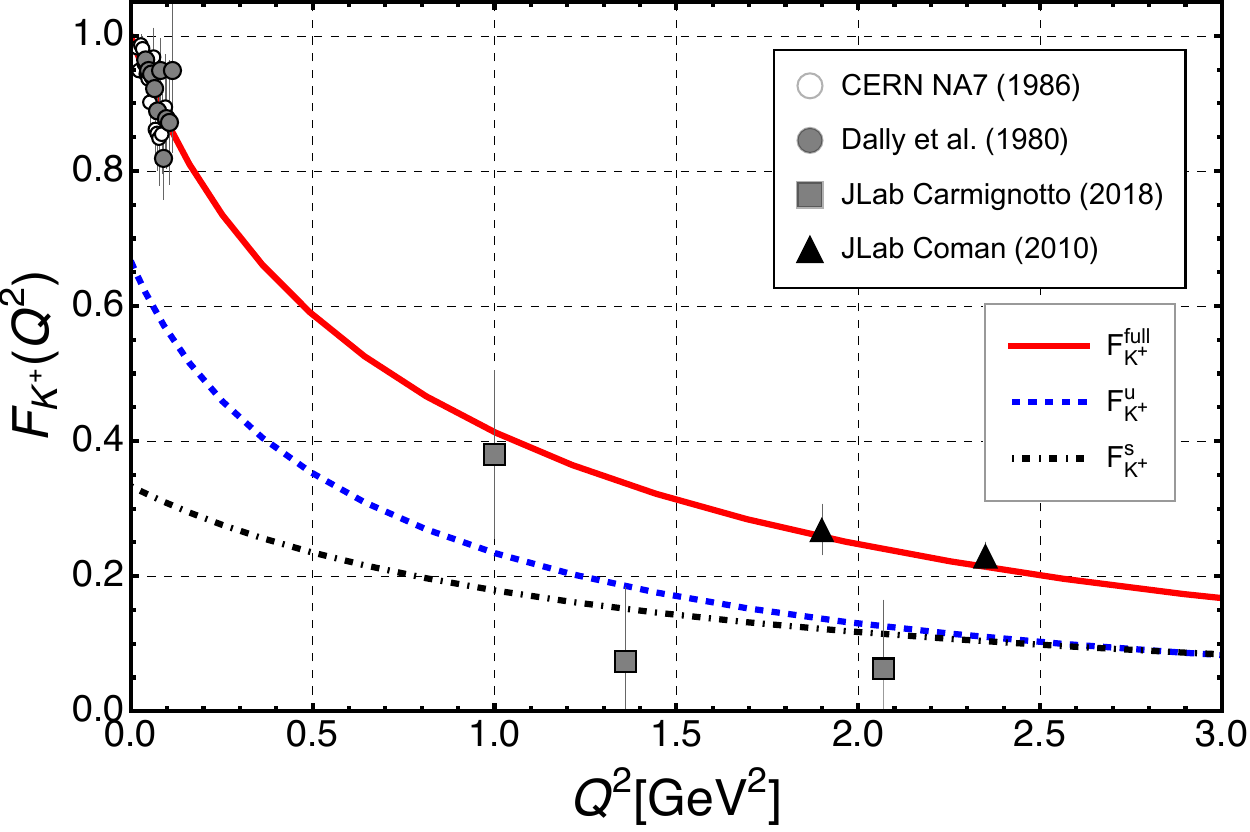}
\includegraphics[width=.45\textwidth]{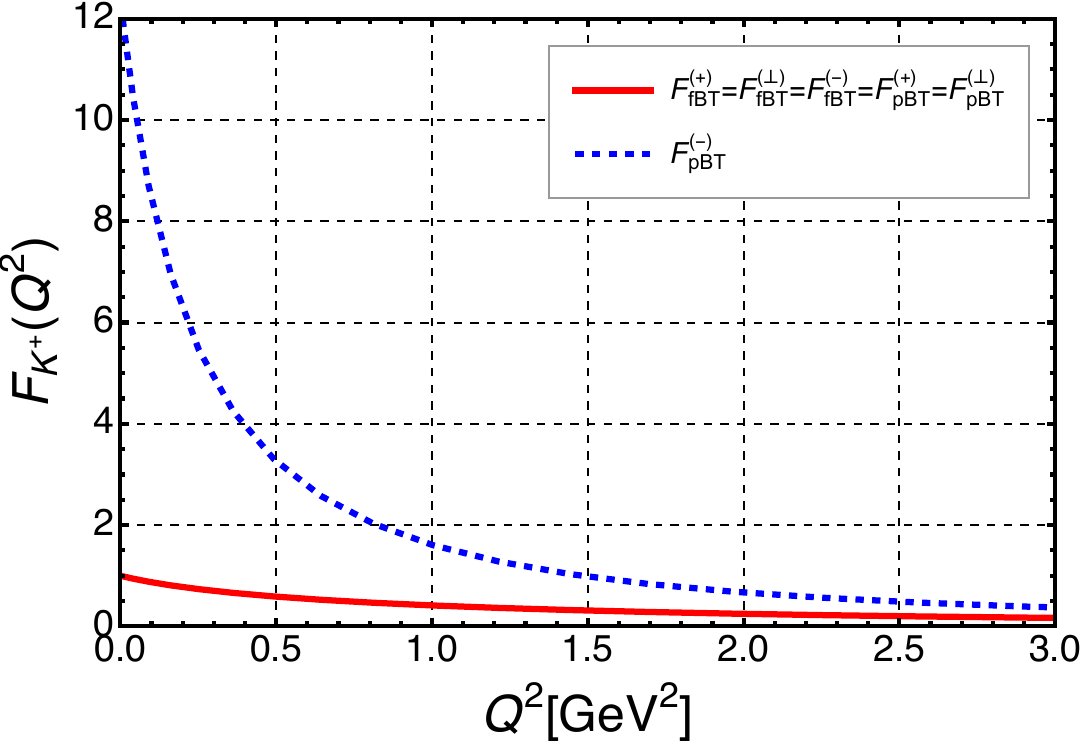}
 \caption{Charged-kaon EMFF $F_{K^+}(Q^2)$. (Left) fBT-LFQM full result (solid red line) with $u$ (dashed blue line) and $s$ (dot-dashed black line) quark contributions, compared with experimental data from CERN NA7~\cite{Amendolia:1986ui}, Dally et al.~\cite{Dally:1980dj}, and JLab~\cite{Carmignotto:2018uqj,JeffersonLabHallA:2009dot}. 
 (Right) Comparison between the component-independent fBT-LFQM result (solid red line) and the pBT-LFQM result extracted from 
 the $\gamma^-$ component (dashed blue line), demonstrating the breakdown of current-component independence in the pBT-LFQM due to
the absence of the effective LF zero-mode correction.\label{fig:KaonEM}}
\end{figure}
In the left panel of Fig.~\ref{fig:KaonEM}, we show our fBT-LFQM prediction for
the charged-kaon EMFF $F_{K^+}(Q^2)$ and compare it with the available experimental
data~\cite{Amendolia:1986ui,Dally:1980dj,Carmignotto:2018uqj,JeffersonLabHallA:2009dot}. 
The solid curve shows the full result, while the dashed and dot-dashed curves
show the separate $u$- and $s$-quark contributions, respectively. The full fBT-LFQM
result is in reasonable agreement with the existing low- and intermediate-$Q^2$
data from CERN NA7~\cite{Amendolia:1986ui}, Dally et al.~\cite{Dally:1980dj}, and JLab~\cite{Carmignotto:2018uqj,JeffersonLabHallA:2009dot}.
At the largest momentum transfers, however, our curve lies visibly above the two highest-$Q^2$ points of Ref.~\cite{Carmignotto:2018uqj}. 
It should be kept in mind that those values are not obtained from an elastic measurement: $F_{K^+}$ is extracted from the longitudinal part of the separated $K^+$ electroproduction cross section using a model for the $t$-channel dynamics, and the associated model dependence is sizable in this $Q^2$ range. On the model side, the hadronic input used here is a valence-only Gaussian LFWF at a low scale, which lacks the perturbative hard-gluon tail that is expected to become increasingly
relevant as $Q^2$ increases. A quantitative description of the large-$Q^2$ region would therefore require going beyond the valence approximation at the model scale, whereas our model provides a reasonable description of the low- and intermediate-$Q^2$ region relevant for the forward-limit TMD analysis of Secs.~\ref{sec:TMD} and \ref{sec:Num}.
The extraction is independent of the current component $(\mu=+,\perp,-)$ and satisfies charge normalization at $Q^2=0$.
As shown in the right panel of Fig.~\ref{fig:KaonEM}, the pBT-LFQM results extracted from the $\gamma^+$ and $\gamma^\perp$ components coincide with the fBT-LFQM
current-component-independent result, whereas the result extracted from the $\gamma^-$ component (dashed) shows a significant 
deviation. This deviation quantifies the effective LF zero-mode correction missing from the minus-component extraction in the pBT-LFQM formulation.

 \begin{figure}[t]
  \centering
  \includegraphics[width=0.8\textwidth]{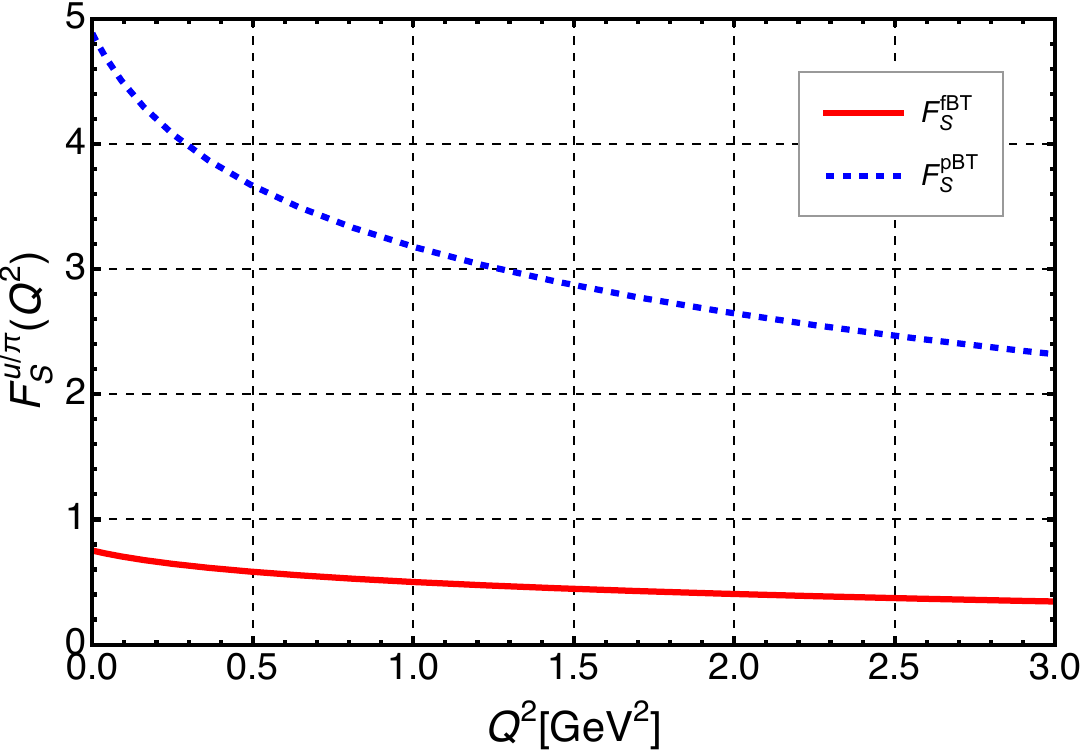}
  \caption{The $u$-quark contribution to the scalar form factor
  $F_S^{u/\pi}(Q^2)$ of the $\pi^+$ as obtained in the fBT-LFQM
  (solid line) and pBT-LFQM (dashed line).
  \label{fig:FpiScalar}}
\end{figure}

  \begin{figure}
 	\centering
 	\includegraphics[width=0.45\textwidth]{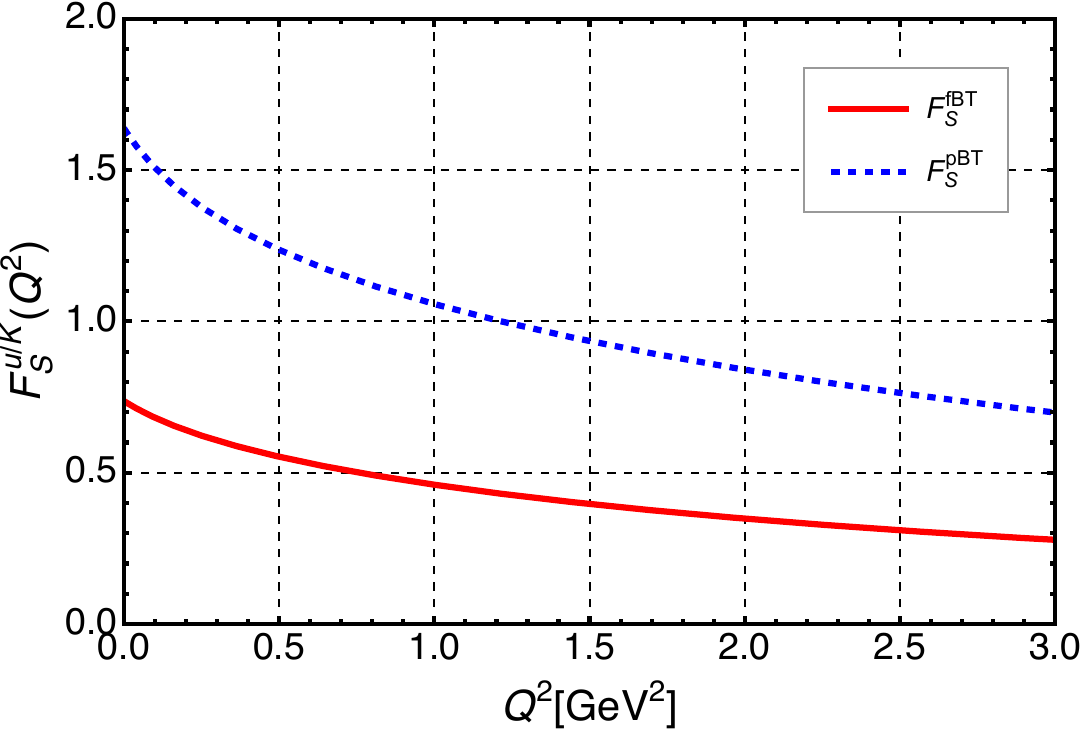}
    \includegraphics[width=0.45\textwidth]{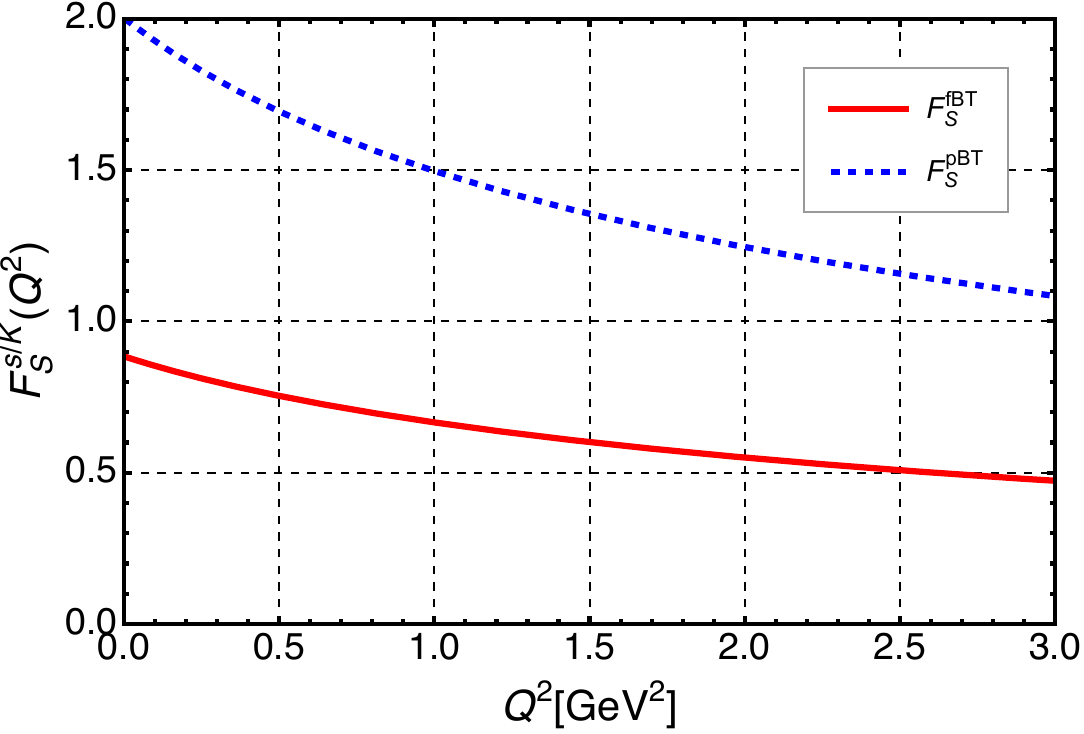}
 	\caption{The $u$-quark (left) and $s$-quark (right) contributions to the scalar form factor $F_S^{q/K}(Q^2)$ 
of the $K^+$, obtained in the fBT-LFQM (solid) and pBT-LFQM (dashed).\label{fig:FKScalar}}  	
 \end{figure}

In Fig.~\ref{fig:FpiScalar}, we show the $u$-quark contribution to the
scalar form factor $F_S^{u/\pi}(Q^2)$ of the $\pi^+$, obtained in
the fBT-LFQM (solid) and pBT-LFQM (dashed).
In the fBT-LFQM, the consistent
replacement $M_\pi\to M_0(x,\bm{k}_\perp)$ is implemented in the scalar Lorentz
prefactor at the integrand level. As a result, the fBT and pBT extractions of
$F_S^{u/\pi}(Q^2)$ differ significantly, and this sizable difference signals the effective LF zero-mode correction missing from the pBT
extraction in the scalar channel.

This difference should be understood in light of the covariant BS-model analysis
presented in Appendix~\ref{app:scalarZM}, where we explicitly demonstrate that the
scalar form factor receives a nonvanishing explicit LF zero-mode contribution.
While the BS-model result provides a clear diagnostic of
the missing LF contribution, as discussed in Appendix~\ref{app:scalarZM},
the fBT--pBT difference is interpreted, at the level of the observable, as an
effective LF zero-mode correction missing from
the pBT extraction. In the fBT-LFQM, this correction is effectively accounted for
through the consistent $M\to M_0$ replacement in the Lorentz structure.
In fact, the fBT-LFQM result gives $F_S^{\rm fBT}(0)<1$ numerically, while the pBT-LFQM
extraction yields a value well above unity. This qualitative pattern mirrors the
behavior seen for the EMFF extracted from the $(-)$ component in the pBT-LFQM,
where the absence of the corresponding effective correction in the pBT extraction
produces a systematic enhancement relative to the fBT result
over the full $Q^2$ range, although its quantitative impact decreases as $Q^2$ increases.
The detailed derivation of the scalar zero mode in the BS model and the distinction
between the explicit BS-model LF zero-mode term and the effective pBT correction are
discussed in Appendix~\ref{app:scalarZM}. The same Appendix also explains the
BS-to-fBT-LFQM correspondence at the valence level through the standard matching
of the vertex and spin structures, supplemented by the BT-consistent replacement
$M^{(\prime)}\to M_0^{(\prime)}$ in the Lorentz structures.

In Fig.~\ref{fig:FKScalar}, we show the $u$-quark (left) and $s$-quark (right)
contributions to the scalar form factor $F_S^{q/K}(Q^2)$ of the $K^+$, obtained
in the fBT-LFQM (solid) and pBT-LFQM (dashed).\footnote{For notational
brevity, $u/K$ and $s/K$ used below label the valence $u$ and $\bar s$
contributions, respectively, in the $K^+=u\bar s$.}
The $s$-quark contribution is
larger than the $u$-quark contribution, reflecting SU(3) flavor-symmetry breaking.
As in the pion case, the pBT-LFQM results are systematically larger than the
fBT-LFQM results, signaling the sensitivity of the scalar channel to LF zero-mode
effects. However, compared with the pion case, the discrepancy for the light
$u$-quark contribution in the kaon is quantitatively reduced. Thus, the kaon
provides a useful system for examining how flavor-symmetry breaking and the
spectator-quark environment affect the effective fBT--pBT correction in the scalar channel.
For clarity, this correction can be expressed in a unified way for both the $(-)$ component
of the vector current and the scalar current. Using Eqs.~\eqref{eq:19} and~\eqref{eq:19con},
it is written as
\bea\label{eq:19zero}
F_{\rm eff}^{[\Gamma]}(Q^2)
&\equiv&
F_{\rm fBT}^{[\Gamma]}(Q^2)
-
F_{\rm pBT}^{[\Gamma]}(Q^2)
\nonumber\\
&=&
\int [d^3{\bm p}_1]\,
\phi'(x,{\bm k}'_\perp)\phi(x,{\bm k}_\perp) S^{[\Gamma]}
\left[
\frac{1}
     {{\cal P}^{[\Gamma]}_{\rm fBT}}
-
\frac{1}
     {{\cal P}^{[\Gamma]}_{\rm pBT}}
\right],
\eea
where $\Gamma=\{\gamma^-,\mathbf{1}\}$, and the spin factors and Lorentz
prefactors are given in Table~\ref{tab:helicity}.
We emphasize that $F_{\rm eff}^{[\Gamma]}$ is defined entirely within the
BT-based LFQM as the difference between two extractions of the same observable
that differ only in the treatment of the Lorentz prefactor. It is therefore a
quantity defined within the on-shell valence framework and should not be
identified with the explicit BS-model LF zero-mode term given below in
Eq.~\eqref{eq:FSLFZM-copy}. Recovering the fBT result within the pBT extraction
would require adding the difference defined in Eq.~\eqref{eq:19zero} as an
external correction, since the on-shell valence framework provides no
independent zero-mode operator that generates it.
In contrast, in the fBT formulation,
the corresponding correction is already accounted for through the
BT-consistent Lorentz prefactor, so that no separate effective zero-mode correction needs to be
added.
Thus, the fBT-LFQM
effectively accounts for the correction missing from the pBT extraction through
the BT-consistent implementation of the Lorentz structure, without introducing
any explicit zero-mode operator.

On the other hand, as explicitly shown in Appendix~\ref{app:scalarZM}, in the
manifestly covariant BS model, which is based on off-mass-shell quark and antiquark
degrees of freedom ($M\neq M_0$), the explicit LF zero-mode contribution to the form factor
can be expressed in terms of the LF valence wave functions
$\chi\propto 1/(M^2-M_0^2)$ as
\be\label{eq:FSLFZM-copy}
[F_{\rm Z.M.}^{[\Gamma]}]^{\rm BS}(Q^2)
=
\frac{N_c}{16\pi^3}\int^1_0\frac{dx}{1-x}\int d^2{\bm k}_\perp\,
\chi'(x,{\bm k}'_\perp)\chi(x,{\bm k}_\perp)
\frac{\hat{{\cal S}}_{\rm Z.M.}^{[\Gamma]}}{{\cal P}^{[\Gamma]}_{\rm BS}},
\ee
where $\hat{{\cal S}}_{\rm Z.M.}^{[\Gamma]}$ is the process-dependent effective
zero-mode operator arising from the spin-trace structure.
Here ${\cal P}^{[\Gamma]}_{\rm BS}$ denotes the Lorentz prefactor of the
covariant BS model, which is built from the physical meson masses and therefore
coincides with the one used in the pBT-LFQM, ${\cal P}^{[\Gamma]}_{\rm BS}
={\cal P}^{[\Gamma]}_{\rm pBT}$ of Table~\ref{tab:helicity}; the two extractions differ
only in the wave functions to which this prefactor is attached (see Eq.~\eqref{eq:PBSpBT}).
We stress that Eq.~\eqref{eq:FSLFZM-copy} is an object of the off-shell BS model:
it is tied to the off-shell propagation and to the vertex regularization of that
model, neither of which has a counterpart in the on-shell valence LFQM. It should
therefore not be identified with the effective difference
$F_{\rm eff}^{[\Gamma]}$ defined in Eq.~\eqref{eq:19zero}, nor can the BS-model
zero-mode operator be imported into the Gaussian LFQM.

In other words, the physical observable is obtained differently in the two frameworks.
In the covariant BS model, it is evaluated with the full matrix element as
\be\label{eq:BS-full}
F_{\rm BS}^{[\Gamma]}(Q^2)
=
\frac{1}{{\cal P}^{[\Gamma]}}
\big\langle P'\big| \bar q(0)\Gamma q(0)\big|P\big\rangle_{\rm full},
\ee
where
$\langle P'| \bar q(0)\Gamma q(0)|P\rangle_{\rm full}$ includes both the valence
and explicit LF zero-mode contributions. This highlights a key difference between the two
approaches: in the field-theoretic BS framework, LF zero modes are encoded in the
matrix element itself, whereas in the on-shell BT-based LFQM
the effective difference defined in Eq.~\eqref{eq:19zero} is accounted for
through the BT-consistent Lorentz prefactors entering the fBT-LFQM extraction.
One should note that 
${\cal F}_{\rm fBT}^{[\Gamma]}(Q^2)$ and~${\cal F}_{\rm BS}^{[\Gamma]}(Q^2)$ are evaluated in the two different frameworks which differ in their dynamical contents and thus no equality between
${\cal F}_{\rm fBT}^{[\Gamma]}$ and ${\cal F}_{\rm BS}^{[\Gamma]}$ can be implied.
The BS analysis of
Appendix~\ref{app:scalarZM} serves as the field-theoretic diagnosis showing
that the $\gamma^-$ and scalar projections are indeed sensitive to LF zero modes.
What Appendix~\ref{app:scalarZM} does establish is a mapping of the BS
\emph{valence} integrand onto the fBT-LFQM one through the standard matching of
the vertex and spin structures, supplemented by the BT-consistent replacement
$M^{(\prime)}\to M_0^{(\prime)}$ in the Lorentz structures, without evaluating,
identifying with, or importing the explicit BS-model LF zero-mode term.

\section{TMDs and PDFs}
\label{sec:TMD}
\subsection{Formal definitions of TMDs and PDFs}
For the pseudoscalar meson state $|P\rangle$ defined in Eq.~\eqref{eq:KP}, 
the valence–quark TMDs are defined via the quark–quark 
correlator~\cite{Metz09,Lorce:2015,Lorce:2016,Puhan:2023ekt,Zhu23}
\footnote{In constituent–quark models without explicit 
gluon degrees of freedom, the Wilson line $\mathcal{W}(0,z)$ in the QCD definition reduces to the identity in color space.} 
\be\label{eq:PhiC}
\Phi^{[\Gamma]}_q (x, {\bm k}_\perp)
=\frac{1}{2}\int [\dd^3 z] e^{i p_1 \cdot z}\langle P| \bar{q}(0)\Gamma {\cal W}(0,z) q(z)|P\rangle|_{z^+=0}, 
\ee
where $[\dd^3z]\equiv \dd z^- \dd^2{\bm z}_\perp/[2(2\pi)^3]$.
Here $p_1$ denotes the quark momentum conjugate to $z$, with
$p_1^+=xP^+$ and ${\bm p}_{1\perp}={\bm k}_\perp$ in the frame
${\bm P}_\perp=0$. Since $z^+=0$, the phase depends only on $p_1^+$ and ${\bm p}_{1\perp}$.
The field $q(z)$ denotes the quark field operator, and $\Gamma$ specifies the Dirac projection of the correlator 
$\Phi_q^{[\Gamma]} (x, {\bm k}_\perp)$, with the subscript $q$ denoting flavor.
Integrating both sides of Eq.~\eqref{eq:PhiC} over $x$ and $\bm k_\perp$ yields the forward matrix element
\be\label{eq:LFforward}
\int\!\dd x\int\!\dd^2\bm k_\perp\;\Phi^{[\Gamma]}_q (x,\bm k_\perp)
=\frac{1}{2 P^+}\,\big\langle P\big| \bar{q}(0)\,\Gamma\, q(0)\big|P\big\rangle
\ee 
and the derivation is given in Appendix~\ref{app:forward}.

In the conventional LF parametrization, the unpolarized $T$–even TMDs are 
represented through the following projections~\cite{Lorce:2015,Lorce:2016},
which correspond to the standard decomposition used in constituent quark models:\footnote{For a spin–0 target these correspond, 
respectively, to twist–2 ($f_1$), twist–3 ($f^{\perp}$ and $e$), and twist–4 ($f_4$) unpolarized structures.}
\begin{subequations}\label{eq:f1f4}
\begin{align}
\Phi^{[\gamma^+]}_q (x, {\bm k}_\perp) &= f^q_1 (x, {\bm k}_\perp), \label{eq:f1f4a} \\
\Phi^{[\gamma^j_\perp]}_q (x, {\bm k}_\perp) &= \frac{{\bm k}^j_\perp}{P^+}\, f^{\perp q} (x, {\bm k}_\perp), \qquad (j=1,2) \label{eq:f1f4b} \\
\Phi^{[\gamma^-]}_q (x, {\bm k}_\perp) &= \frac{M^2}{(P^+)^2}\, f^q_4 (x, {\bm k}_\perp), \label{eq:f1f4c} \\
\Phi^{[\mathbf{1}]}_q (x, {\bm k}_\perp) &= \frac{M}{P^+}\, e^q (x, {\bm k}_\perp). \label{eq:f1f4d}
\end{align}
\end{subequations}
Here, $f_1^q$, $f^{\perp q}$, and $f_4^q$ originate from vector projections, while $e^q$ is associated with the scalar projection.
Among these projections, the $\gamma^-$ projection for $f_4^q$ and the scalar
projection for $e^q$ are zero-mode sensitive. In the
covariant BS framework, the full matrix element
$\langle P| \bar q(0)\Gamma q(0)|P\rangle_{\rm full}$
includes both valence and explicit LF zero-mode contributions, as discussed in
Appendix~\ref{app:scalarZM}. Therefore, the standard decomposition in
Eq.~\eqref{eq:f1f4} can be maintained. In the on-shell BT-based LFQM, however,
the effective LF zero-mode correction missing from the pBT extraction is
incorporated, at the level of the observable, through the BT-consistent
Lorentz prefactors rather than through an explicit nonvalence contribution
to the matrix element itself.
Consequently, a consistent extraction of $f_4^q$ and $e^q$
requires the fBT prescription discussed above.

Integrating a generic TMD  over transverse momentum yields the corresponding (collinear) PDF,
\be\label{eq:PDFTMD}
f(x) = \int \dd^2 {\bm k}_\perp f(x, {\bm k}_\perp).
\ee
Unlike $f_1^q$, $e^q$, and $f_4^q$, the $f^{\perp q}$ projection is proportional to $\bm k_\perp^{\,j}$ and
$\int\!\mathrm{d}^2\bm k_\perp\, \bm k_\perp^{\,j}\, f(x,\bm k_\perp)=0$ by azimuthal symmetry. Nevertheless, we define the
formal collinear integral
$f^{\perp q}(x)\equiv \int\!\mathrm{d}^2\bm k_\perp\, f^{\perp q}(x,\bm k_\perp)$, following Refs.~\cite{Lorce:2015,Lorce:2016}.

In both the pBT- and fBT-LFQMs, the 
$\Gamma=\gamma^+$ and $\Gamma=\gamma^j_\perp$ projections (Eqs.~\eqref{eq:f1f4a}–\eqref{eq:f1f4b}) 
require no effective LF zero-mode correction,
so that $f_1^q$ and $f^{\perp q}$ are determined entirely by the on–shell (valence) correlator. 

In our conventions,\footnote{The overall sign of $f^{\perp q}(x,\bm k_\perp)$ depends on the definition 
of ${\bm k}_{1\perp}=\pm{\bm k}_\perp$, leading to $f^{\perp q}(x,\bm k_\perp) = \pm\frac{1}{x}\, f^q_1 (x,\bm k_\perp)$.}
\begin{align}
f^q_1 (x,\bm k_\perp) &= \frac{\lvert \phi(x,\bm k_\perp)\rvert^2}{2(2\pi)^3},\nonumber\\
f^{\perp q}(x,\bm k_\perp) &= \frac{1}{x}\, f^q_1 (x,\bm k_\perp).
\label{eq:fperprel}
\end{align}
These relations are characteristic of the valence constituent structure of the LFQM and should not be 
interpreted as general QCD identities.

Within the present valence LFQM framework, the twist-2 PDF $f_1^q(x)$ satisfies the number and momentum sum rules,
\begin{align}
\label{eq:Nq}
\int_0^1\!dx\, f_1^q(x) &= N_q,\nonumber\\
\sum_q \int_0^1\!dx\, x\, f_1^q(x) &= 1,
\end{align}
where the sum runs over the valence quark and antiquark flavors, and $N_q$
denotes the corresponding valence count, e.g., $N_u=N_{\bar s}=1$ for $K^+(u\bar s)$.

\subsection{Extraction of TMDs from $\Gamma=\{\gamma^-, \mathbf{1}\}$}
The TMD projections in Eq.~\eqref{eq:f1f4} must be consistent with the forward–limit relation in Eq.~\eqref{eq:LFforward}, 
which is tied to EM charge normalization through the form factor condition $F^{[\gamma^\mu]}(0)=1$.

In the pBT-LFQM, applying the $\gamma^-$ component of Eq.~\eqref{eq:LFforward} formally yields
\begin{equation}
\label{eq:LFforward-con}
\int\!\mathrm{d}x\int\!\mathrm{d}^2\bm k_\perp\;
\frac{2P^+}{\mathcal P^{[\gamma^-]}_{\rm pBT}(0)}
\Phi^{[\gamma^-]}_q (x,\bm k_\perp)
= \frac{\big\langle P\big| \bar{q}(0) \gamma^- q(0)\big|P\big\rangle}{\mathcal P^{[\gamma^-]}_{\rm pBT}(0)}
=F^{[\gamma^-]}_{\rm pBT}(0)\stackrel{\rm formal}{=}1,
\end{equation}
with $\mathcal P^{[\gamma^-]}_{\rm pBT}(0)= 2 P^- = 2M^2/P^+$ the kinematic projector for the $\gamma^-$ current in the pBT-LFQM realization.
Using Eq.~\eqref{eq:f1f4c}, one obtains
\begin{equation}
\label{eq:LFforward2}
\int\!\mathrm{d}x\int\!\mathrm{d}^2\bm k_\perp\; f^q_4 (x,\bm k_\perp)=F^{[\gamma^-]}_{\rm pBT}(0)\stackrel{\rm formal}{=}1.
\end{equation}
From the matrix element
$\langle P|\bar q(0)\gamma^- q(0)|P\rangle_{\rm BT}$ given in Table~\ref{tab:forward}
and from Eq.~\eqref{eq:LFforward}, one finds
\be
\Phi^{[\gamma^-]}_q(x,\bm k_\perp)
=
\frac{\bm k_\perp^2 + m_1^2}{(xP^+)^2}
\, \frac{|\phi(x,\bm k_\perp)|^2}{2 (2\pi)^3},
\ee
and thus the explicit relation between $f^q_4$ and $f^q_1$ is formally given by
\begin{equation}
\label{eq:Phi4-con}
f^q_4 (x,\bm k_\perp)=\frac{(P^+)^2}{M^2}\,\Phi^{[\gamma^-]}_q (x,\bm k_\perp)\stackrel{\rm formal}{=}
\frac{\bm k_\perp^2+m_1^2}{x^2 M^2}\, f^q_1(x,\bm k_\perp).
\end{equation}
As already demonstrated in the EM form-factor analysis, the $\gamma^-$ projection is zero-mode sensitive. Equation~\eqref{eq:Phi4-con} therefore represents the pBT valence result without the effective LF zero-mode correction defined as the fBT--pBT difference.
Consequently, the sum rule
\begin{equation}
\label{eq:Nqf4}
\int_0^1\!\mathrm{d}x\, f_4^q(x)\equiv \int\!\mathrm{d}x\int\!\mathrm{d}^2\bm k_\perp\, f^q_4 (x,\bm k_\perp)=N_q
\end{equation}
is generally violated unless this correction is included.
We note that our normalization in Eq.~\eqref{eq:Nqf4} differs from that in
Refs.~\cite{Lorce:2015,Lorce:2016}, where
$2\!\int_0^1\!dx\, f_4^q(x)=N_q$.\footnote{This difference originates from the choice of LF $+\!/\!-$ components:
we use a convention in which, at $\bm P_\perp=\mathbf 0$, $M^2=P^+P^-$, while
Refs.~\cite{Lorce:2015,Lorce:2016} adopt the alternative normalization yielding $M^2=2\,P^+P^-$.
Equivalently, this corresponds to the difference between defining $a^\pm=a^0\pm a^3$ and $a^\pm=(a^0\pm a^3)/\sqrt{2}$.}

On the other hand, in the fBT-LFQM the forward-limit relation is constructed so that the EM form-factor normalization is satisfied identically. 
Within the pBT-to-fBT comparison, replacing $\mathcal P^{[\gamma^-]}_{\rm pBT}(0)$ by the BT-consistent prefactor $\mathcal P^{[\gamma^-]}_{\rm fBT}(0)$ 
incorporates this correction. Equation~\eqref{eq:LFforward} then leads to
\begin{equation}
\label{eq:LFforward-BT}
\int\!\mathrm{d}x\int\!\mathrm{d}^2\bm k_\perp\;\frac{2P^+}{{\cal P}^{[\gamma^-]}_{\rm fBT}(0)}\,\Phi^{[\gamma^-]}_q (x,\bm k_\perp)
=F^{[\gamma^-]}_{\rm fBT}(0)=1,
\end{equation}
with $\mathcal P^{[\gamma^-]}_{\rm fBT}(0)= 2{\cal M}^2_{\rm BT}/ P^+$.
Likewise, the exact relation between $f^q_4$ and $f^q_1$ in the BT-based LFQM is given by
\begin{equation}
\label{eq:Phi4-BT}
f^q_4 (x,\bm k_\perp)=\frac{(P^+)^2}{{\cal M}^2_{\rm BT}}\,\Phi^{[\gamma^-]}_q (x,\bm k_\perp)
=\frac{\bm k_\perp^2+m_1^2}{x^2 {\cal M}^2_{\rm BT}}\,f^q_1(x,\bm k_\perp),
\end{equation}
the sum rule in Eq.~\eqref{eq:Nqf4} is satisfied automatically
and the extraction is manifestly current-component independent.

Positivity inequalities also provide an important 
consistency check for the TMDs~\cite{Lorce:2015,Lorce:2016}
which require $f^q_1 (x, {\bm k}_\perp)\geq 0$ and $f^q_4 (x, {\bm k}_\perp)\geq 0$.
In contrast, the sign of $f^{\perp q} (x, {\bm k}_\perp)$ depends on the transverse momentum
assignment: choosing ${\bm k}_{1\perp}={\bm k}_\perp$ leads to a positive distribution,
while ${\bm k}_{1\perp}=-{\bm k}_\perp$ results in a negative one.

For the scalar current case, the forward-limit matrix element in the pBT-LFQM
also formally satisfies
$\langle P|\bar q(0){\bf 1} q(0)| P\rangle_{\rm BT}\stackrel{\rm formal}{=} 2 M F_S^q(0)$.
Equation~\eqref{eq:LFforward} therefore gives, for $\Gamma=\mathbf{1}$,
\begin{align}\label{eq:LFforwardS}
\int\!\mathrm{d}x \int\!\mathrm{d}^2 \bm k_\perp \,
\Phi^{[\mathbf{1}]}_q(x,\bm k_\perp)
&=\frac{1}{2P^+}\,
\langle P|\bar q(0){\bf 1} q(0)|P\rangle_{\rm BT}
\stackrel{\rm formal}{=}
\frac{M}{P^+}\,F_S^q(0).
\end{align}

Combining this relation with the expression for
$\langle P|\bar q(0) {\bf 1} q(0)|P\rangle_{\rm BT}$
in Table~\ref{tab:helicity}, one finds
\begin{equation}\label{eq:Phis}
\Phi^{[\mathbf{1}]}_q(x,{\bm k}_\perp)
=
\frac{M}{P^+}\, e^q(x,{\bm k}_\perp)
\stackrel{\rm formal}{=}
\frac{m_1}{xP^+}\, f_1^q(x,{\bm k}_\perp).
\end{equation}
Thus, in the pBT-LFQM one formally obtains the relation~\cite{Lorce:2015,Lorce:2016}
\begin{equation}\label{eq:eq-pBT}
e^q(x,{\bm k}_\perp)
\stackrel{\rm formal}{=}
\frac{m_1}{xM}\, f_1^q(x,{\bm k}_\perp).
\end{equation}
However, a central finding of the present work is that the forward matrix 
element $\langle P|\bar{q}(0)q(0)|P\rangle$, which governs the normalization 
of $e^q$ through Eq.~\eqref{eq:LFforward}, is itself zero-mode sensitive,
as shown in Appendix~\ref{app:scalarZM}.
Consequently, Eq.~\eqref{eq:eq-pBT} represents only the valence pBT contribution.
The BT-consistent extraction within the BT-based LFQM requires the
fBT prescription, in which $M\to M_0(x,\bm{k}_\perp)$ is implemented consistently
at the integrand level in the scalar Lorentz prefactor $2M$. This yields
\be\label{eq:eq-fBT}
e^q(x,{\bm k}_\perp)
=
\frac{m_1}{xM_0}\,
f_1^q(x,{\bm k}_\perp),
\ee
with the difference from Eq.~\eqref{eq:eq-pBT} again given by the
corresponding fBT--pBT difference.

It is useful to note that, in full QCD, $e^q(x)$ admits the 
QCD equations-of-motion (EOM)
decomposition~\cite{Anatoli}
$e^q(x)=e^q_{\rm sing}(x)+e^q_{\rm tw3}(x)+e^q_{\rm mass}(x)$,
where the mass term satisfies
$x\,e^q_{\rm mass}(x)=(m_1/M)\,f_1^q(x)$.
In the present valence $q\bar q$ LFQM without explicit gluonic degrees
of freedom, the genuine twist-3 quark--gluon correlation term is absent, 
and no singular distribution localized at $x=0$ is represented explicitly.
The explicit LF zero-mode contribution discussed above refers to the
zero-mode-sensitive scalar matrix element and should not be identified
with the QCD $\delta(x)$ term in $e^q(x)$.
The fBT result in Eq.~\eqref{eq:eq-fBT} may therefore be regarded as
a BT-consistent regular valence-sector relation for $0<x<1$ that is
formally analogous to the EOM mass-term relation, rather than as a
term-by-term realization of the full QCD EOM decomposition.
The replacement $M\to M_0$ should not be interpreted as a modification
of the QCD EOM, but rather as the self-consistent implementation of the
valence-sector LFQM within the BT prescription.

\begin{figure*}[t]
  \centering
    \includegraphics[width=0.45\textwidth]{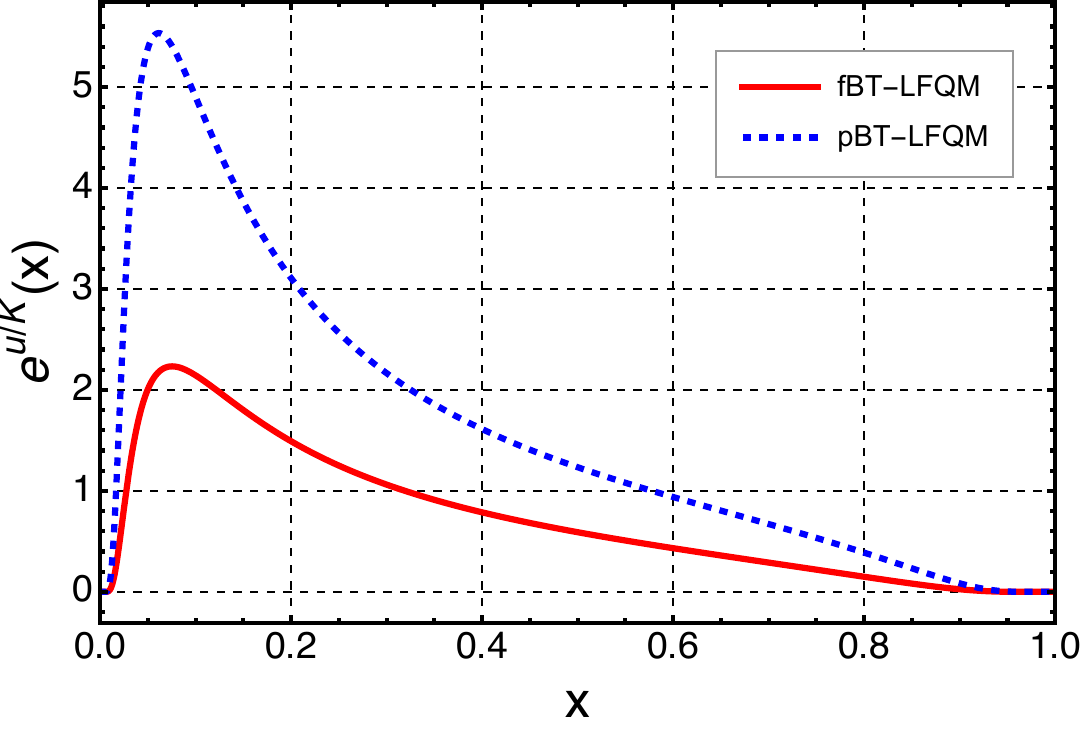}
    \includegraphics[width=0.45\textwidth]{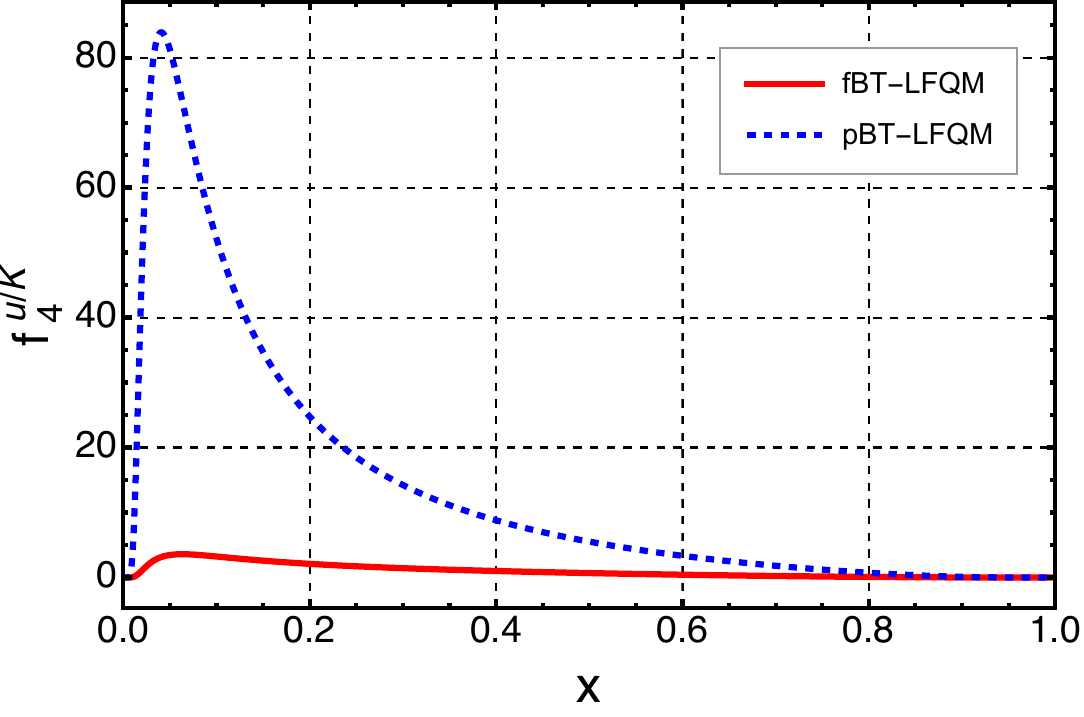}
  \caption{Comparison of $u$-quark PDFs in $K^+$: 
$e^{u/K}(x)$ (left) and $f_4^{u/K}(x)$ (right), shown for the fBT-LFQM (solid) and the pBT-LFQM (dashed).\label{fig:f4eq-ZM}}  
\end{figure*}

\subsection{Internal consistency of the LFQM}
Testing the internal consistency of the LFQM is essential, since the model describes free on–shell quarks and does not explicitly include 
gluonic Fock states. In this work, we contrast the pBT-LFQM with the fBT-LFQM.

As shown in the previous subsections, the pBT-LFQM~\cite{Lorce:2015,Lorce:2016} formally satisfies
\begin{align}\label{eq:QCDrel}
x\, e^q (x, {\bm k}_\perp) &= \frac{m_1}{M}\, f^q_1(x, {\bm k}_\perp), \nonumber\\
x\, f^{\perp q}(x,{\bm k}_\perp) &= f^q_1 (x, {\bm k}_\perp), \nonumber\\
x^2 f^q_4 (x, {\bm k}_\perp) &= \frac{{\bm k}_\perp^2 + m_1^2}{M^2}\, f^q_1 (x, {\bm k}_\perp).
\end{align}
However, the relations involving $e^q$ and $f_4^q$ generally 
fail numerically in the pBT-LFQM unless the
effective LF zero-mode corrections defined as the corresponding fBT--pBT differences
are properly accounted for.
These discrepancies arise in the pBT extraction when the physical meson
mass is retained in the Lorentz prefactor for the zero-mode-sensitive
projections, particularly for $\Gamma=\gamma^-$ and $\Gamma=\mathbf{1}$.

Consistent with this observation, our recent  analysis of the $\rho$-meson decay constants and 
beyond–leading–twist DAs~\cite{hhk8-g2bj}
showed that the QCD EOM relation
$f^{\rm S}_\rho = f^{\perp}_\rho- \frac{2 m}{M}\, f^{\parallel}_\rho$
between the scalar ($f^{\rm S}_\rho$), vector/longitudinal ($f^{\parallel}_\rho$), and tensor/transverse ($f^{\perp}_\rho$) decay constants is violated in the pBT-LFQM but restored in the fBT-LFQM. We therefore employ 
the fBT-LFQM formulation whenever component independence and EOM consistency are essential, 
notably for the $\gamma^-$ and scalar projections.

In the fBT-LFQM, analogous consistency relations hold with the following replacements:
\begin{align}\label{eq:QCDrelBT}
x\, e^q (x, {\bm k}_\perp) &= \frac{m_1}{M_0}\, f^q_1(x, {\bm k}_\perp), \nonumber\\
x\, f^{\perp q}(x,{\bm k}_\perp) &= f^q_1 (x, {\bm k}_\perp), \nonumber\\
x^2 f^q_4 (x, {\bm k}_\perp) &= \frac{{\bm k}_\perp^2 + m_1^2}{{\cal M}^2_{\rm BT}}\, f^q_1 (x, {\bm k}_\perp).
\end{align}
These relations illustrate that the fBT-LFQM preserves the EOM-like valence
relations for both the scalar ($\Gamma=\mathbf{1}$) and $\gamma^-$ channels,
where the LF zero-mode effects are essential, while maintaining BT consistency and current-component independence
through the uniform use of the BT kinematic masses $M_0$ and ${\cal M}_{\rm BT}$ inside
the $(x,\bm{k}_\perp)$ integral, rather than the external hadron mass $M$ in the
Lorentz prefactors.

In Fig.~\ref{fig:f4eq-ZM}, we compare the $u$-quark PDFs
$f^{u/K}_4(x)$ and $e^{u/K}(x)$ obtained in the fBT-LFQM (solid) and pBT-LFQM
(dashed). The fBT-LFQM result satisfies the $f^{u/K}_4$ sum rule in
Eq.~\eqref{eq:Nqf4}, whereas the pBT-LFQM result does not. 
This difference reflects the effective LF zero-mode correction missing from
the pBT extraction of $f^{u/K}_4(x)$, and a similar pattern is observed for
$e^{u/K}(x)$.
Because the scalar channel does not correspond to a conserved charge, its
normalization depends on the mass entering the Lorentz prefactor
($M$ vs.\ $M_0$). The fBT-LFQM implements the replacement $M\to M_0$
consistently under the $(x,\bm{k}_\perp)$ integral and is therefore
self-consistent with the definition in Eq.~\eqref{eq:massfactFS}.

In the following numerical section, we present our fBT-LFQM results for the TMDs and PDFs.

\begin{figure*}[t]
\centering
\includegraphics[width=0.24\textwidth]{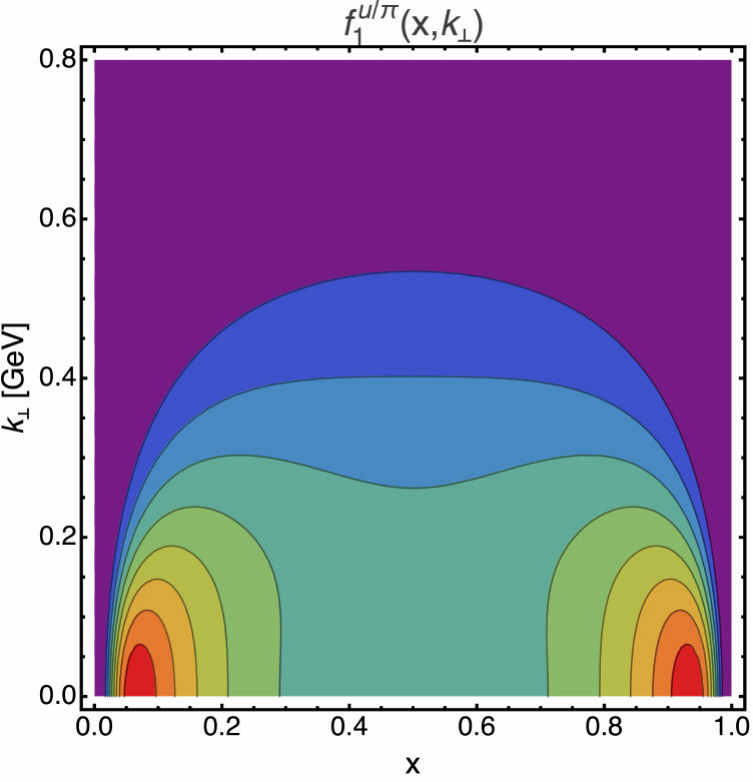}
\includegraphics[width=0.24\textwidth]{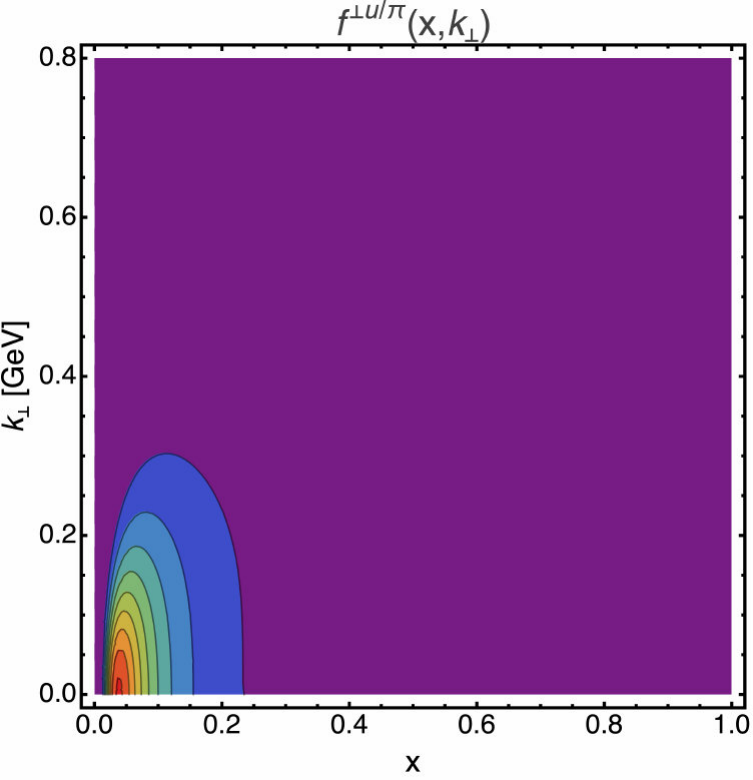}
\includegraphics[width=0.24\textwidth]{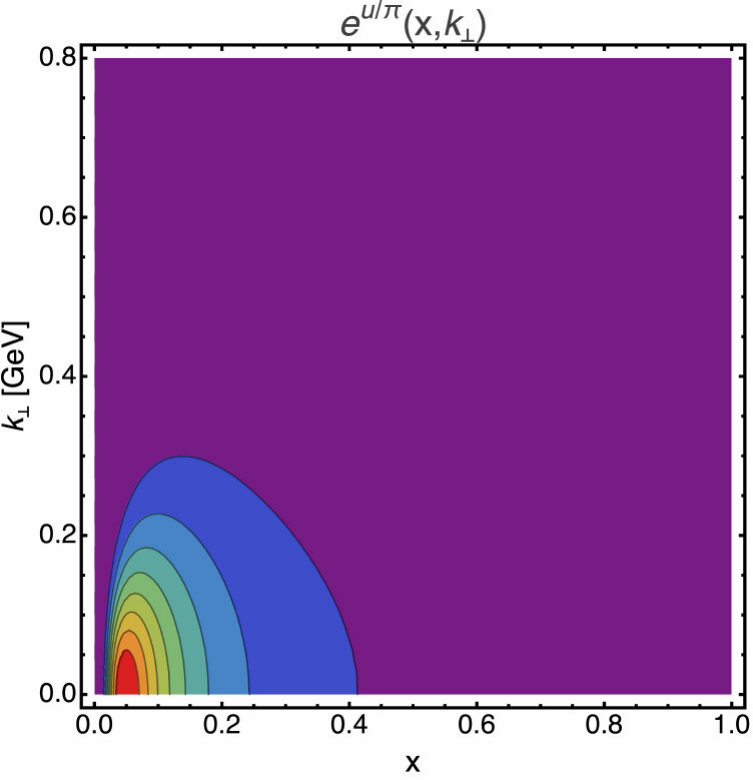}
\includegraphics[width=0.24\textwidth]{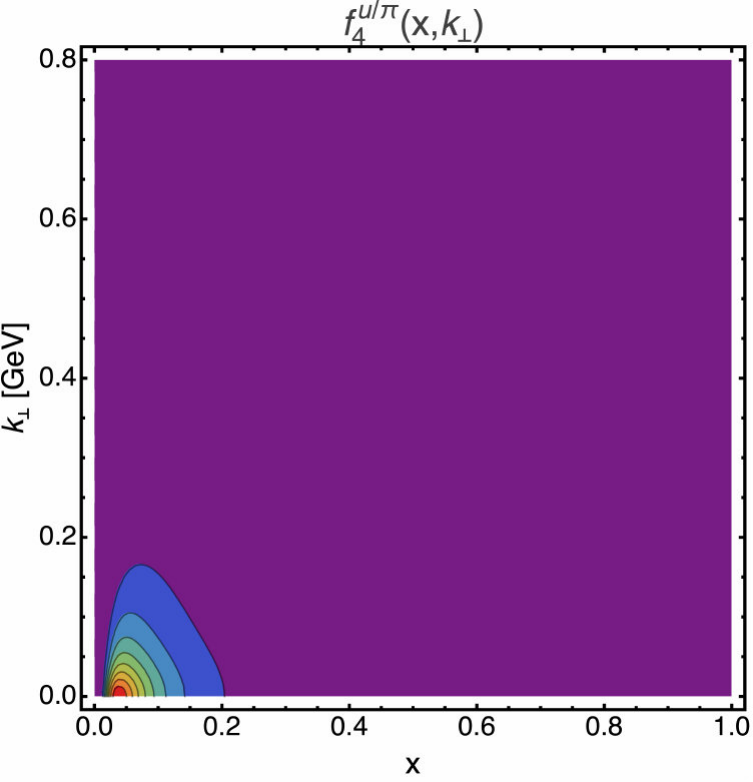}
\includegraphics[width=0.24\textwidth]{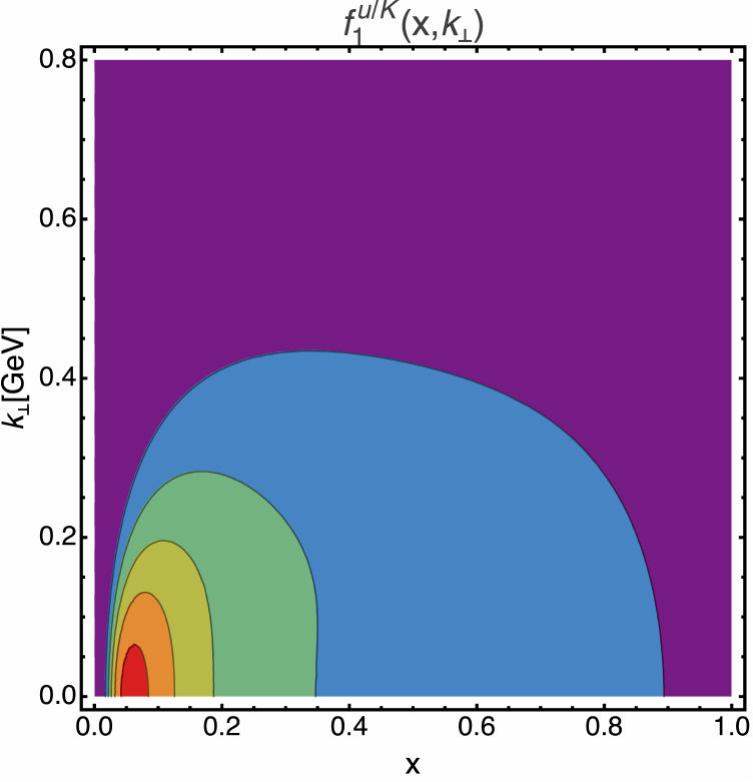}
\includegraphics[width=0.24\textwidth]{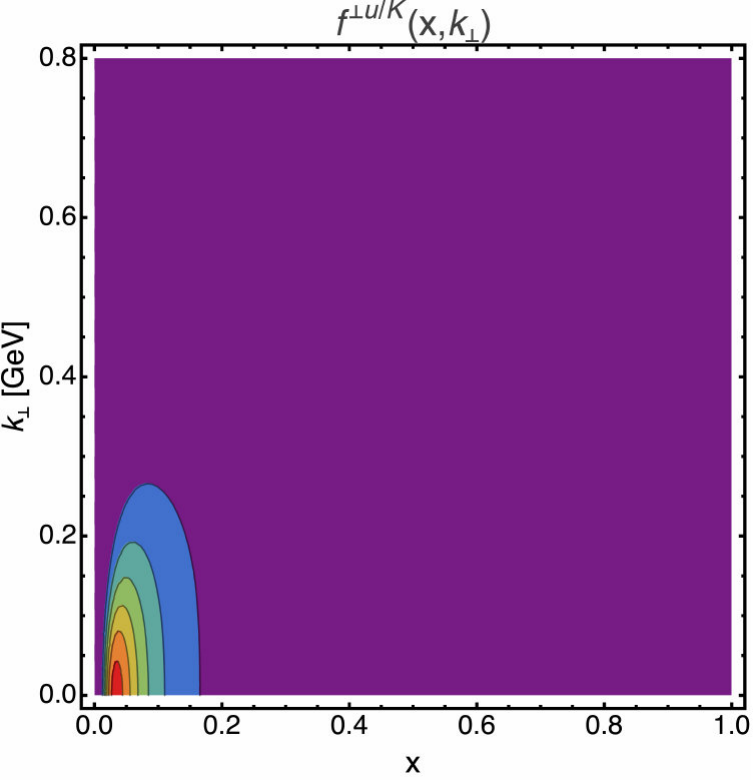}
\includegraphics[width=0.24\textwidth]{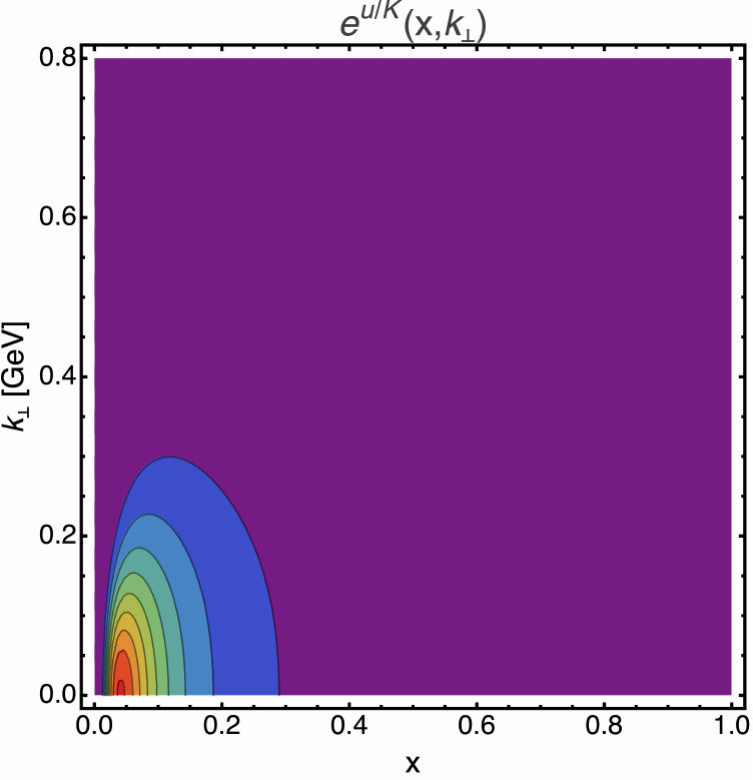}
\includegraphics[width=0.24\textwidth]{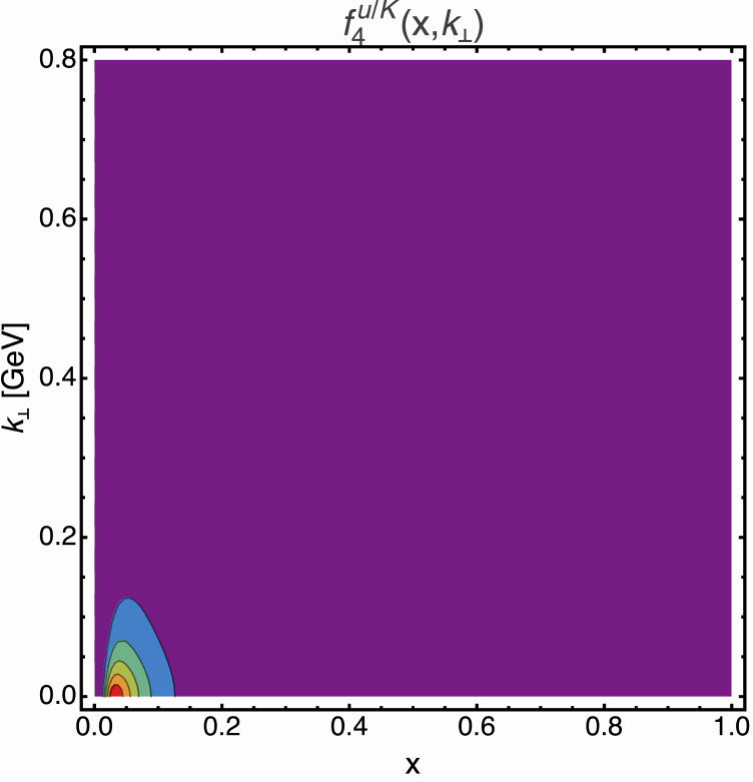}
\includegraphics[width=0.24\textwidth]{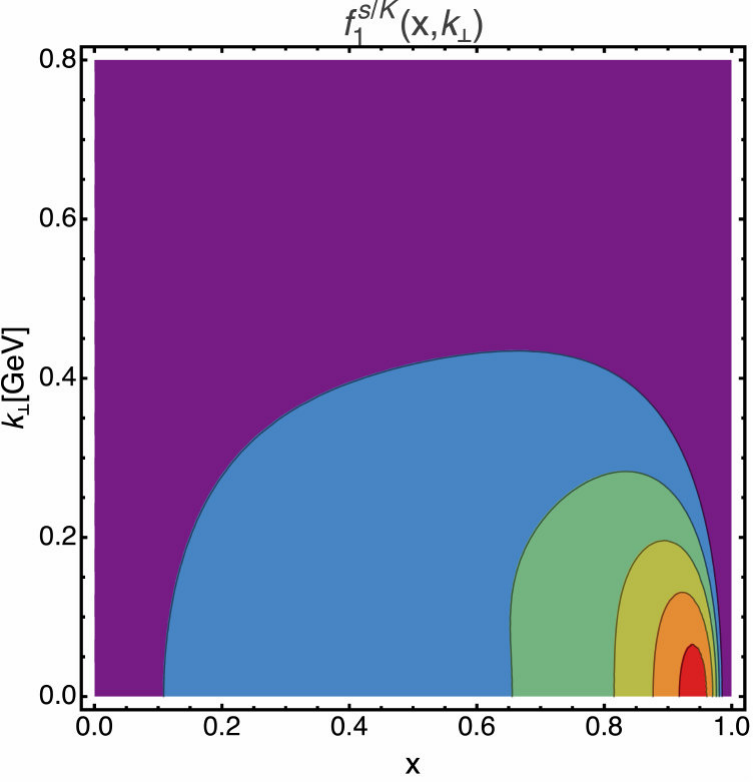}
\includegraphics[width=0.24\textwidth]{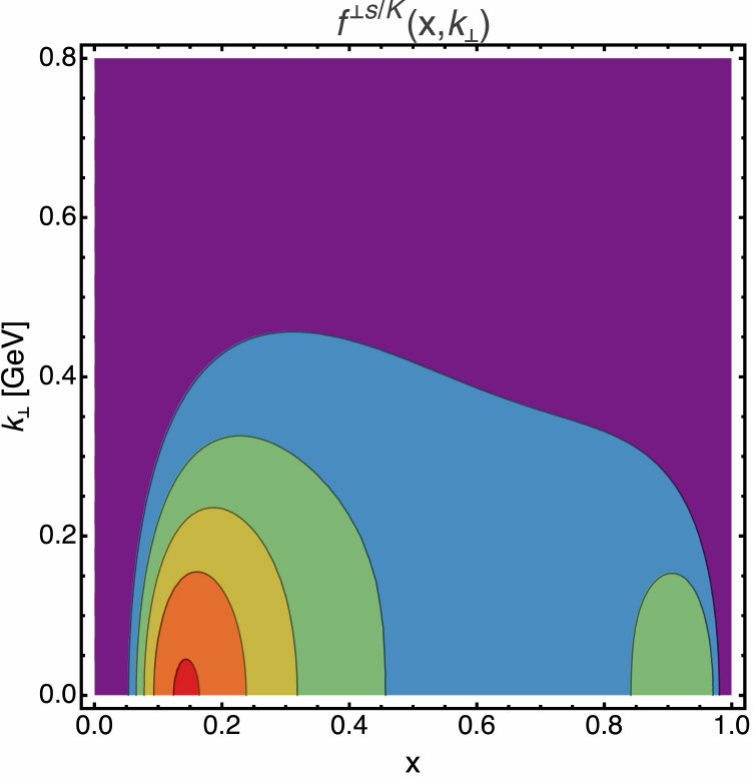}
\includegraphics[width=0.24\textwidth]{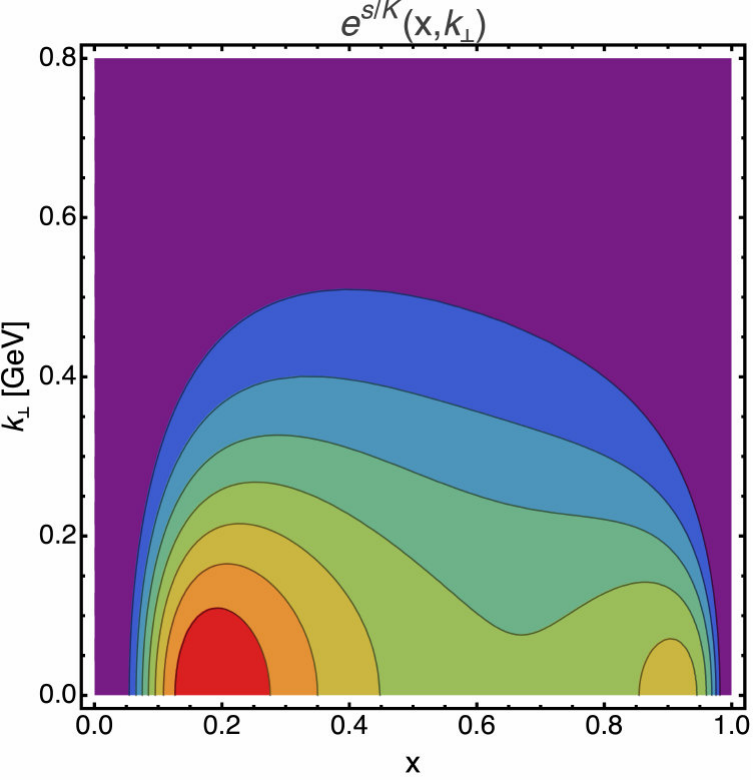}
\includegraphics[width=0.24\textwidth]{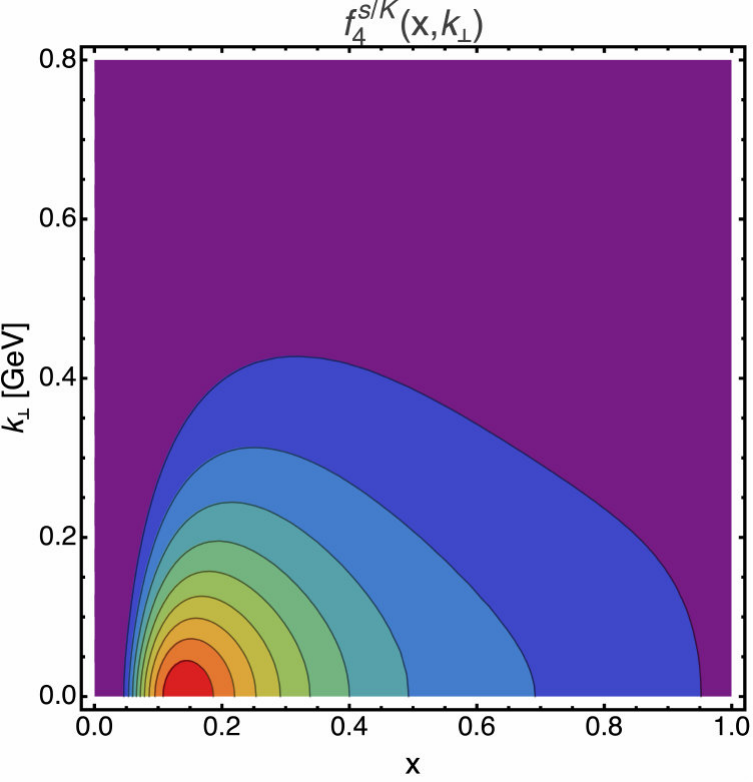}
\caption{Contour maps of the unpolarized T-even TMDs $(f_1^q,\,f^{\perp q},\,e^q,\,f_4^q)$ for
$u$ in $\pi^+$ (top), $u$ in $K^+$ (middle), and $s$ in $K^+$ (bottom), shown in the $(x,k_\perp)$
plane (color: red = high, violet = low).\label{Fig:TMD-contour}}
\end{figure*}

\begin{figure*}[t]
\centering
\includegraphics[width=0.4\textwidth]{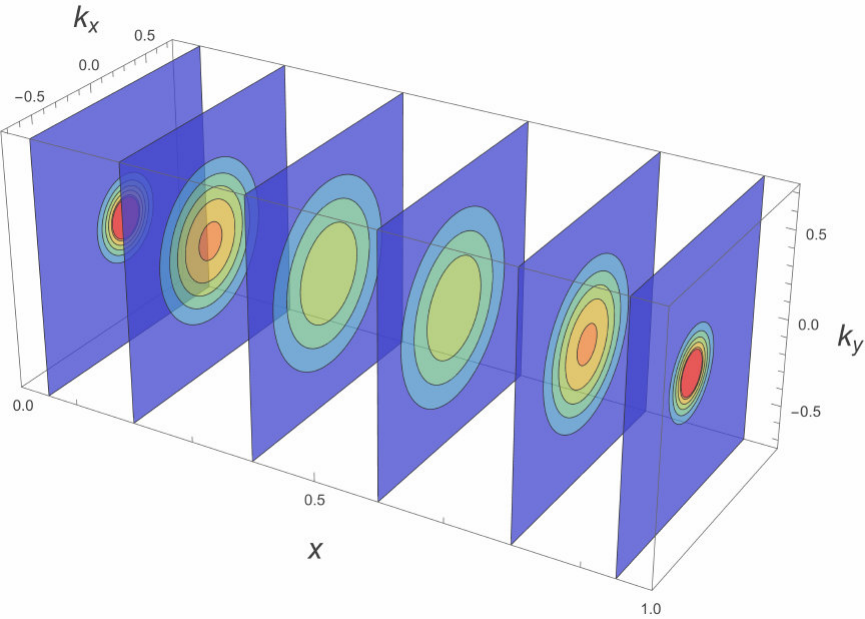}
\includegraphics[width=0.4\textwidth]{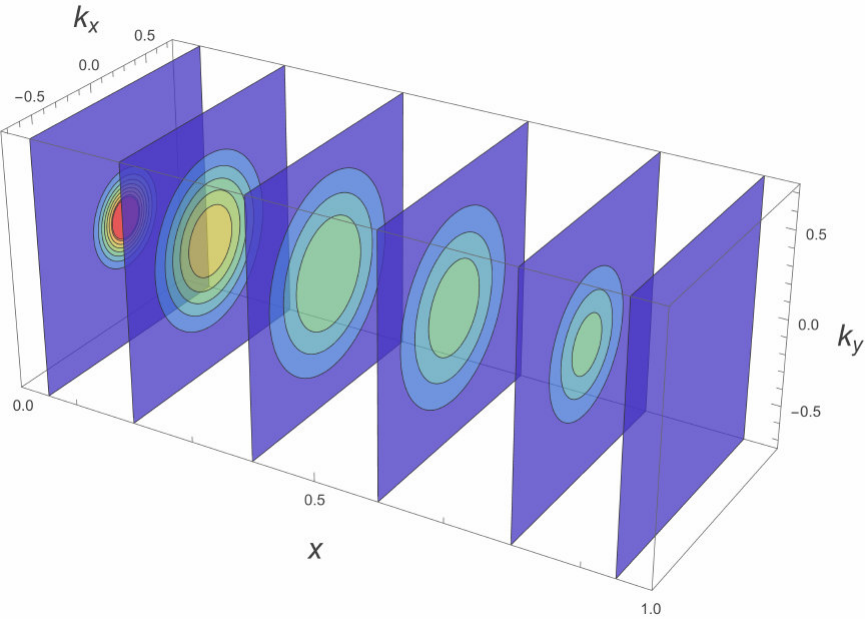}
\caption{Three-dimensional visualizations of the twist-2 TMD 
$f_1^q(x,\bm{k}_\perp)$ for the $u$ quark in the 
$\pi^+$ (left) and $K^+$ (right). Each vertical slice 
at fixed $x$ shows the transverse-momentum profile in 
the $(k_x,k_y)$ plane.\label{fig:3DTMD}}
\end{figure*}

\begin{figure*}[t]
\centering
\includegraphics[width=0.4\textwidth]{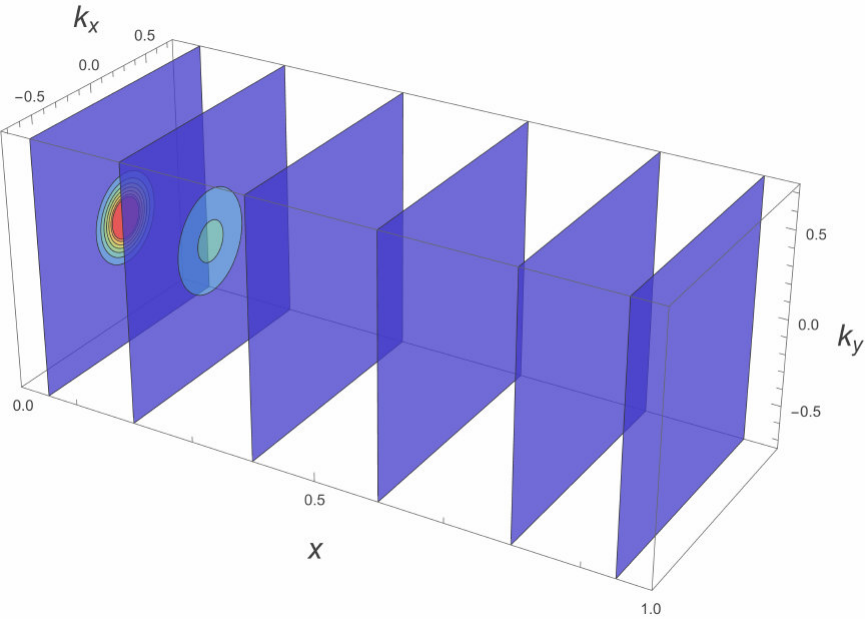}
\includegraphics[width=0.4\textwidth]{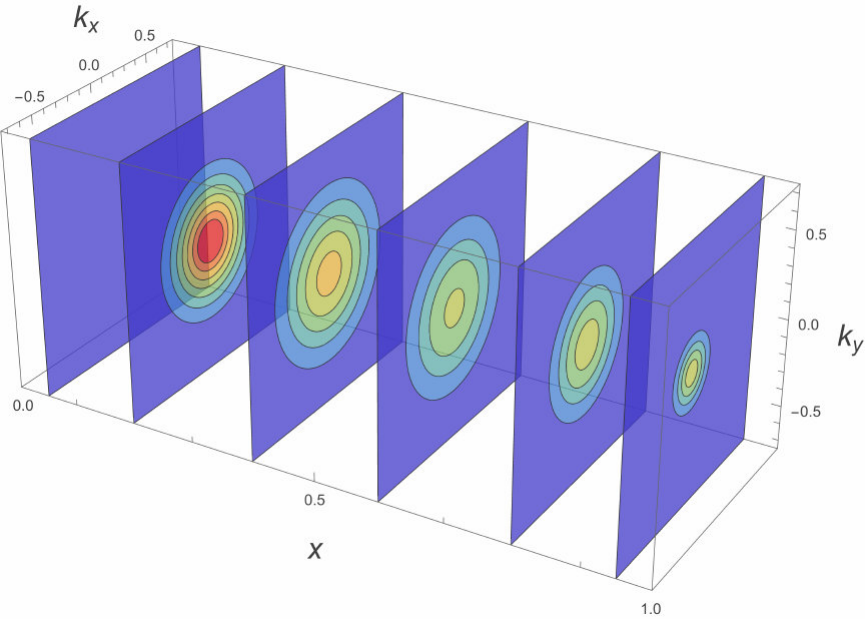}
\caption{Three-dimensional visualizations of the twist-3 TMD 
$e^{q/K}(x,\bm{k}_\perp)$ for the $u$ quark (left) and $s$ quark (right) in the $K^+$.\label{fig:3DscalarTMD}}
\end{figure*}

\section{Numerical results}
\label{sec:Num}
\subsection{TMDs and Transverse Moments}

Figure~\ref{Fig:TMD-contour} shows contour maps of the unpolarized T-even TMDs
$f(x,\bm{k}_\perp)$ 
with $f \in\{\allowbreak f_1^q,\,\allowbreak 
f^{\perp q},\,\allowbreak e^q,\,\allowbreak f_4^q\}$
for the $u$ quark in $\pi^+$ (top) and $K^+$ (middle),
and for the $s$ quark in $K^+$ (bottom).
The distributions are shown in the $(x,k_\perp)$ plane,
where $k_\perp \equiv |\bm{k}_\perp|$ and the color scale runs from violet (low) to red (high).

The twist–4 distribution $f_4^q(x,k_\perp)$ is associated with the
$\gamma^-$ projection and therefore involves the BT kinematic mass
${\cal M}_{\rm BT}$, which arises from the BT-consistent Lorentz prefactor and
contains an explicit angular dependence at the integrand level through a $\cos^2\theta$ term.
The results shown here correspond to the unpolarized TMD defined after azimuthal averaging over the transverse angle,
so that the final $f_4^q(x,k_\perp)$ depends only on $x$ and $k_\perp$.\footnote{To ensure the same normalization convention as 
for the angle–independent unpolarized TMDs ($f_1^q$, $f^{\perp q}$, $e^q$), the twist–4 distribution is defined after azimuthal averaging,
$f_4^q(x,k_\perp)=\frac{1}{2\pi}\int_0^{2\pi} d\theta f_4^q(x,k_\perp,\theta)$.
With this definition, $f_4^q(x,k_\perp)$ enters the transverse integral $\int d^2\bm{k}_\perp$ in the same way as the lower–twist TMDs, 
leading to the standard relation $\int d^2\bm{k}_\perp f_4^q = \pi\int d k_\perp^2 f_4^q$, and ensuring dependence only on $k_\perp^2$ as required for unpolarized TMDs.}

For the twist–2 pion TMD $f^{u/\pi}_1$, the $(x,k_\perp)$ contour
is symmetric under $x\!\leftrightarrow\!1-x$ and shows two highest–density (red) regions at small $k_\perp$ 
near the endpoints of $x$. As $k_\perp$ increases, the distribution decreases symmetrically about $x=\tfrac12$, 
reflecting SU(2) isospin symmetry with $m_u=m_d$.
After integration over transverse momentum, the corresponding PDF becomes single-peaked around $x=\tfrac12$.
In the kaon, SU(3) flavor breaking shifts the distributions:
$f^{u/K}_1$ is concentrated at smaller $x$, while $f^{s/K}_1$ shifts toward larger $x$,
leading to contour distributions shifted toward smaller (larger) $x$
for the $u$ ($s$) quark.

The higher-twist TMDs $e^q$ and $f_4^q$ inherit additional flavor dependence through
the BT-consistent mass factors $m_1/M_0$ and $(\bm{k}_\perp^2+m_1^2)/{\cal M}_{\rm BT}^2$,
which encode the LF zero-mode-sensitive structure of the scalar and $\gamma^-$ projections.
For the $u$ quark in both $\pi^+$ and $K^+$, these distributions are concentrated at small $x$ and small $k_\perp$.
For the $s$ quark in $K^+$, the support extends to larger $x$ and broader $k_\perp$.
The twist-4 distribution $f_4^{s/K}(x,k_\perp)$ is therefore broader and decreases smoothly with increasing $x$ and $k_\perp$.

Figure~\ref{fig:3DTMD} displays 3D visualizations of the twist-2 TMD $f_1^q(x,\bm{k}_\perp)$ for the $u$ quark 
in the $\pi^+$ (left panel) and $K^+$ (right panel). Each vertical slice at a fixed value of $x$ represents 
the transverse-momentum profile in the $(k_x,k_y)$ plane.

The most striking difference between the two distributions reflects SU(3) flavor-symmetry breaking. For the 
pion, the distribution is symmetric about $x=1/2$, with prominent lobes appearing in the small- and large-$x$ regions
at small transverse momentum. This $x\leftrightarrow 1-x$ symmetry follows from the equal constituent masses in the pion, $m_u=m_d$, 
which gives $f_1^{u/\pi}(x,\bm{k}_\perp) = f_1^{u/\pi}(1-x,\bm{k}_\perp)$.
The two-lobe structure is governed by the competition between the Jacobian factor and the Gaussian suppression in the LFWF. 
Since $f_1^q(x,\bm{k}_\perp) \propto |\phi(x,\bm{k}_\perp)|^2 
\propto (\partial k_z/\partial x)\,\exp\!\left[-(\bm{k}_\perp^2+k_z^2)/\beta^2\right]$,
the Jacobian enhances the near-endpoint regions, while the Gaussian factor suppresses the exact endpoints where 
$|k_z|\to\infty$. In the pion case, this competition is symmetric under $x\leftrightarrow 1-x$, leading to two 
symmetric lobes.

In contrast, for the kaon, the distribution is strongly asymmetric: a single dominant lobe appears at small but 
nonzero $x$ and rapidly diminishes toward larger $x$. Although the Jacobian also grows as $x\to1$, the heavier 
strange antiquark mass causes the internal longitudinal momentum to increase rapidly in this region. For the $u$ quark in the $K^+$, 
$k_z \approx \frac{1}{2}M_0 \sim \frac{1}{2}\sqrt{(\bm{k}_\perp^2 + m_s^2)/(1-x)}$
as $x\to 1$, so that $e^{-k_z^2/\beta^2} \sim \exp\!\left(-(\bm{k}_\perp^2+m_s^2)/[4\beta^2(1-x)]\right) \to 0$.
This exponential suppression overwhelms the power-like growth of the Jacobian and removes the large-$x$ lobe.%
\footnote{More precisely, both the pion and kaon share 
the same convergence form in the $x\to 1$ limit, 
$\frac{\partial k_z}{\partial x}\,e^{-k_z^2/\beta^2}
\sim \epsilon^{-3/2}\,e^{-C/\epsilon}\to 0$ as 
$\epsilon=1-x\to 0$, where $C = ({\bm k}_\perp^2 + m_{\rm spec}^2)/(4\beta^2)$
with $m_{\rm spec}$ being the spectator antiquark mass.
For the kaon, $C^{(K)}=({\bm k}_\perp^2+m_s^2)/(4\beta^2)$ is larger than the corresponding pion value
$C^{(\pi)}=({\bm k}_\perp^2+m_u^2)/(4\beta^2)$ because $m_s>m_q$. Thus, the Gaussian suppression is stronger 
in the kaon. The large-$x$ lobe visible in the pion is the mirror image of the small-$x$ lobe under the 
$x\leftrightarrow 1-x$ symmetry, which is absent in the kaon due to $m_u\neq m_s$.}
At small $x$, the corresponding suppression involves the lighter active $u$-quark mass, allowing the dominant 
lobe to remain at small but nonzero $x$. Thus, the single-lobe structure of $f_1^{u/K}$ arises from the interplay between 
the Jacobian enhancement and the Gaussian suppression, modulated by the constituent mass difference $m_s>m_u$.
These features are fully consistent with the 2D contour maps shown in Fig.~\ref{Fig:TMD-contour}, where 
$f_1^{u/\pi}(x,\bm{k}_\perp)$ exhibits a symmetric two-lobe structure in $x$, 
whereas $f_1^{u/K}(x,\bm{k}_\perp)$ is concentrated toward smaller values of $x$.

Figure~\ref{fig:3DscalarTMD} shows 3D visualizations of the
twist-3 scalar TMD $e^{q/K}(x,\bm{k}_\perp)$ for the $u$ quark (left panel)
and $s$ quark (right panel) in the $K^+$. 
The overall behavior again reflects SU(3) flavor-symmetry breaking in the
kaon. The $u$-quark distribution is concentrated in the small-$x$ region,
whereas the $s$-quark distribution is shifted toward larger values of $x$.
This complementary pattern follows from the unequal constituent masses in the
$K^+$: the lighter $u$ quark tends to carry a smaller longitudinal momentum
fraction, while the heavier strange quark carries a larger fraction. Apart
from this shift in the longitudinal distribution, the transverse-momentum
profiles show a similar Gaussian-like falloff, as expected from the common
radial LFWF structure.
\begin{table*}[t]
\centering
\setlength{\tabcolsep}{9pt}
\renewcommand{\arraystretch}{1.5}
\begin{tabular}{lllcccc}
\hline\hline
Model & Hadron & Quantity & $f^u_{1}$ & $e^u$ & $f^{\perp u}$ & $f^u_{4}$ \\
\hline
\multirow{6}{*}{This Work}
  & \multirow{3}{*}{$\pi^{+}(u\bar d)$}
    & $\langle k_\perp\rangle$ & 0.324 & 0.261 & 0.291 & 0.269 \\
  & & $\langle k_\perp^2\rangle^{1/2}$ & 0.366 & 0.301 & 0.335 & 0.317 \\
  & & $R_G$ & 1 & 0.976 & 0.982 & 0.956 \\
  & \multirow{3}{*}{$K^{+}(u\bar s)$}
    & $\langle k_\perp\rangle$ & 0.344 & 0.274 & 0.299 & 0.284 \\
  & & $\langle k_\perp^2\rangle^{1/2}$ & 0.389 & 0.317 & 0.345 & 0.336 \\
  & & $R_G$ & 1 & 0.974 & 0.977 & 0.954 \\
\hline
\multirow{6}{*}{LCQM~\cite{Puhan:2023ekt}}
  & \multirow{3}{*}{$\pi^{+}(u\bar d)$}
    & $\langle k_\perp\rangle$ & 0.22 & 0.18 & 0.21 & 0.21 \\
  & & $\langle k_\perp^2\rangle^{1/2}$ & 0.26 & 0.22 & 0.25 & 0.24 \\
  & & $R_G$ & 0.96 & 0.95 & 0.96 & 0.95 \\
  & \multirow{3}{*}{$K^{+}(u\bar s)$}
    & $\langle k_\perp\rangle$ & 0.26 & 0.21 & 0.23 & 0.23 \\
  & & $\langle k_\perp^2\rangle^{1/2}$ & 0.30 & 0.25 & 0.27 & 0.27 \\
  & & $R_G$ & 0.98 & 0.96 & 0.96 & 0.96 \\
\hline
\multirow{6}{*}{LFHM~\cite{Puhan:2023ekt}}
  & \multirow{3}{*}{$\pi^{+}(u\bar d)$}
    & $\langle k_\perp\rangle$ & 0.24 & 0.21 & 0.23 & 0.22 \\
  & & $\langle k_\perp^2\rangle^{1/2}$ & 0.27 & 0.24 & 0.26 & 0.25 \\
  & & $R_G$ & 1.00 & 0.99 & 0.99 & 0.99 \\
  & \multirow{3}{*}{$K^{+}(u\bar s)$}
    & $\langle k_\perp\rangle$ & 0.24 & 0.21 & 0.22 & 0.22 \\
  & & $\langle k_\perp^2\rangle^{1/2}$ & 0.27 & 0.24 & 0.25 & 0.25 \\
  & & $R_G$ & 1.00 & 0.99 & 0.99 & 0.99 \\
\hline
\multirow{3}{*}{LFCM~\cite{Lorce:2016} }
  & \multirow{3}{*}{$\pi^{+}(u\bar d)$}
    & $\langle k_\perp\rangle$ & 0.28 & 0.26 & 0.26 & 0.30 \\
  & & $\langle k_\perp^2\rangle^{1/2}$ & 0.32 & 0.30 & 0.30 & 0.33 \\
  & & $R_G$ & 0.99 & 0.99 & 0.99 & 0.98 \\
\hline
\multirow{2}{*}{BLFQ \cite{Zhu23}}
  & \multirow{2}{*}{$\pi^{+}(u\bar d)$}
    & $\langle k_\perp\rangle$ & 0.26 & 0.26 & 0.25 & -- \\
  & & $\langle k_\perp^2\rangle^{1/2}$ & 0.30 & 0.30 & 0.29 & -- \\
  & & $R_G$ & 0.98 & 0.98 & 0.97 & -- \\
\hline
\end{tabular}
\caption{ Values of $\langle k_\perp\rangle$ and $\langle k_\perp^2\rangle^{1/2}$ (in GeV), and the ratio $R_G$, 
for the unpolarized $T$-even $u$-quark TMDs
in the $\pi^+$ and $K^+$. For $f^q_1$, the exact Gaussian form of the $x$-integrated
transverse profile gives $R_G=1$ analytically.\label{tab:TMDu-piK}}
\end{table*}
\begin{table*}[t]
\centering
\setlength{\tabcolsep}{8pt}
\renewcommand{\arraystretch}{1.5}
\begin{tabular}{lllcccc}
\hline\hline
Model & Hadron & Quantity & $f^s_{1}$ & $e^s$ & $f^{\perp s}$ & $f^s_{4}$ \\
\hline
\multirow{3}{*}{This Work}
  & \multirow{3}{*}{$K^{+}(u\bar s)$}
    & $\langle k_\perp\rangle$ & 0.344 & 0.310 & 0.340 & 0.303 \\
  & & $\langle k_\perp^2\rangle^{1/2}$ & 0.389 & 0.352 & 0.384 & 0.350 \\
  & & $R_G$ & 1 & 0.993 & 0.999 & 0.978 \\
\hline
\multirow{3}{*}{LCQM~\cite{Puhan:2023ekt}}
  & \multirow{3}{*}{$K^{+}(u\bar s)$}
    & $\langle k_\perp\rangle$ & 0.26 & 0.21 & 0.23 & 0.18 \\
  & & $\langle k_\perp^2\rangle^{1/2}$ & 0.30 & 0.25 & 0.27 & 0.21 \\
  & & $R_G$ & 0.98 & 0.96 & 0.96 & 0.97 \\
\hline
\multirow{3}{*}{LFHM~\cite{Puhan:2023ekt}}
  & \multirow{3}{*}{$K^{+}(u\bar s)$}
    & $\langle k_\perp\rangle$ & 0.23 & 0.20 & 0.21 & 0.19 \\
  & & $\langle k_\perp^2\rangle^{1/2}$ & 0.26 & 0.23 & 0.24 & 0.22 \\
  & & $R_G$ & 1 & 0.98 & 0.99 & 0.97 \\
\hline
\end{tabular}
\caption{ Values of $\langle k_\perp\rangle$ and $\langle k_\perp^2\rangle^{1/2}$ (in GeV), and the ratio $R_G$, 
for the unpolarized $T$-even $s$-quark TMDs in the $K^+$. For $f^q_1$, the exact Gaussian form
of the $x$-integrated transverse profile gives $R_G=1$ analytically.\label{tab:TMDs-K}}
\end{table*}

For a generic TMD $f(x,\bm{k}_\perp)$, the $n$th transverse moment ($n\ge 1$) is defined as
\begin{equation}
\label{eq:k-moments-def}
\langle k_\perp^{\,n}\rangle
=\frac{\displaystyle \int_0^1\!dx\int d^2\bm{k}_\perp\, k_\perp^{\,n}\, f(x,\bm{k}_\perp)}
{\displaystyle \int_0^1\!dx\int d^2\bm{k}_\perp\, f(x,\bm{k}_\perp)}.
\end{equation}
If the $k_\perp$ dependence is isotropic and Gaussian in the transverse plane, $f(x,{\bm k}_\perp)\propto
\exp[- {\bm k}_\perp^{2}/(2\sigma^{2})]$, one obtains
$\langle k_\perp\rangle=\sigma\sqrt{\pi/2}$ and $\langle k_\perp^{2}\rangle=2\sigma^{2}$, implying
\begin{equation}
\label{eq:RG}
R_G \equiv \frac{2}{\sqrt{\pi}}\,
\frac{\langle k_\perp\rangle}{\langle k_\perp^2\rangle^{1/2}} = 1.
\end{equation}
We use $R_G$ as a simple indicator of how closely the transverse–momentum distribution follows a Gaussian shape:
$R_G=1$ for an exact Gaussian profile, while deviations from unity quantify non-Gaussian behavior. 
This diagnostic analysis provides useful  means to assess the accuracy of the Gaussian Ansatz often adopted to describe the transverse–momentum dependence in DIS and related analyses~\cite{Boffi09,DALESIO2008394,Sch10}.

In our fBT-LFQM, at the model scale, the Gaussian LFWF
$\phi(x,{\bm k}_\perp)$ is given by Eq.~\eqref{eq:8}, and the twist-2 TMD is
expressed as
$f^q_1(x,{\bm k}_\perp)\propto|\phi(x,\bm{k}_\perp)|^2$.
Using the variable transformation
$(\partial k_z/\partial x)\,dx=dk_z$,
the $x$ integration of $f_1^q$ can be converted into an integral
over $k_z$, so that its $x$-integrated transverse profile is obtained as the
transverse projection of a three-dimensional isotropic Gaussian in
$(k_x,k_y,k_z)$. In particular, the $x$-integrated transverse-momentum profile
$f_1^q({\bm k}_\perp)\equiv
\int_0^1\!dx\,f_1^q(x,{\bm k}_\perp)$ is a purely 2D Gaussian,
\begin{equation}\label{eq:gauss2d}
\int_{-\infty}^{\infty}\!dk_z\;
e^{-({\bm k}_\perp^2+k_z^2)/\beta^2}
\;\propto\;
e^{-{\bm k}_\perp^2/\beta^2},
\end{equation}
so that
$f_1^q({\bm k}_\perp)\propto
\exp(-{\bm k}_\perp^2/\beta^2)$.
Comparing with the generic form
$\exp(-{\bm k}_\perp^2/2\sigma^2)$ gives
$\sigma=\beta/\sqrt{2}$, hence
\[
\langle k_\perp^2\rangle=\beta^2,\qquad
\langle k_\perp\rangle=\frac{\beta}{2}\sqrt{\pi},
\]
and consequently $R_G=1$ holds exactly for this ansatz.
Thus, for the $x$-integrated twist-2 profile at the model scale,
the intrinsic root-mean-square (RMS) transverse width
$\langle k_\perp^2\rangle^{1/2}$ is directly determined by the variational
parameter $\beta$.
This exact Gaussian form is a property of the unevolved model TMD
and should not be interpreted as being preserved under QCD TMD evolution.

The exact Gaussian result for $f_1^q({\bm k}_\perp)$, and hence $R_G=1$, relies on the
Jacobian factor $\sqrt{\partial k_z/\partial x}$ in $\phi(x,{\bm k}_\perp)$. 
With this factor included, the $x$–integration
becomes an integral over $k_z$ and yields $f_1^q({\bm k}_\perp)\propto e^{- {\bm k}_\perp^2/\beta^2}$. 
If the Jacobian factor is omitted, an additional weight $(\partial x/\partial k_z)$ remains inside the $k_z$ integral, 
yielding a $k_\perp$–dependent prefactor. The transverse profile then ceases to be purely Gaussian, and in general $R_G\neq 1$.

\begin{figure}[htbp]
\centering
\includegraphics[width=0.48\textwidth]{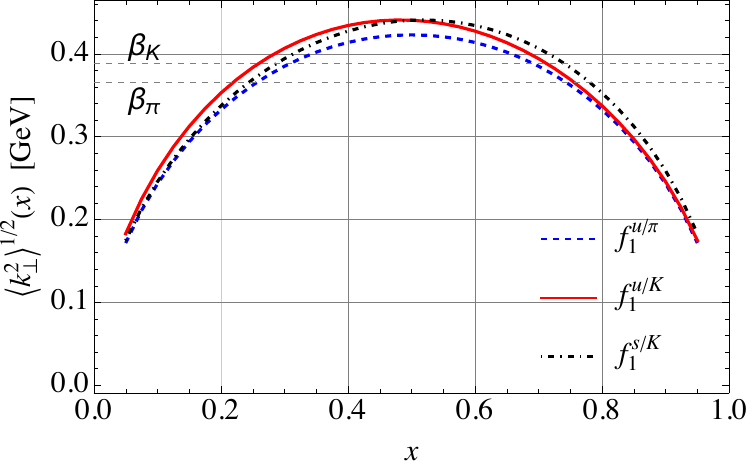}
\hfill
\includegraphics[width=0.48\textwidth]{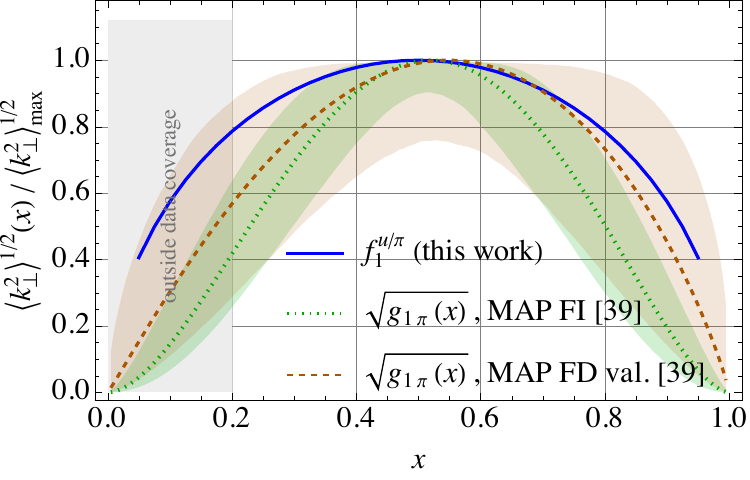}
\caption{(Left) $x$-dependent transverse width
$\langle k_\perp^2\rangle^{1/2}(x)$ of the twist-2 TMD
$f_1^q(x,\bm{k}_\perp)$ for the $u$ quark in $\pi^+$ (dashed) and for the
$u$ (solid) and $s$ (dot-dashed) quarks in $K^+$. The horizontal dashed
lines indicate the $x$-integrated widths
$\beta_\pi\equiv\beta_{q\bar q}$ and $\beta_K\equiv\beta_{q\bar s}$
quoted in Tables~\ref{tab:TMDu-piK} and~\ref{tab:TMDs-K}.
(Right) Our pion transverse width from the left panel compared with the intrinsic
width $\sqrt{g_{1\pi}(x)}$ of the pion TMD extraction of
Ref.~\cite{Rossi:2025plb}, obtained from Eq.~\eqref{eq:g1pi} with its
flavor-independent (FI) and flavor-dependent valence (FD val.)
parameters, each curve normalized to its own maximum. The shaded region
$x<0.2$ lies outside the kinematic coverage of the Drell--Yan data used
there. See the text for the meaning of the bands.}
\label{fig:widthx}
\end{figure}
It should be emphasized that $R_G$ and the widths quoted in
Tables~\ref{tab:TMDu-piK} and~\ref{tab:TMDs-K} characterize the
$x$-integrated transverse profile, for which the exact Gaussian form of
Eq.~\eqref{eq:gauss2d} holds with the constant width $\beta$. At fixed
$x$, by contrast, the LFWF of Eq.~\eqref{eq:8} depends on
$\bm{k}_\perp$ both through $k_z(x,\bm{k}_\perp)$ and through the
Jacobian $\partial k_z/\partial x$, so that the effective width
\begin{equation}
\langle k_\perp^2\rangle^{1/2}(x)
=\left[\frac{\int d^2\bm{k}_\perp\, k_\perp^2\,
  f_1^q(x,\bm{k}_\perp)}
 {\int d^2\bm{k}_\perp\, f_1^q(x,\bm{k}_\perp)}\right]^{1/2}
\label{eq:widthx}
\end{equation}
is genuinely $x$ dependent. This is shown in the left panel of Fig.~\ref{fig:widthx}. 
The width varies substantially with $x$, being maximal near
$x\simeq 1/2$ and suppressed toward both endpoints,
where the rapid growth of $M_0(x,\bm{k}_\perp)$ enhances the
Gaussian suppression at large $k_\perp$. 
The $x$-integrated twist-2 RMS widths listed in Tables~\ref{tab:TMDu-piK} and~\ref{tab:TMDs-K} are
recovered as the $f_1^{q/H}(x)$-weighted averages $(H=\pi,K)$,
\be\label{eq:xweighted-width}
\int_0^1 dx\, f_1^{q/H}(x)
\left\langle k_\perp^2\right\rangle_{q/H}(x)
= \langle k_\perp^2\rangle_{q/H}=
\beta_H^2,
\ee
so that the $x$-integrated profile and the fixed-$x$ widths are mutually
consistent.
For the pion the curve is symmetric under $x\leftrightarrow
1-x$, while in the kaon the $u$- and $s$-quark widths are exact mirror
images of each other,
$\langle k_\perp^2\rangle^{1/2}_{s/K}(x)
=\langle k_\perp^2\rangle^{1/2}_{u/K}(1-x)$,
which follows from the invariance of $M_0$ under the simultaneous
interchange $x\leftrightarrow 1-x$ and $m_1\leftrightarrow m_2$.
Phenomenological extractions of the pion TMDs~\cite{Vladimirov:2019bfa,
Cerutti:2022lmb,Barry:2023qqh,Rossi:2025plb} likewise introduce an
explicitly $x$-dependent width in the nonperturbative transverse
structure. 
In the MAP parametrizations~\cite{Cerutti:2022lmb,Rossi:2025plb}, the
Fourier-transformed TMD PDF $\hat f_1^{\,q}(x,\bm{b}_T;\mu,\zeta)$ is
factorized into a perturbatively calculable part, matched onto the
collinear PDF at small $b_T$, and a nonperturbative factor $f_{\rm NP}$
fitted to data. Its intrinsic transverse structure is taken as a Gaussian
in $b_T$, $f_{\rm NP}\propto\exp[-g_{1\pi}(x)\bm{b}_T^2/4]$, with
\be\label{eq:g1pi}
g_{1\pi}(x)=N_\pi\,
\frac{x^{\sigma_\pi}(1-x)^{\alpha_\pi^2}}
     {\hat x^{\sigma_\pi}(1-\hat x)^{\alpha_\pi^2}},
\qquad \hat x=0.1,
\ee
so that, upon Fourier transformation, $g_{1\pi}(x)$ plays the role of the
intrinsic transverse-momentum variance
$\langle k_\perp^2\rangle_{\rm NP}(x)$ associated with the
nonperturbative Gaussian factor.%
\footnote{For a two-dimensional Fourier transform, the transverse
second moment associated with a normalized $b_T$-space factor satisfies
$\langle k_\perp^2\rangle_{\rm NP}(x)
=-\big[\nabla_b^2 f_{\rm NP}/f_{\rm NP}\big]_{\bm{b}_T=0}$.
For the intrinsic Gaussian factor
$f_{\rm NP}=e^{-g_{1\pi}(x)\bm{b}_T^2/4}$, the two-dimensional
Laplacian gives
$\nabla_b^2 f_{\rm NP}|_{0}=-g_{1\pi}(x)$, hence
$\langle k_\perp^2\rangle_{\rm NP}(x)=g_{1\pi}(x)$.
Including the Collins--Soper term of Eq.~(8) of
Ref.~\cite{Rossi:2025plb}, the nonperturbative factor reads
$f_{\rm NP}
=\exp\big[-\tfrac{\bm{b}_T^2}{4}
\big(g_{1\pi}(x)+g_2\ln(\zeta/Q_0)\big)\big]$.
Thus, the Collins--Soper evolution contributes an $x$-independent
additive term to the transverse-momentum variance; the curves shown
correspond to the intrinsic width $g_{1\pi}(x)$ itself.}

The right panel of Fig.~\ref{fig:widthx} compares the $x$ dependence of
$\sqrt{g_{1\pi}(x)}$, evaluated with the parameters of the most recent
extraction of Ref.~\cite{Rossi:2025plb} for both its flavor-independent
(FI) fit and the valence channel of its flavor-dependent (FD) fit, with
our pion result
$\langle k_\perp^2\rangle_{u/\pi}^{1/2}(x)$, each curve being normalized
to its own maximum. Over the region covered by the Drell--Yan data
($x\gtrsim0.2$), the extracted widths and our model result exhibit a
similar qualitative $x$ dependence, with a broad maximum around
$x\simeq1/2$.

This comparison is limited in four respects. First, the two quantities
are different objects: ours is the second moment of the full model TMD
at $\mu_0$, whereas $g_{1\pi}(x)$ characterizes only the intrinsic
nonperturbative Gaussian factor of the MAP parametrization, so that only
their $x$-dependent shapes are compared. Second, the bands shown here are
intended only to illustrate the sensitivity to the quoted $1\sigma$
uncertainties of $\sigma_\pi$ and $\alpha_\pi$. They should not be
interpreted as the full uncertainty bands of
Ref.~\cite{Rossi:2025plb}, since parameter correlations are not
published and hence are not included. Third, the extracted width would
be symmetric under $x\leftrightarrow1-x$ for
$\sigma_\pi=\alpha_\pi^2$, a condition not imposed in the fit but
compatible with the quoted parameter uncertainties, so that the
apparent asymmetry should not be overinterpreted. Fourth, a quantitative
comparison of the full distributions would require TMD evolution from
$\mu_0$ to the scale of the data, which deserves further investigations.

In Table~\ref{tab:TMDu-piK} we summarize our predictions for 
$\langle k_\perp\rangle$, $\langle k_\perp^2\rangle^{1/2}$, and the dimensionless ratio 
$R_G$ for the unpolarized $T$-even $u$-quark TMDs in $\pi^+$ and $K^+$, together with other light-front
model results from Refs.~\cite{Lorce:2016,Puhan:2023ekt,Zhu23}. 
Table~\ref{tab:TMDs-K} reports the corresponding quantities for the unpolarized $T$-even $s$-quark 
TMDs in $K^+$, compared with Ref.~\cite{Puhan:2023ekt}.

The mean and RMS transverse moments,
$\langle k_\perp\rangle$ and $\langle k_\perp^2\rangle^{1/2}$, exhibit the general hierarchy
$f_1^q > f^{\perp q} > \{e^q,\, f_4^q\}$ across the unpolarized $T$-even TMDs,
with the leading-twist distribution being the broadest in every case.
For the $s$ quark in $K^+$, however, the twist-3 distribution $f^{\perp s}$
is nearly degenerate with $f_1^s$.
The relative ordering of $e^q$ and $f_4^q$ is flavor dependent:
for the $u$ quark in both $\pi^+$ and $K^+$ one finds
$f_4^q > e^q$, whereas for the $s$ quark in $K^+$ the two are nearly degenerate,
with $e^s$ slightly larger.
This ordering follows from the $\bm{k}_\perp$ dependence of the kinematic factors
relating the higher-twist TMDs to $f_1^q$ in Eq.~\eqref{eq:QCDrelBT}, rather than from
twist counting itself, since $\langle k_\perp\rangle$ and $\langle k_\perp^2\rangle^{1/2}$
are normalized ratios and are therefore insensitive to any overall suppression factor.
The factor $1/x$ relating $f^{\perp q}$ to $f_1^q$ carries no $\bm{k}_\perp$ dependence
but reweights the distribution toward smaller $x$, where the intrinsic width
$\langle k_\perp^2\rangle^{1/2}(x)$ of Fig.~\ref{fig:widthx} is smaller, so that
$f^{\perp q}$ is narrower than $f_1^q$.
For $e^q$, the additional factor
$m_1/M_0(x,\bm{k}_\perp)$ decreases monotonically with $k_\perp$
at fixed $x$ and therefore provides an additional transverse suppression
beyond the small-$x$ reweighting. For $f_4^q$, by contrast, the growth of
$\mathcal{M}_{\rm BT}^2$ in the denominator of
$(\bm{k}_\perp^2+m_1^2)/\mathcal{M}_{\rm BT}^2$ is partially compensated by
the explicit $\bm{k}_\perp^2$ growth in the numerator. Consequently, the net
transverse suppression of $f_4^q$ is weaker than that of $e^q$.
For the heavier $s$ quark, $m_s/M_0$ varies more slowly with $k_\perp$, which weakens the
suppression of $e^s$ and restores $e^s \gtrsim f_4^s$.
Since $M_0$ depends strongly on $x$ and $\bm{k}_\perp$, however, this hierarchy need not be
manifest after integration over phase space, and the relative magnitudes of $e^q$ and
$f^{\perp q}$ in the $(x,k_\perp)$ contour plots reflect this kinematic averaging.

\begin{figure*}[t]
\centering
\includegraphics[width=0.24\textwidth]{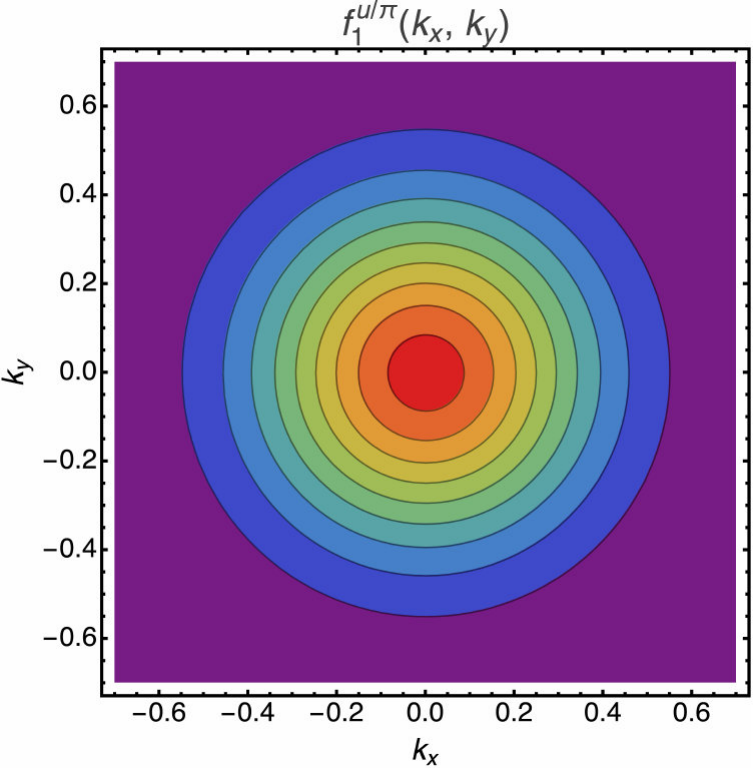}
\includegraphics[width=0.24\textwidth]{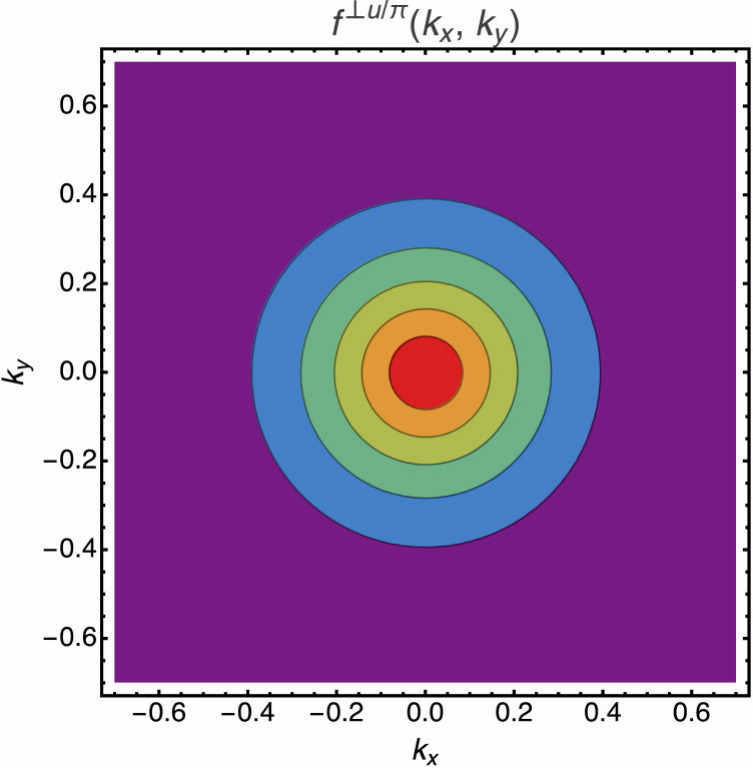}
\includegraphics[width=0.24\textwidth]{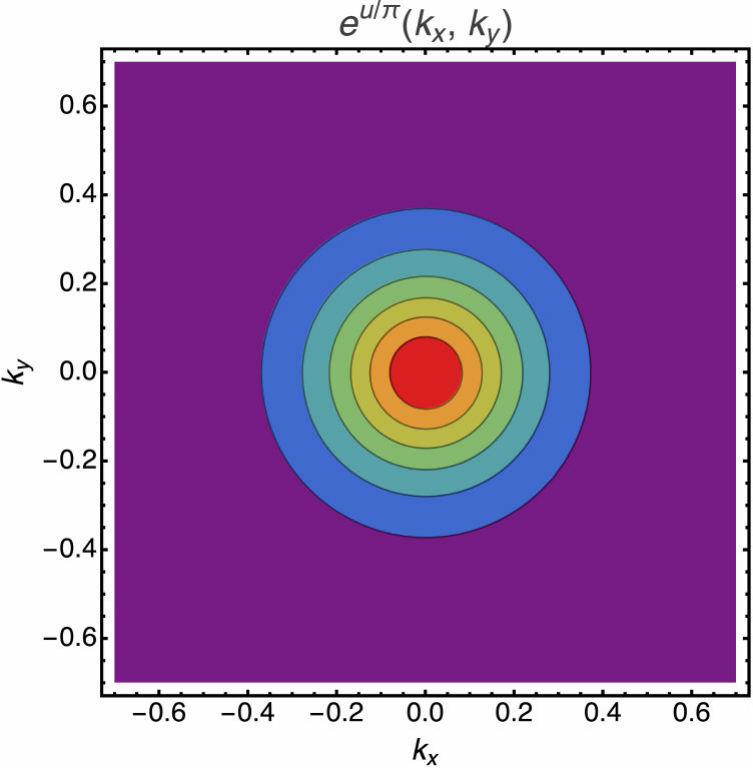}
\includegraphics[width=0.24\textwidth]{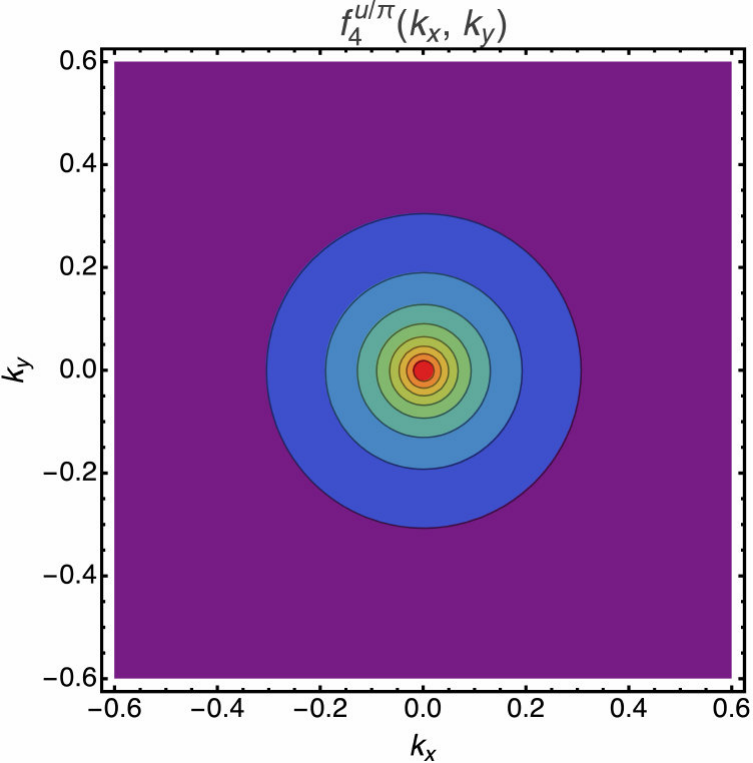}
\includegraphics[width=0.24\textwidth]{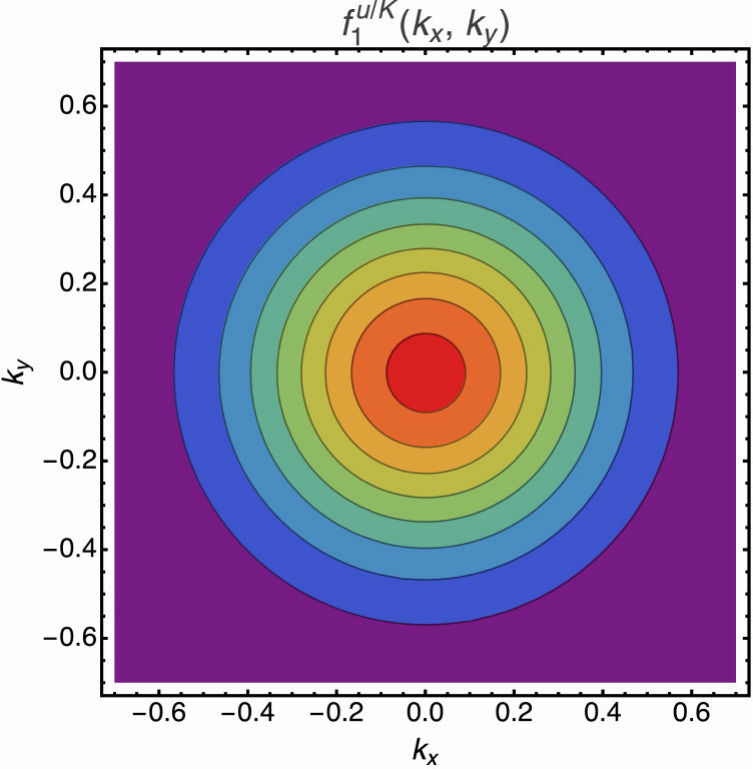}
\includegraphics[width=0.24\textwidth]{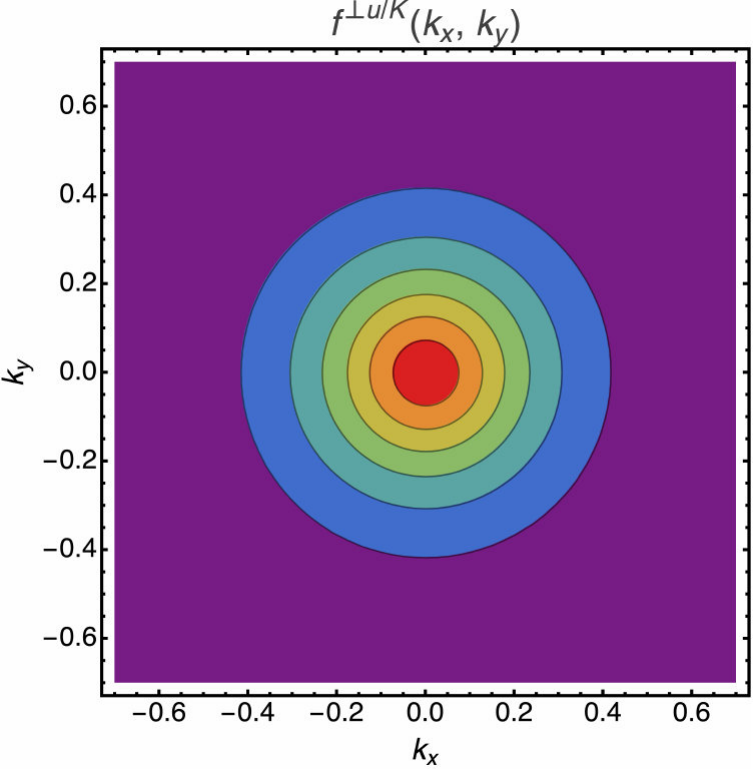}
\includegraphics[width=0.24\textwidth]{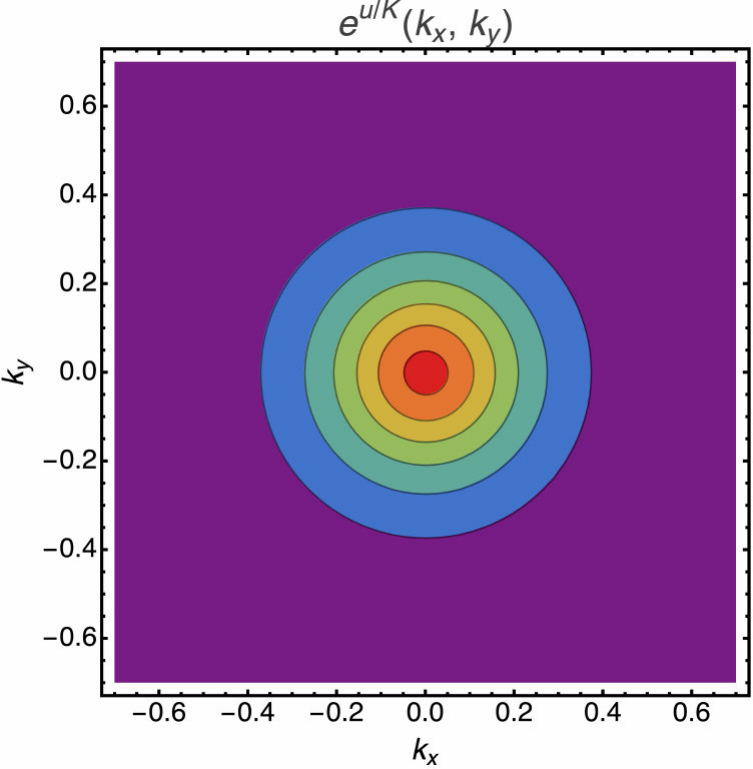}
\includegraphics[width=0.24\textwidth]{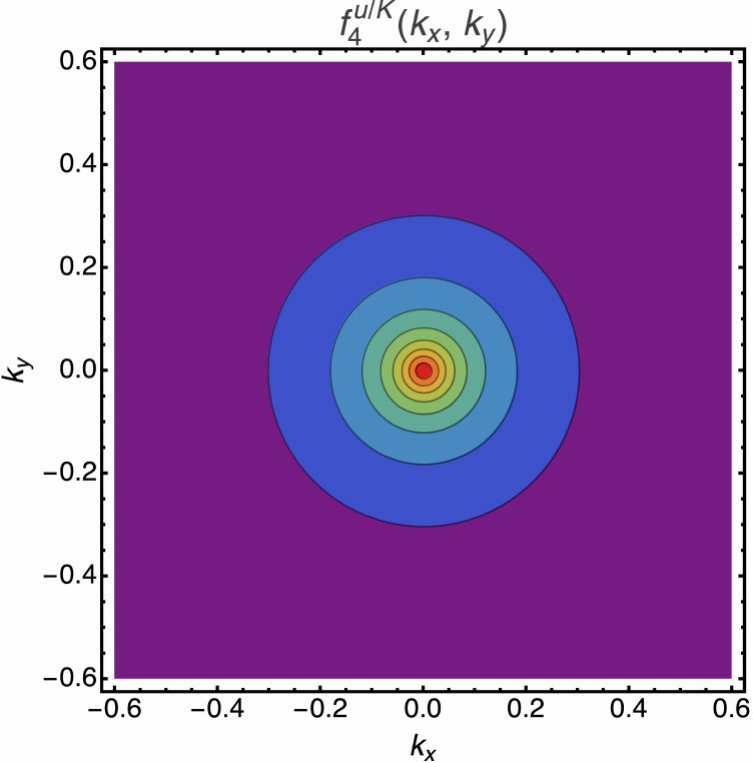}
\includegraphics[width=0.24\textwidth]{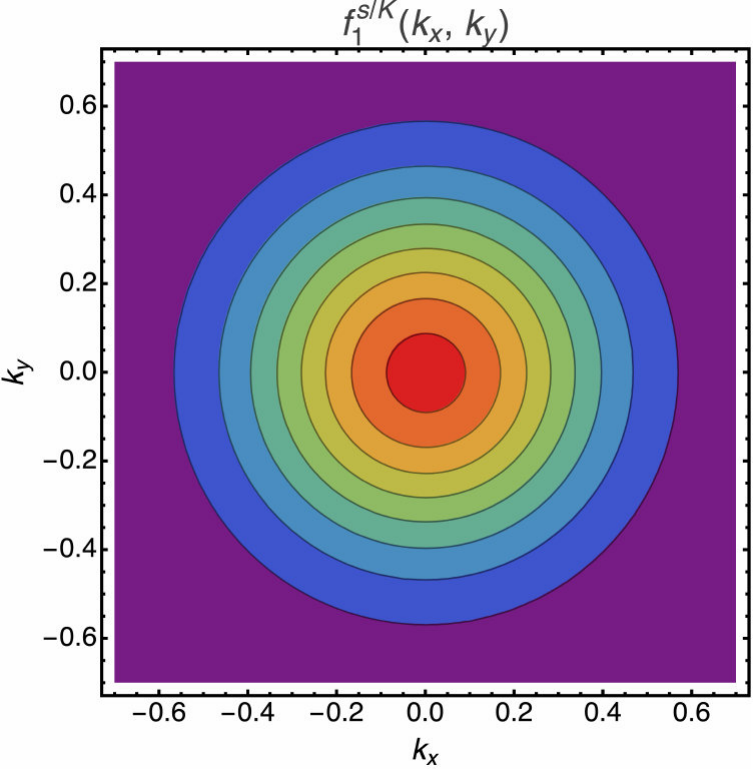}
\includegraphics[width=0.24\textwidth]{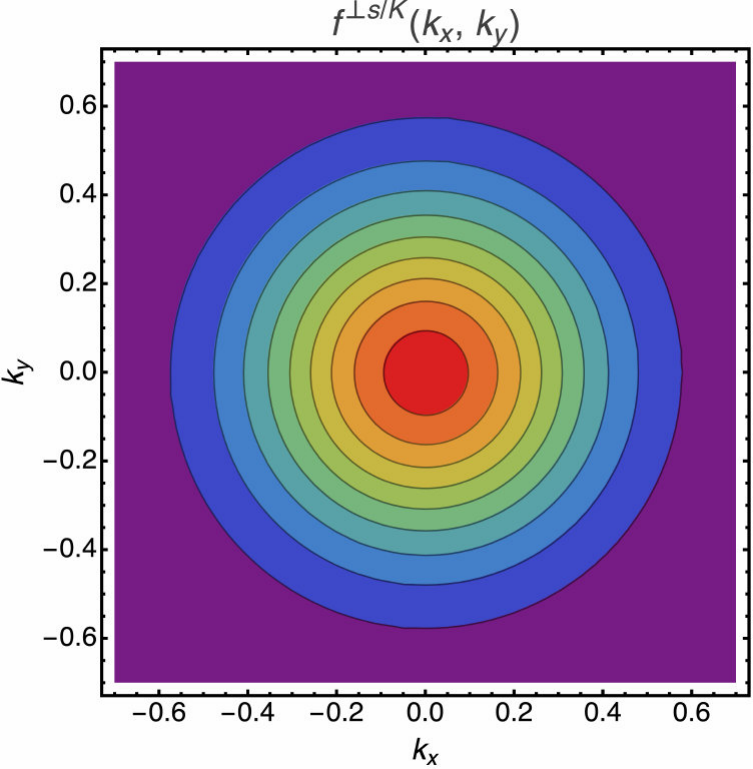}
\includegraphics[width=0.24\textwidth]{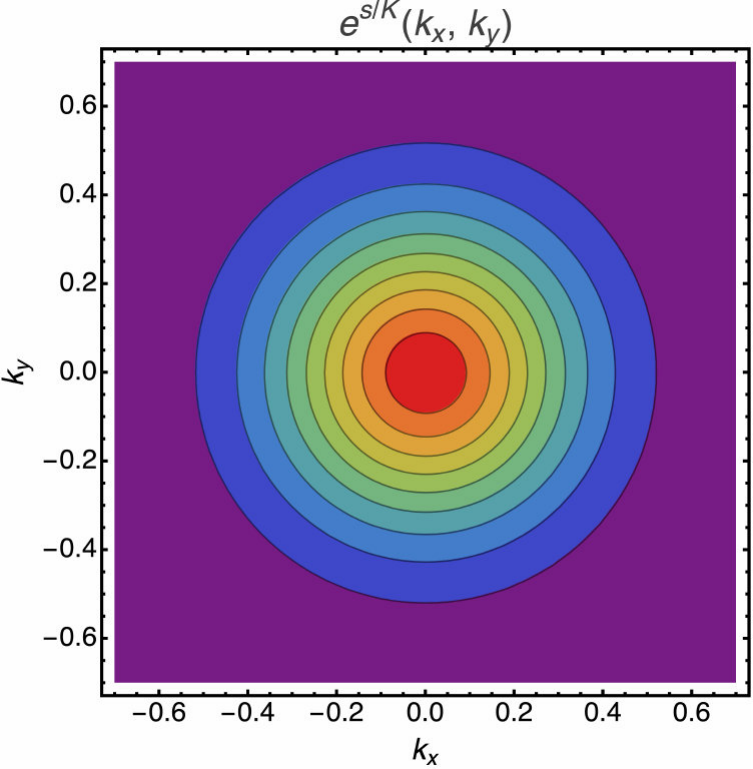}
\includegraphics[width=0.24\textwidth]{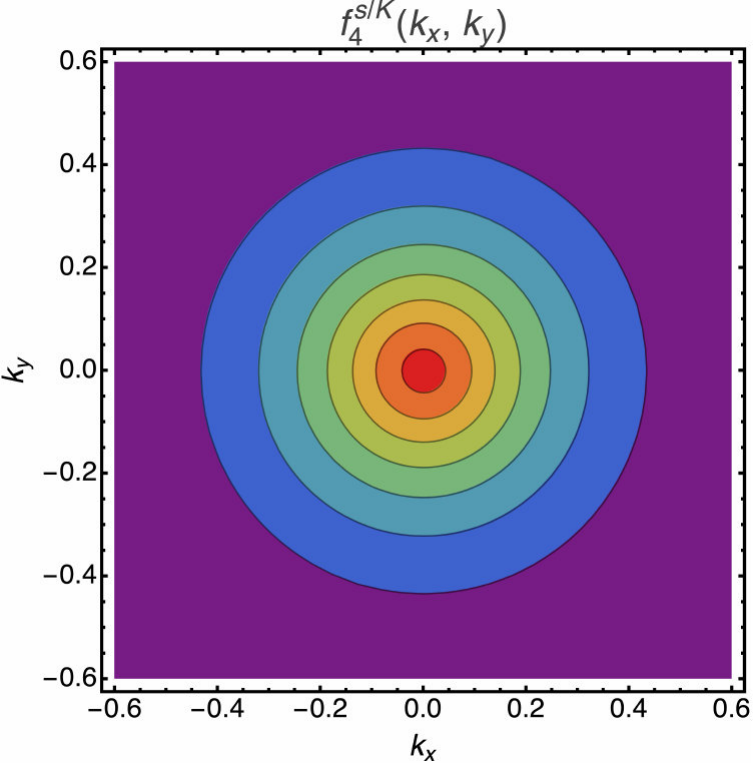}
\caption{
The $u$-quark TMDs $(f^u_1,\, f^{\perp u},\, e^u,\, f^u_4)$ in the $(k_x,k_y)$ plane for the $\pi^+$ (top row) and $K^+$ (middle row), 
and the corresponding $s$-quark TMDs $(f^s_1,\, f^{\perp s},\, e^s,\, f^s_4)$ for the $K^+$ (bottom row). 
Here $k_x$ and $k_y$ are given in units of GeV.
All distributions are integrated over $x$, and $f_4^q$ is additionally averaged over the azimuthal angle $\theta$.\label{Fig:TMDkxky-contour}
}

\end{figure*}

\begin{figure*}
\centering
\includegraphics[width=0.32\textwidth]{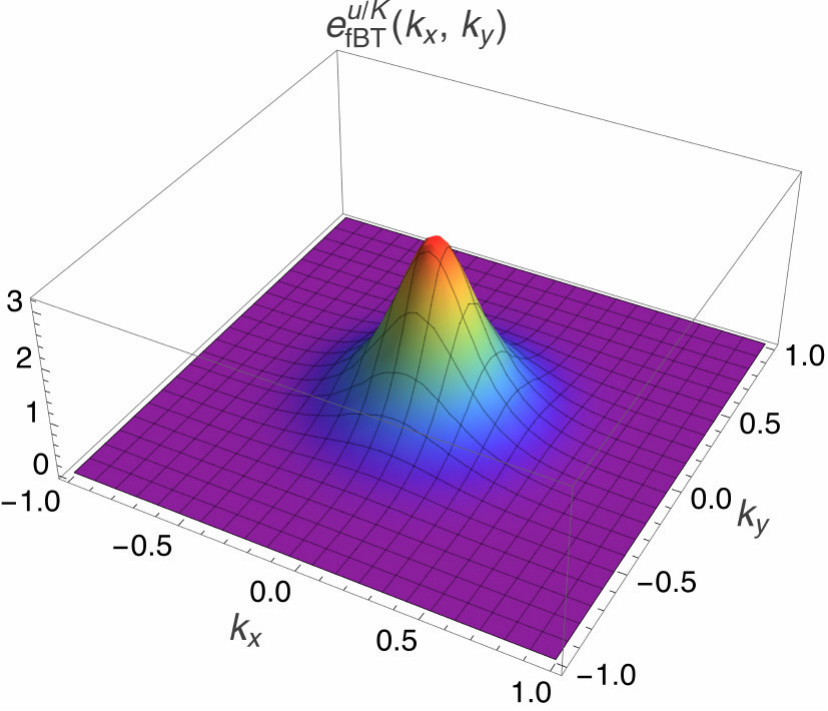}
\includegraphics[width=0.32\textwidth]{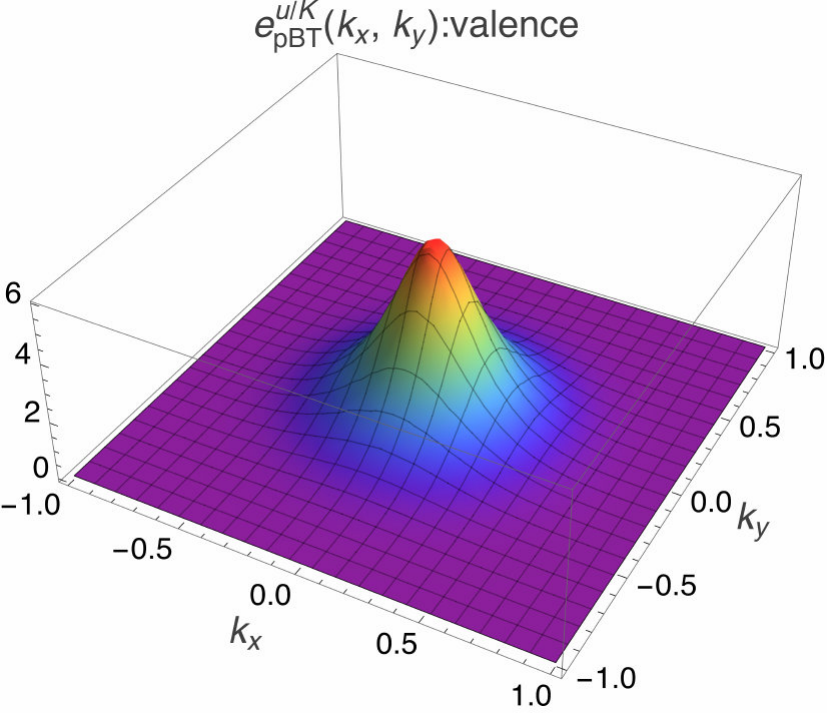}
\includegraphics[width=0.32\textwidth]{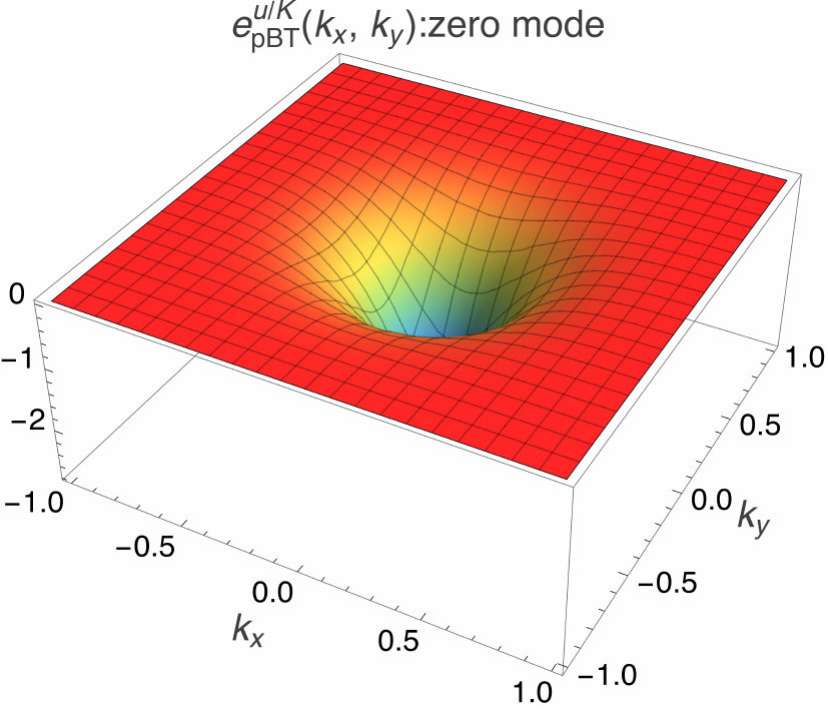}

\vspace{2\baselineskip} 

\includegraphics[width=0.32\textwidth]{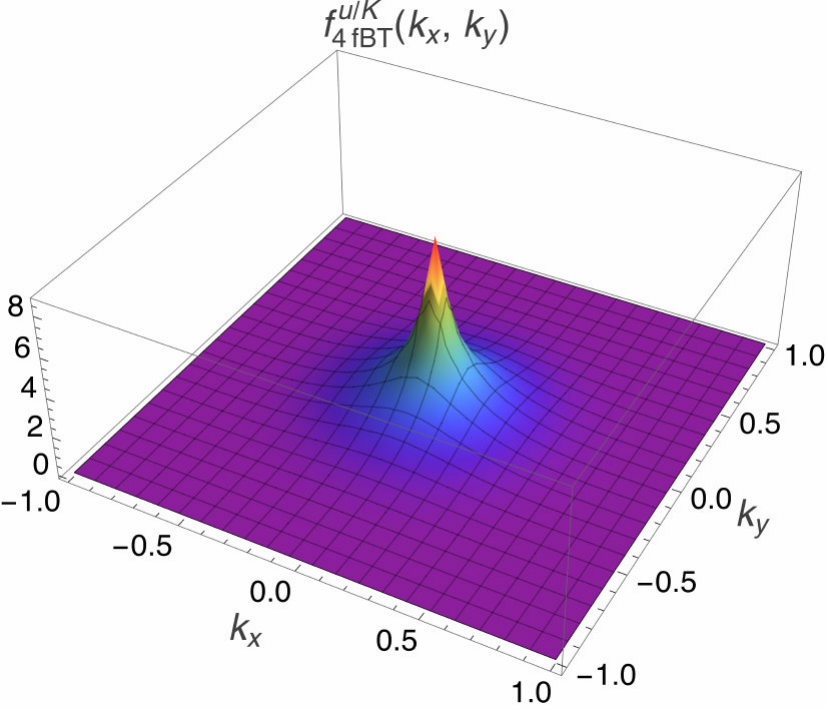}
\includegraphics[width=0.32\textwidth]{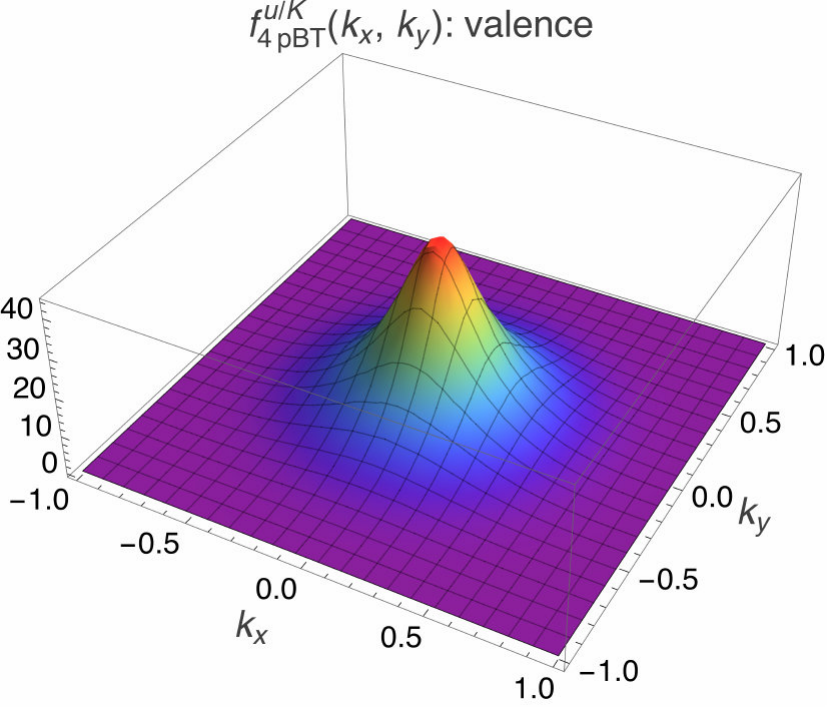}
\includegraphics[width=0.32\textwidth]{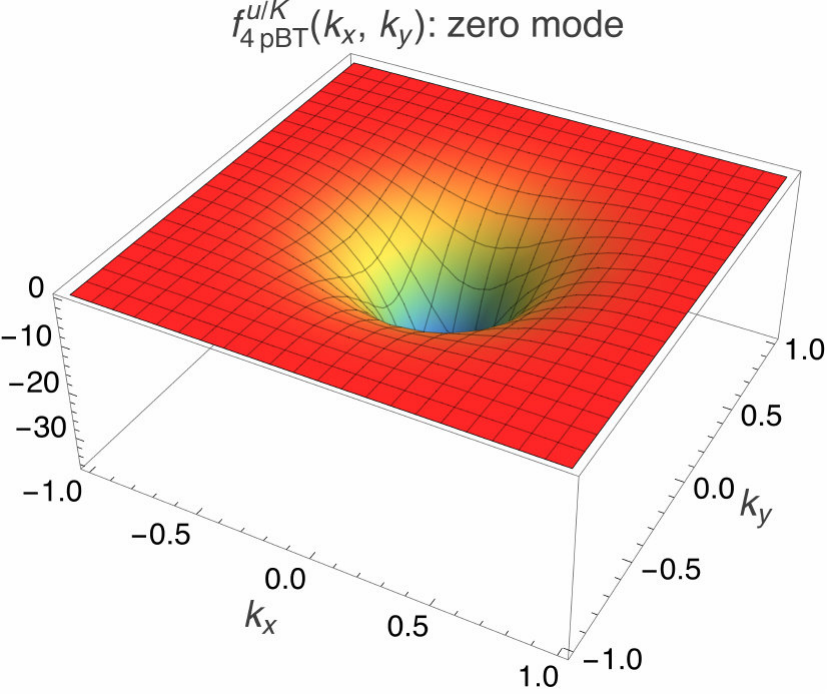}
\caption{
Effective LF zero-mode corrections to
$e^u(k_x,k_y)$ (top row) and $f_4^u(k_x,k_y)$ (bottom row) for the $K^+$.
Shown are the full fBT-LFQM result (left), the pBT-LFQM valence result
(middle), and the corresponding
effective correction defined as the fBT--pBT difference (right).
By construction, adding this effective correction to the pBT-LFQM valence
result exactly reproduces the fBT-LFQM result.
\label{Fig:TMDkxky-contour3D}
}
\end{figure*}

Beyond twist ordering, two SU(3)-breaking trends are robust.
First, across mesons the kaon moments 
exceed the corresponding pion moments for a fixed flavor (e.g., $u$ in $K^+$ vs.\ $u$ in $\pi^+$).
Second, within a given $K^+$ the $s$-quark moments exceed those of the $u$-quark. 
The first trend is largely driven by a larger variational scale in the kaon, 
$\beta_{q\bar s}>\beta_{q\bar q}$. This implies a tighter coordinate–space wave function and, 
by Fourier conjugacy, broader momentum–space profiles. 
The second trend arises even at fixed $\beta_{q\bar s}$ from mass–asymmetric LF kinematics, skewed $x$ distributions, 
larger typical $M_0(x,\bm k_\perp)$, and the Jacobian $\partial k_z/\partial x$, together with flavor–dependent projector factors. 
Finally, the twist–4 distribution $f_4^q$ (see Eq.~\eqref{eq:QCDrelBT}) is the
one most strongly affected by the BT mass structure, through the ratio
$(\bm{k}_\perp^2+m_1^2)/\mathcal{M}_{\rm BT}^2$.
Although the BT kinematic mass ${\cal M}_{\rm BT}^2$ contains an explicit
angular dependence at the integrand level through
$|{\bm k}_\perp|^2\cos^2\theta$, the unpolarized TMD shown here is defined
after azimuthal averaging in the $|{\bm q}_\perp|\to 0$ forward limit.
Consequently, the resulting TMD depends only on
$k_\perp=|{\bm k}_\perp|$ and is rotationally symmetric in the transverse plane.
This partial compensation modulates the Gaussian falloff inherited from
$f_1^q$ rather than suppressing it overall, and produces the largest deviation
from a Gaussian profile among the unpolarized $T$-even TMDs, as measured by
$R_G$, distinguishing the transverse profile of $f_4^q$ from those of the
lower-twist distributions.

The quark transverse–momentum profiles $f^q(\bm k_\perp)$ 
in the $(k_x,k_y)$ plane are shown in Fig.~\ref{Fig:TMDkxky-contour} for the $u$ quark in $\pi^+$ (top), the $u$ quark in 
$K^+$ (middle), and the $s$ quark in $K^+$ (bottom).
All four transverse–momentum profiles are azimuthally symmetric in $(k_x,k_y)$ and peak at $\bm k_\perp=\mathbf 0$, 
reflecting the underlying Gaussian LFWF that governs the radial falloff.
Moreover, the overall transverse extent of each contour directly correlates with the RMS width $\langle k_\perp^2\rangle^{1/2}$: 
distributions with a more localized high-density core (smaller red region) are more concentrated at low $k_\perp$ and therefore yield smaller transverse moments, 
whereas broader contours correspond to larger RMS widths. 
This trend is clearly visible when comparing the profiles of different twists and flavors.

The twist–2 $s$-quark profile $f^{\,s/K}_1(\bm k_\perp)$ is identical to the $u$-quark profile $f^{\,u/K}_1(\bm k_\perp)$
in $K^+$, since both are governed by the common Gaussian width parameter $\beta_{q\bar s}$.
In contrast, the higher-twist profiles $(f^{\perp\,s/K},\,e^{\,s/K},\,f^{\,s/K}_4)$ in $K^+$ exhibit larger RMS transverse widths 
than their $u$-quark counterparts. This difference arises from flavor-dependent kinematic factors in the twist–3 and twist–4 projectors such as 
$m_1/M_0$ and $({\bm k}_\perp^2+m_1^2)/\mathcal M_{\rm BT}^2$. 
This behavior is also reflected in the $(k_x,k_y)$ profiles shown in
Fig.~\ref{Fig:TMDkxky-contour}: among the unpolarized $T$-even TMDs, $f_4^q$
shows the largest deviation from a purely Gaussian shape as measured by $R_G$,
while the smallest mean and RMS transverse moments are found for $e^q$ in the
$u$-quark sector of both $\pi^+$ and $K^+$, and for $f_4^s$ in the $s$-quark
sector of $K^+$, as shown in Tables~\ref{tab:TMDu-piK} and
\ref{tab:TMDs-K}.

In Fig.~\ref{Fig:TMDkxky-contour3D}, we show the
effective LF zero-mode corrections to $e^u(k_x,k_y)$ (top row) and
$f_4^u(k_x,k_y)$ (bottom row) for the $K^+$.
Shown are the full fBT-LFQM result (left), the pBT-LFQM valence result
(middle), and the corresponding
effective correction defined as the fBT--pBT difference
(right). In both channels, adding this effective correction to the
pBT-LFQM valence result exactly reproduces the fBT-LFQM result.

All three results exhibit rotational symmetry in the transverse plane,
depending only on ${\bm k}_\perp^2=k_x^2+k_y^2$. For both $e^u$ and $f_4^u$,
the pBT-LFQM valence result produces an enhanced positive peak centered at
$|{\bm k}_\perp|=0$, whereas the effective correction
appears as a negative component localized near the origin. Quantitatively,
the effective correction is substantially larger for $f_4^u$ than for
$e^u$: both the excess in the pBT-LFQM valence result and the compensating
negative correction are more pronounced in the $f_4^u$ channel. This
indicates a quantitatively stronger fBT--pBT difference for the $\gamma^-$
projection than for the scalar projection within the present model. The
effective correction therefore compensates the excess pBT-LFQM valence
contribution at small ${\bm k}_\perp$, yielding the corresponding fBT-LFQM
result in both channels.

\subsection{ PDFs and Longitudinal Moments}
\begin{figure*}
\centering
\includegraphics[width=0.45\textwidth]{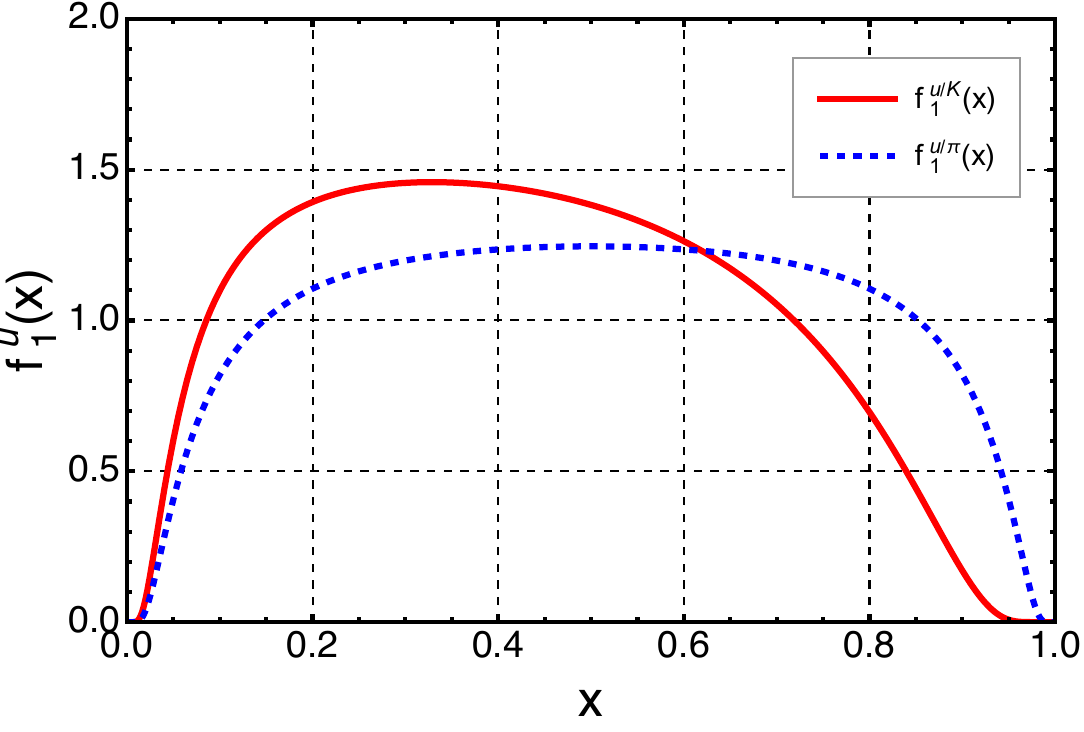}
\includegraphics[width=0.45\textwidth]{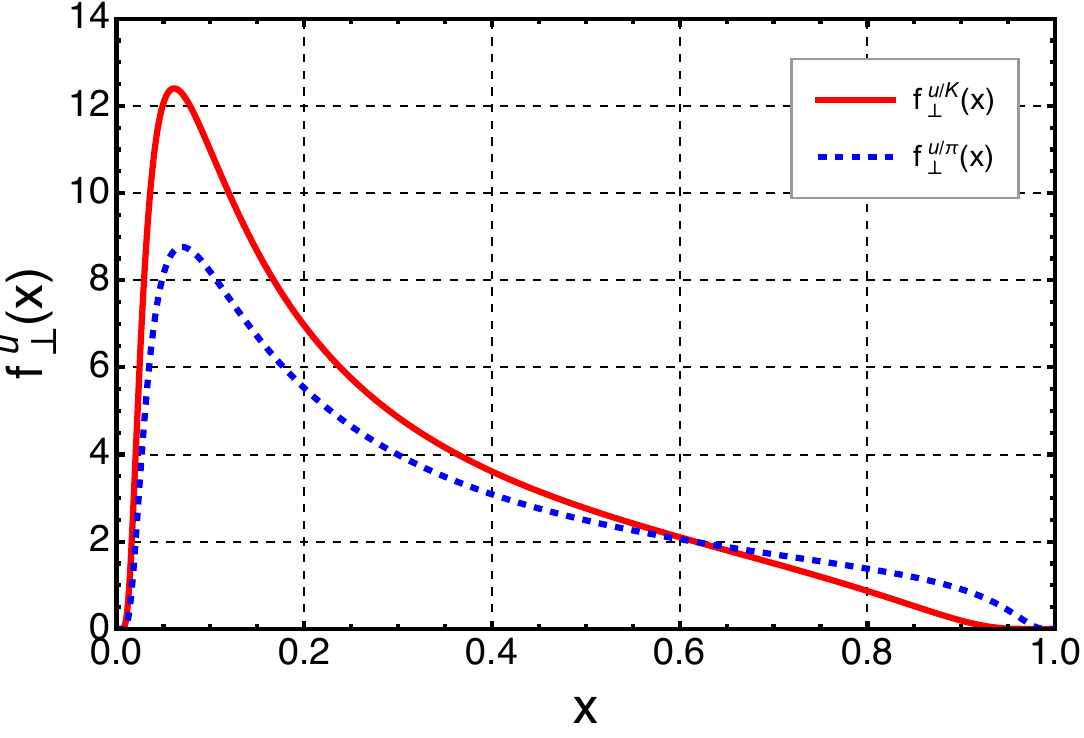}
\includegraphics[width=0.45\textwidth]{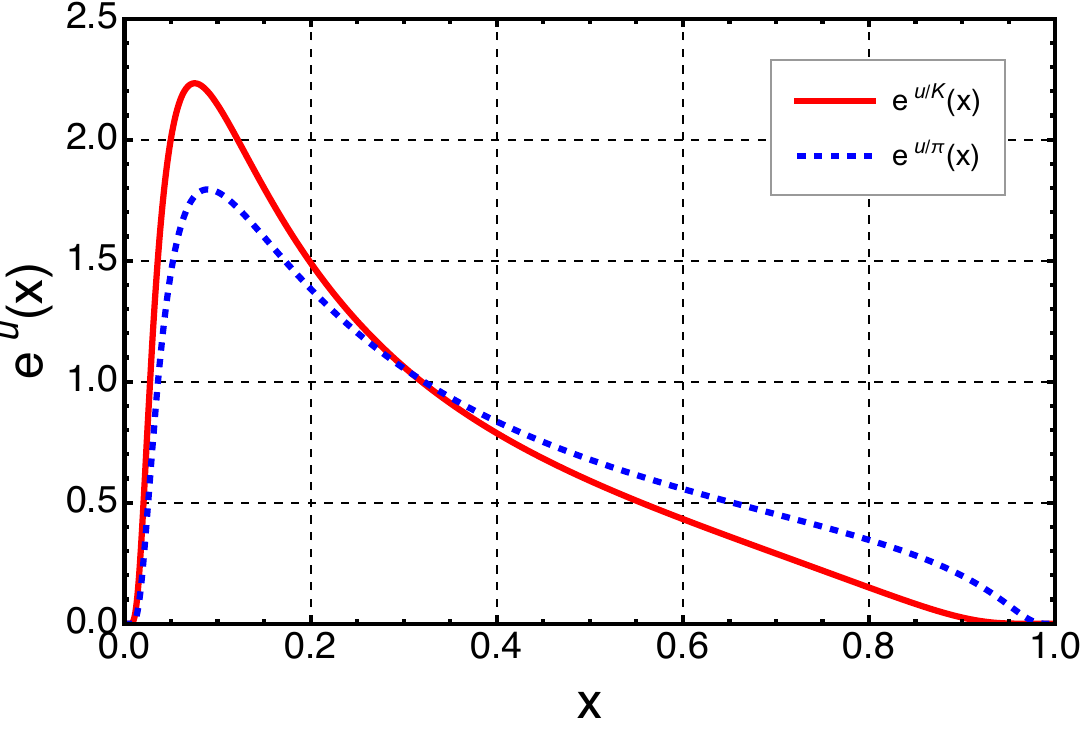}
\includegraphics[width=0.45\textwidth]{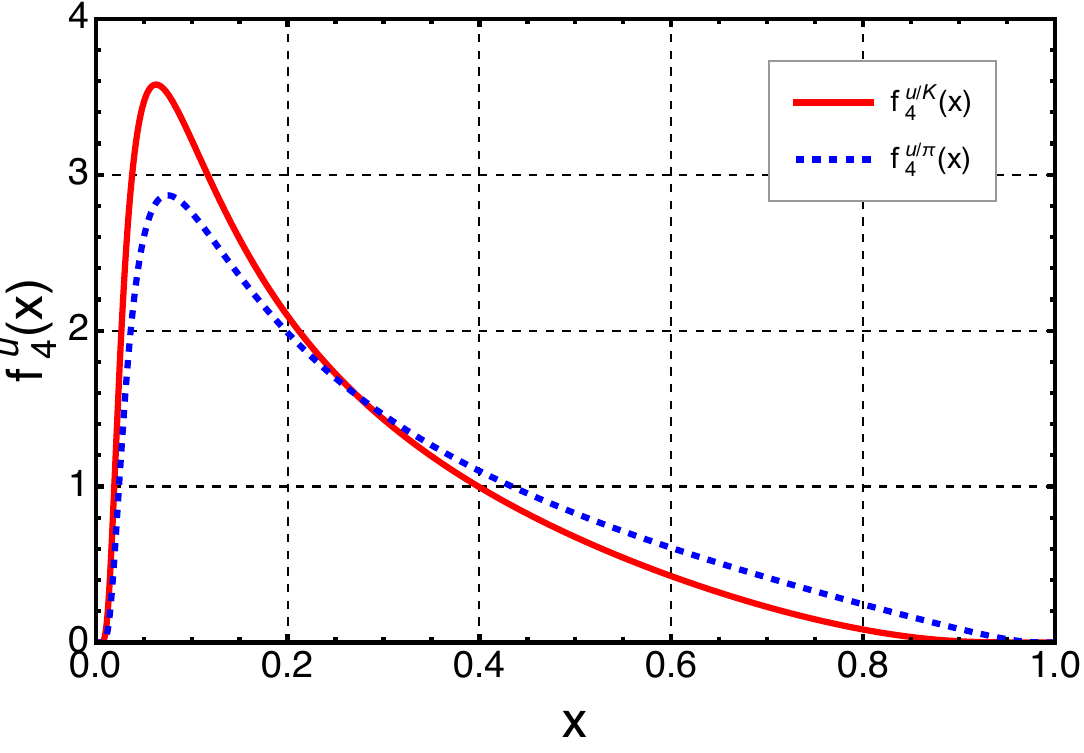}
\includegraphics[width=0.45\textwidth]{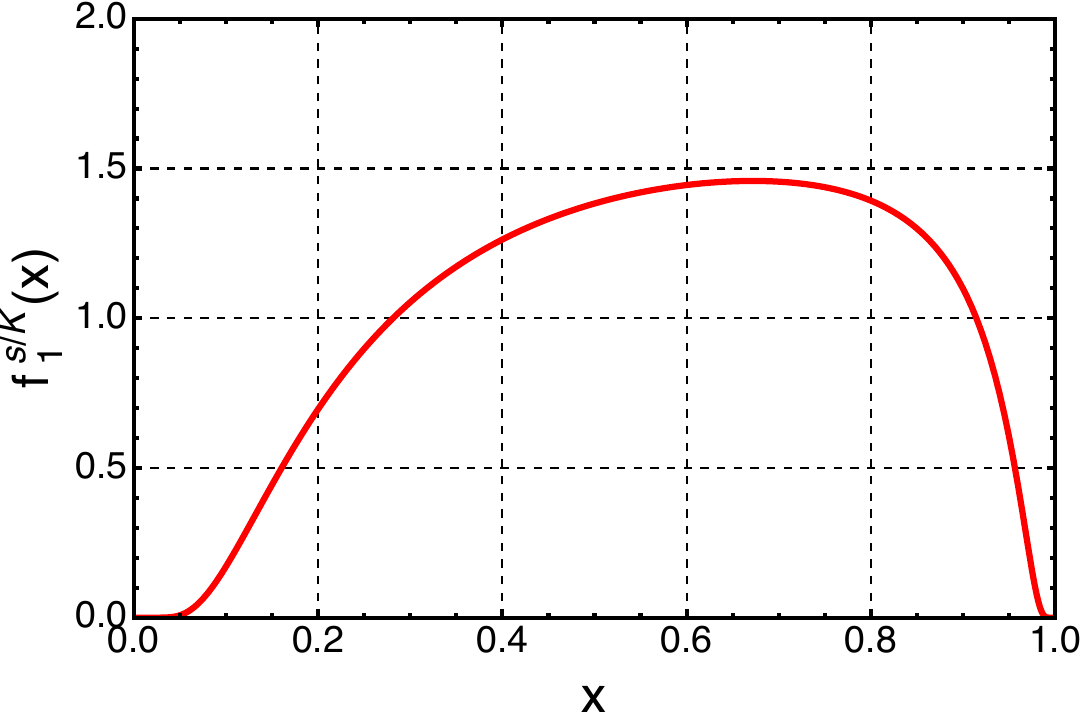}
\includegraphics[width=0.45\textwidth]{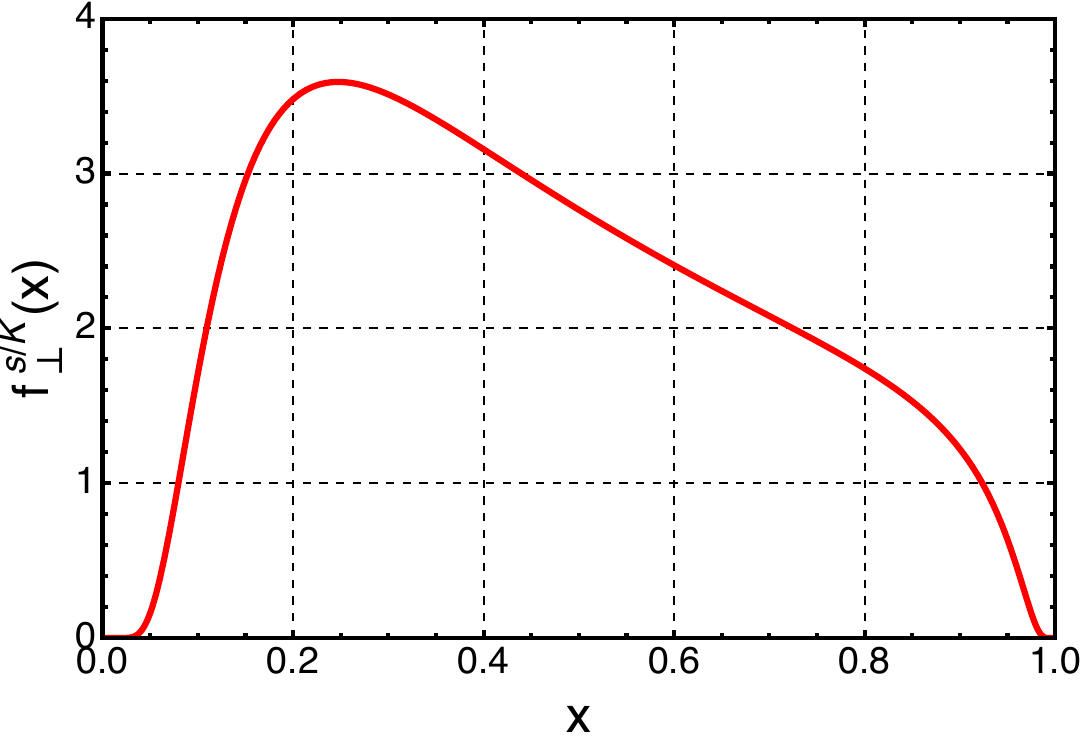}
\includegraphics[width=0.45\textwidth]{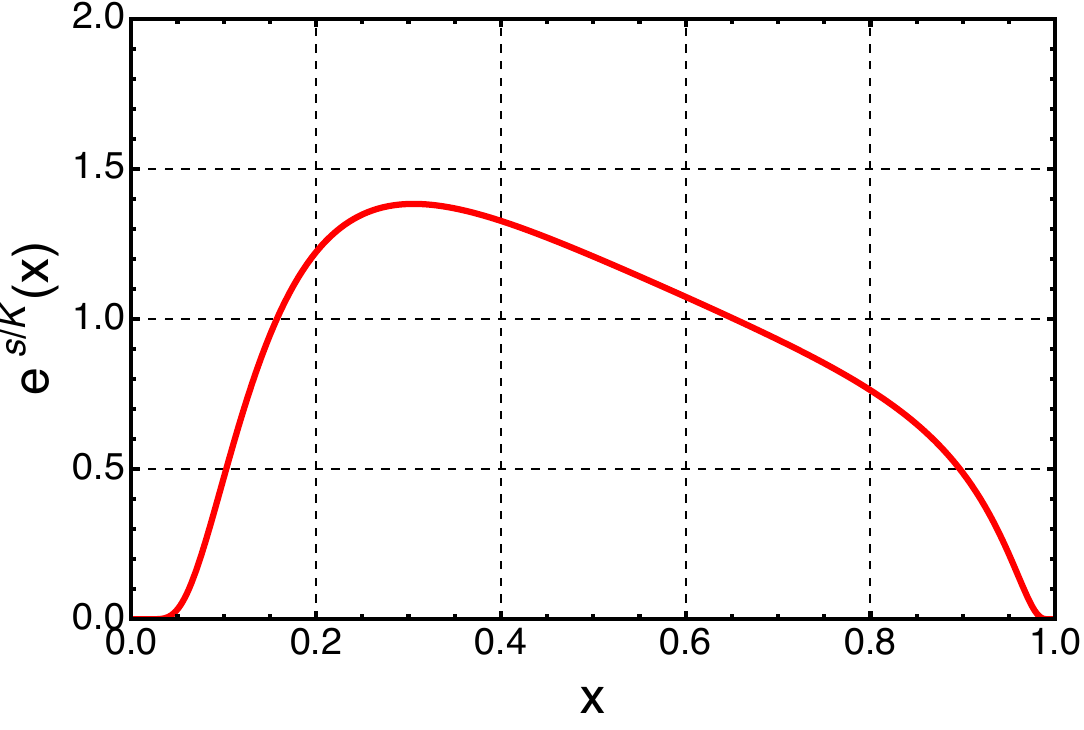}
\includegraphics[width=0.45\textwidth]{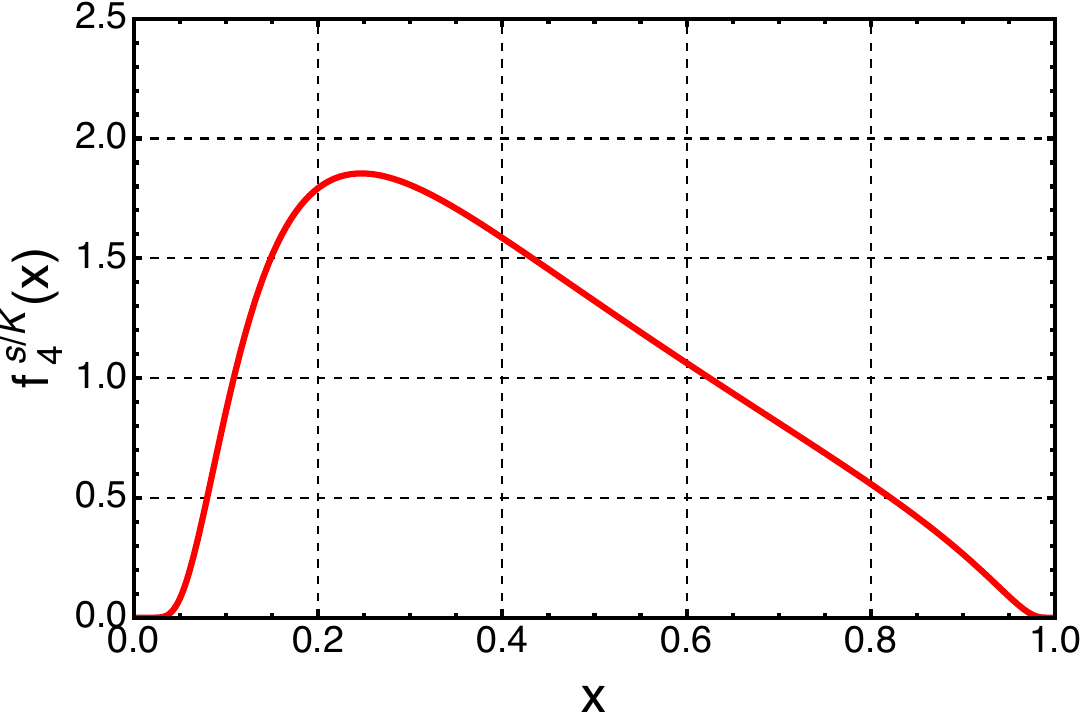}
\caption{The $u$-quark PDFs $(f^u_1, f^{\perp u}, e^u, f^u_4)$ (top row) for the $\pi^+$ (dashed lines) and $K^+$ (solid lines), 
and the $s$-quark PDFs $(f^s_1, f^{\perp s}, e^s, f^s_4)$ (bottom row) for the $K^+$. \label{Fig:PDF-comparison}}
\end{figure*}

Figure~\ref{Fig:PDF-comparison} shows the twist-2, twist-3, and twist-4 PDFs $f^q_1(x)$, $f^{\perp q}(x)$, $e^q(x)$, and $f^q_4(x)$ 
for the $u$- and $s$-quark sectors in the $\pi^+$ and $K^+$ mesons. 
The upper panels compare the $u$-quark distributions in $K^+$ (solid lines) and $\pi^+$ (dashed lines), 
while the lower panels show the $s$-quark distributions in $K^+$.

For twist–2, the pion distribution $f^{u/\pi^+}_1(x)$ is symmetric in $x$, reflecting $\mathrm{SU}(2)$ flavor symmetry. 
In $K^+$, explicit $\mathrm{SU}(3)$ breaking 
leads to asymmetric momentum sharing: $f^{s/K^+}_1(x)$ is shifted toward larger $x$,
while $f^{u/K^+}_1(x)$ is skewed toward smaller $x$.
For higher twists, the $u$–quark distributions in both $\pi^+$ and $K^+$ remain concentrated toward smaller $x$. In $K^+$, 
the $s$–quark higher–twist PDFs are shifted toward larger $x$ and exhibit more pronounced large–$x$ tails
than the corresponding $u$–quark distributions, although their peaks still lie below $x=\tfrac12$. 
This behavior follows from the struck-quark-mass-dependent
kinematic factors, with $m_1=m_u$ for $f^u$ and $m_1=m_s$ for $f^s$, such as
$m_1/M_0$ and $(\bm{k}_\perp^2+m_1^2)/{\cal M}_{\rm BT}^2$, together with the
$M_0$ dependence of the Gaussian LF wave function. For a heavier struck quark,
the small-$x$ region is more strongly suppressed through the growth of $M_0^2$,
which biases the $s$-quark distributions toward larger $x$. This enhancement
toward large $x$ is ultimately countered by the endpoint growth
$M_0^2\propto(1-x)^{-1}$ as $x\to1$.

A distinct small–$x$ feature appears for the twist–3 transverse distribution. 
In our fBT-LFQM one has 
the model identity $x\,f^{\perp q}(x,\bm{k}_\perp)=f_1^q(x,\bm{k}_\perp)$, which implies $f^{\perp q}(x)=f_1^q (x) /x$,
after integrating over ${\bm k}_\perp$.
Thus, within the present model, $f^{\perp q} (x)$ is kinematically enhanced at small $x$ relative to $f_1^q(x)$, while its large–$x$ falloff follows that of $f_1^q$. 
Overall, these trends illustrate how mass asymmetries and higher–twist 
kinematic factors shape the partonic structure of light pseudoscalar mesons.

In summary, Fig.~\ref{Fig:PDF-comparison} reveals two robust trends. 
(i) For the $u$–quark, the higher–twist PDFs $f^{\perp u}(x)$, $e^{\,u}(x)$, and $f_4^{\,u}(x)$ 
are shifted toward smaller $x$ and are more suppressed near $x\to 1$ than the leading–twist distribution $f_1^u(x)$. 
(ii) For the $s$–quark in $K^+$, all three higher–twist PDFs are comparatively suppressed at small $x$ but enhanced toward larger $x$ 
relative to the corresponding $u$–quark PDFs, consistent with the heavier 
struck-quark mass entering kinematic factors and biasing the $s$ channel toward larger $x$.

We also note that, in our fBT--LFQM, the twist--4 PDF $f_4^q(x)$ satisfies the sum rule in Eq.~\eqref{eq:Nqf4}, whereas 
in the pBT-LFQM this sum rule is not recovered unless the
effective LF zero--mode correction defined as the fBT--pBT difference is properly included.
The scalar channel shows an analogous sensitivity. As demonstrated through the scalar
form-factor analysis in Appendix~\ref{app:scalarZM}, the scalar matrix element
in the covariant BS model receives a nonvanishing explicit LF zero-mode contribution,
demonstrating the zero-mode sensitivity of the extraction of the twist-3 PDF $e^q(x)$.

In the pBT-LFQM, $e^q(x)$ is obtained with the physical meson mass $M$ in the
scalar Lorentz prefactor, and therefore represents only the valence contribution
unless the corresponding effective LF zero-mode correction is added explicitly. In contrast, the
fBT-LFQM implements the BT replacement $M\to M_0$ consistently under the
$(x,\bm{k}_\perp)$ integral,
so that the resulting fBT--pBT difference is interpreted as the
effective scalar LF zero-mode correction missing from the pBT extraction.

Numerically, the twist-3 PDF $e^q(x)$ obtained from  the fBT-LFQM (pBT-LFQM) yields
\bea
\int_0^1 dx\, e^{u/\pi^+}(x)=0.752 ~(4.883),\nonumber\\
\int_0^1 dx\, e^{u/K^+}(x)=0.737 ~(1.639), \nonumber\\
\int_0^1 dx\, e^{s/K^+}(x)=0.885 ~(2.003).
\eea
The large differences between the fBT- and pBT-LFQM results therefore provide
a quantitative illustration of the
effective scalar LF zero-mode correction defined as the fBT--pBT difference.
They mirror the pattern observed for $f_4^q(x)$.

\begin{table}[t]
\centering
\renewcommand{\arraystretch}{1.5}
\setlength{\tabcolsep}{2pt}
\begin{tabular}{c|ccc|ccc}
\hline\hline
 PDF &$\langle x \rangle^{u/\pi}$ &
 $\langle x \rangle^{u/K}$ &
$\langle x \rangle^{s/K}$ &
$\langle x^{-1} \rangle^{u/\pi}$ &
$\langle x^{-1}  \rangle^{u/K}$ &
$\langle x^{-1} \rangle^{s/K}$ \\
\hline
$f^q_1(x)$ &0.5 & 0.429 & 0.571 & 3.108 & 3.679 & 2.199 \\
$e^q(x)$ & 0.255 & 0.205 & 0.419 & 4.325 & 5.184 & 2.534 \\
$f^{\perp q}(x)$ & 1 & 1 & 1 & 20.597 & 28.638 & 6.845 \\
$f^q_4(x)$ &0.290 & 0.241 & 0.417 & 6.795 & 8.208 & 3.327 \\
\hline\hline
\end{tabular}
\caption{First ($\langle x\rangle$) and inverse ($\langle x^{-1}\rangle$) $x$–moments of the four unpolarized 
$T$–even PDFs $f_1^q(x)$, $e^q(x)$, $f^{\perp q}(x)$, and $f_4^q(x)$ for the
$u$ quark in $\pi^+$ and the $u$ and $s$ quarks in $K^+$ within the fBT-LFQM.\label{tab:singlecol}}
\end{table}

\subsection{QCD evolution of the twist-2 PDFs}

In Table~\ref{tab:singlecol}, we list the first ($\langle x\rangle$) and inverse ($\langle x^{-1}\rangle$) $x$–moments of the 
unpolarized $T$–even PDFs $f_1^q(x)$, $e^q(x)$, $f^{\perp q}(x)$, and $f_4^q(x)$ for the
$u$ quark in $\pi^+$ and the $u$ and $s$ quarks in $K^+$ within the fBT–LFQM.
The $n$th moment of a PDF $f^q(x)$ is defined as
\begin{eqnarray}
    \la x^n \ra^q = \int_0^1 dx\; x^n f^q(x).
\end{eqnarray}
We note that the first moment of the twist-3 PDF $f^{\perp q}(x)$ satisfies $\langle x\rangle^q_{f^\perp} =1$ for any quark flavor.
This follows from the TMD relation $x f^{\perp q}(x, {\bm k}_\perp)= f^{q}_1(x, {\bm k}_\perp)$ given by Eq.~\eqref{eq:fperprel},
together with the number sum rule for the twist-2 PDF $f^{q}_1(x)$ given by Eq.~\eqref{eq:Nq}.

The inverse moment ($n=-1$) is finite in the present model owing to its
small-$x$ behavior~\cite{Lorce:2015,Lorce:2016,Brodsky:2007eConf}
and may be viewed as a Weisberger-type $1/x$ moment.
Through the Weisberger relation~\cite{Weisberger}, the flavor-summed
$1/x$ moment is formally related to $\partial M_H^2/\partial m_q^2$, which in QCD connects high-energy parton
structure to low-energy sigma-term physics.
Since $m_q$ in the present LFQM is a constituent-quark mass, however, no
direct identification with a QCD sigma term is made here. In full QCD, the
corresponding moment requires an appropriate small-$x$ regularization
prescription.
From Table~\ref{tab:singlecol} one sees that in the kaon the $s$-quark has a
larger first moment than the $u$-quark, while its inverse moment is smaller.
This pattern reflects a harder $s$-quark distribution that is shifted to
larger $x$ and depleted at very small $x$. The shift raises
$\langle x\rangle$ but reduces the weight in the $1/x$-enhanced region, which
lowers $\langle x^{-1}\rangle$ relative to $u$.

\begin{table*}[t]
\centering
\renewcommand{\arraystretch}{1.5}
\setlength{\tabcolsep}{2pt}
\begin{tabular}{c|c|cccc|cccc} \hline\hline 
 & $\mu^{2}~[\mathrm{GeV}^2]$ & $\langle x \rangle^{u/K}_{\rm val}$ 
& $\langle x^{2} \rangle^{u/K}_{\rm val}$
& $\langle x^{3} \rangle^{u/K}_{\rm val}$ & $\langle x^{4} \rangle^{u/K}_{\rm val}$ & $\langle x \rangle^{s/K}_{\rm val}$ 
& $\langle x^{2} \rangle^{s/K}_{\rm val}$
& $\langle x^{3} \rangle^{s/K}_{\rm val}$ & $\langle x^{4} \rangle^{s/K}_{\rm val}$\\
\hline
This work & 16 & 0.213 & 0.080 & 0.038 & 0.021 & 0.282 & 0.129 & 0.072 & 0.044 \\
 & 27 & 0.204 & 0.075 & 0.036 & 0.019 & 0.271 & 0.121 & 0.066 & 0.041 \\
\hline
DSE approach~\cite{Chen:2016sno} & 27 & 0.28 & 0.11 & 0.048 & -- &  0.36 & 0.17 & 0.092 &--\\
$\chi$CQM~\cite{Watanabe:2018qju} & 27 & 0.23 & 0.091 & 0.045 & -- &  0.24 & 0.096 & 0.049 &-- \\
BLFQ-NJL~\cite{Lan:2019rba} & 27  & 0.201 & 0.077 & 0.037 & 0.021 & 0.228 & 0.094 & 0.049 & 0.029 \\
lattice QCD~\cite{Alexandrou:2021mmi} & 27 & 0.217 & 0.079 & 0.036 & 0.019 & 0.279 & 0.115 & 0.058 & 0.033 \\
\hline\hline
\end{tabular}
\caption{Lowest four Mellin moments of the kaon valence quark PDF $f^q_1(x)$ for the $u$- and $s$-quark sectors, 
evaluated at $\mu^2=16$ and 27 GeV$^2$ after NNLO evolution from the hadronic input scale.
Our results are compared with representative theoretical model approaches~\cite{Chen:2016sno,Watanabe:2018qju,Lan:2019rba,Alexandrou:2021mmi}.\label{tab:Mellin} }
\end{table*}

The valence PDFs at the model scale $\mu_0$ are evolved to higher
scales $\mu$ using perturbative QCD. 
We perform the NNLO DGLAP evolution~\cite{Dok,Gribov,Altarelli} 
using the \textsc{HOPPET} package \cite{Rojo}, which resums
the leading logarithms $\ln(\mu^2/\mu_0^2)$ and consistently generates gluon and
sea--quark distributions from a valence input through
$q\!\to\! qg$, $g\!\to\! (q\bar q, \,gg)$ splittings. Consequently, a structure that is purely
valence at $\mu_0$ acquires the expected QCD radiative content at higher $\mu$.

The LFQM transverse scale reflects the intrinsic nonperturbative width of the wave function,
characterized either by the saturation cutoff $k_\perp^{\max}$
or by the RMS width $\sqrt{\langle \bm k_\perp^2\rangle}=\beta$ of the Gaussian LFWF.
By contrast, the collinear scale $\mu_0$ 
serves as the factorization scale separating nonperturbative input from perturbative evolution.

We adopt a pragmatic matching strategy. The TMD integrals saturate near
$k_\perp^{\max}\sim1~\mathrm{GeV}$, 
while the Gaussian LFWF has an intrinsic RMS width
$\beta\simeq0.37$--$0.39~\mathrm{GeV}$. These quantities characterize the
transverse structure of the nonperturbative LFWF and are conceptually distinct
from the collinear factorization scale. For the collinear evolution, we adopt
$\mu_0=0.6~\mathrm{GeV}$ as a phenomenological model-matching scale at which a
valence-only collinear input is specified for evolution to
$\mu=2~\mathrm{GeV}$.

Applying this setup ($\mu_0=0.6$ GeV, $\al_s(\mu_0)=1$) in the DGLAP evolution, 
we obtain for the pion the first moment of valence-quark PDF at $\mu^2 = 4$ GeV$^2$ as
 \be
     \langle x \rangle^\pi_{\rm val} \equiv 2 \langle x \rangle^{u/\pi}_{\rm val} 
     = 2 \int^1_0 d x \; x f^{u/\pi}_1(x) = 0.562.
     \label{eq:xpival}
 \ee
In our model, the valence quarks in the pion carry about 56\% of the total
momentum at $\mu=2~{\rm GeV}$ and thus retain the majority of the pion
momentum. The remaining momentum is shared by gluons and sea quarks. The same evolution setup is applied to the kaon.

Table~\ref{tab:Mellin} summarizes the lowest four Mellin moments of 
the kaon’s twist–2 valence PDFs for the $u$- and $s$-quark sectors, evaluated at $\mu^{2}=16$ and $27~\mathrm{GeV}^{2}$.
We compare our results with representative theoretical calculations
at $27~\mathrm{GeV}^{2}$, including those from the Dyson--Schwinger equation (DSE) approach~\cite{Chen:2016sno},
the chiral constituent quark model ($\chi$CQM)~\cite{Watanabe:2018qju}, 
basis light-front  quantization with Nambu--Jona-Lasinio interactions  (BLFQ-NJL)~\cite{Lan:2019rba}, and
lattice QCD reconstructed PDFs~\cite{Alexandrou:2021mmi}.
QCD evolution redistributes momentum from valence quarks to gluons and sea quarks.
Accordingly, the valence–quark moments decrease as $\mu^{2}$ increases, while the total momentum is conserved. 
Overall, our results provide a coherent description of the kaon’s valence structure and 
remain broadly consistent with other
theoretical approaches, within the expected model dependence.

\begin{figure}
 	\centering
 	\includegraphics[width=0.65\columnwidth]{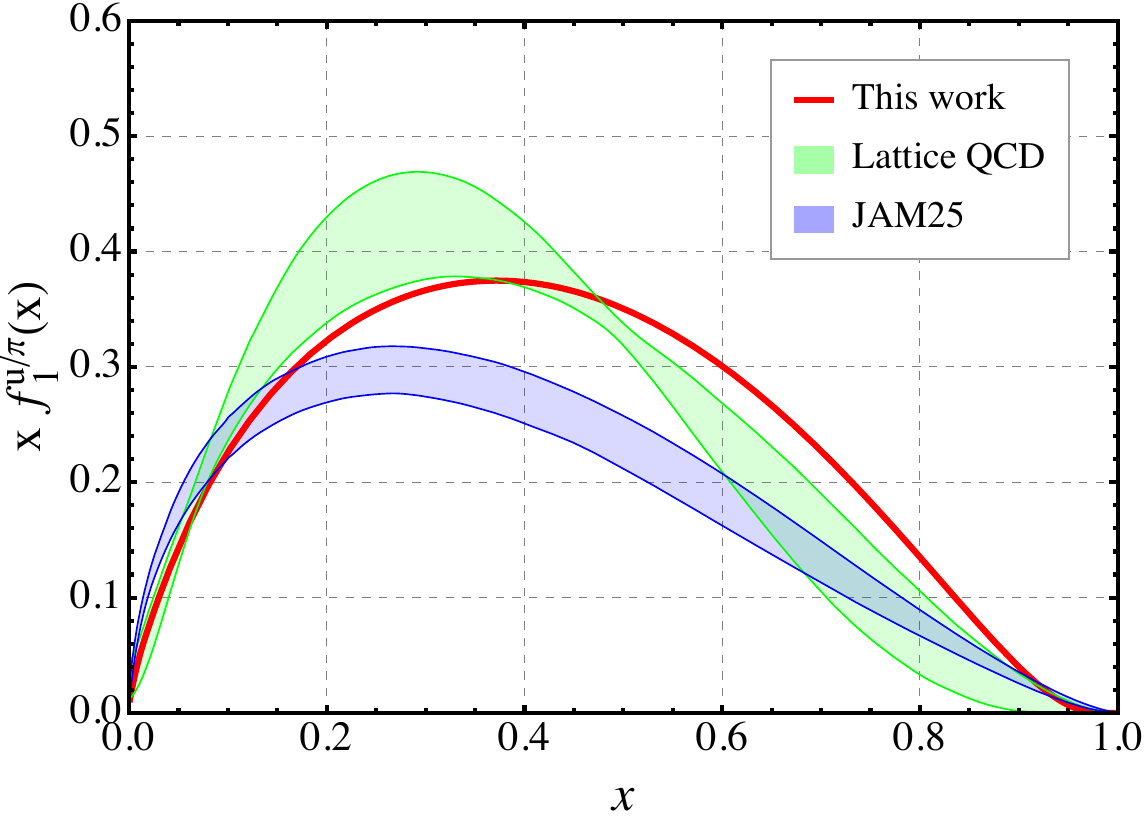}
    \includegraphics[width=0.65\columnwidth]{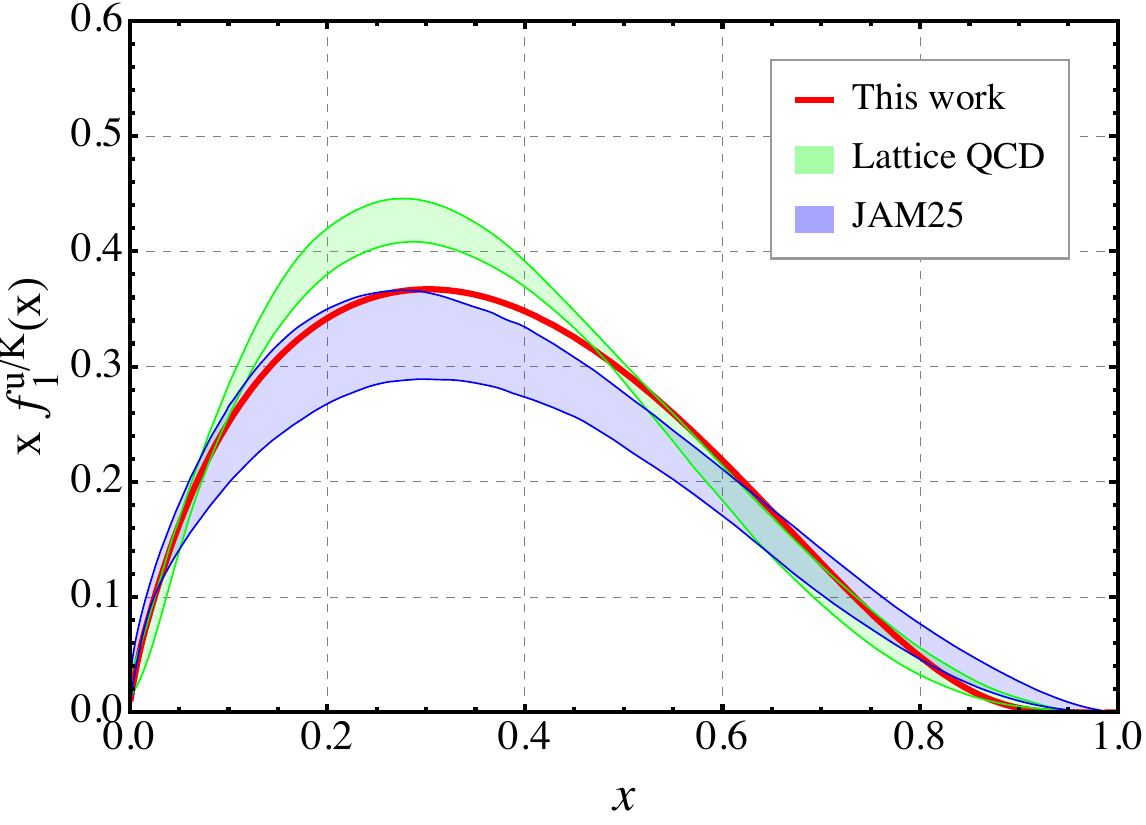}
    \includegraphics[width=0.65\columnwidth]{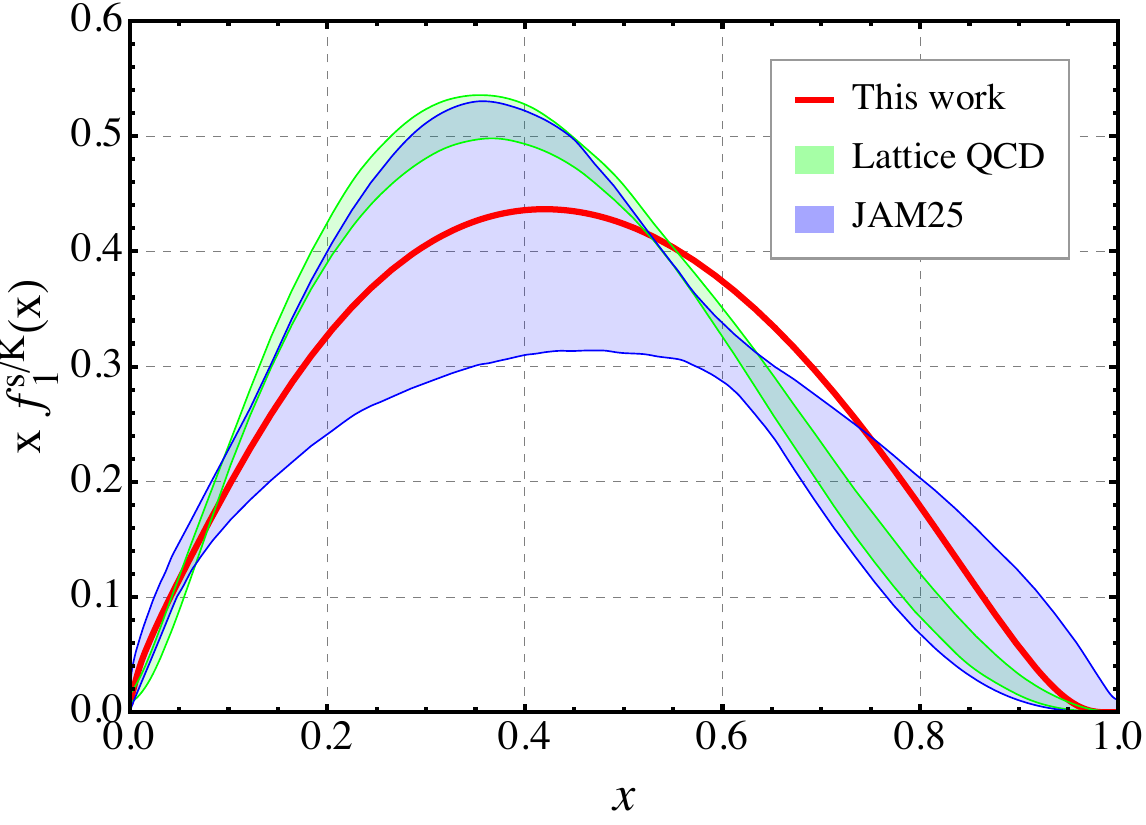}
 	\caption{Valence PDFs of the pion and kaon evolved to $\mu^{2}=27~\mathrm{GeV}^{2}$.
The panels show $x f_{1}^{u/\pi}(x)$ (top), $x f_{1}^{u/K}(x)$ (middle), and $x f_{1}^{s/K}(x)$ (bottom),
compared with the two-state-fit lattice QCD results of Ref.~\cite{Alexandrou:2021mmi} (green bands)
and with the global QCD analysis of Ref.~\cite{Barry:2025wjx} (blue bands).}
 	\label{fig:evolution}
 \end{figure}
 
Figure~\ref{fig:evolution} shows the valence PDFs for the pion and kaon evolved to $\mu^{2}=27~\mathrm{GeV}^{2}$.
The top, middle, and bottom panels display $x f_{1}^{u/\pi}(x)$, $x f_{1}^{u/K}(x)$, and $x f_{1}^{s/K}(x)$, respectively.
Our results (solid red curves) are compared with the lattice QCD reconstructed PDFs of Ref.~\cite{Alexandrou:2021mmi}
(green bands) and with the simultaneous global QCD analysis of pion and kaon PDFs of Ref.~\cite{Barry:2025wjx}
(blue bands, labeled JAM25), the latter provided at the same scale. The two bands are not on the same statistical
footing: the JAM25 bands are $95\%$ credible intervals of the Bayesian posterior, whereas the lattice bands indicate
the uncertainties quoted in Ref.~\cite{Alexandrou:2021mmi} for the reconstructed distributions.
For the kaon, the magnitudes and relative behavior of the $u$- and
$\bar{s}$-valence PDFs are broadly consistent with other lattice QCD
results~\cite{Alexandrou:2021mmi,Salas-Chavira:2021wui,ExtendedTwistedMass:2024kjf}.
In the small-$x$ region, our results agree well with both bands for all three distributions. Around the peak region,
the lattice results tend to peak at slightly smaller $x$ and with somewhat larger amplitudes than our fBT-LFQM
predictions. The comparison with JAM25 is instead flavor dependent: both kaon distributions remain inside the
extracted bands over essentially the entire $x$ range, whereas our pion distribution is harder than the extracted one,
lying above its band outside the small-$x$ region and approaching it again only as $x\to1$.
At larger $x$, visible model-dependent differences appear, especially in the kaon sector. These differences
may reflect the valence-only nature of the present hadronic input
and the absence of explicit higher Fock-state and gluonic degrees of freedom at
the model scale, which in the pion also shows up as a valence momentum fraction larger than the value preferred by
the global fit. Nevertheless, the qualitative features of SU(3) flavor-symmetry breaking
are consistently reproduced: the kaon $u$-valence PDF is softer than
the pion $u$-valence PDF, while the $\bar s$-valence PDF is shifted toward
larger $x$.
An important result of the present analysis is that the SU(3)-breaking pattern
of the valence sector at the hadronic input scale is not washed out by NNLO
evolution. Hence, the evolved distributions preserve a nontrivial imprint of SU(3) 
flavor-symmetry breaking at high scales.

It is instructive to examine the flavor ratio of the valence Mellin
moments,
\begin{equation}
R_{n} \;\equiv\;
\frac{\langle x^{n}\rangle^{s/K}_{\rm val}}
{\langle x^{n}\rangle^{u/K}_{\rm val}} \,,
\label{eq:Rn}
\end{equation}
which provides a particularly stable measure of SU(3) flavor-symmetry
breaking under QCD evolution. At leading and next-to-leading orders,
the valence distributions $q-\bar q$ evolve as flavor non-singlet
combinations with the same anomalous dimension, so that the common
evolution factor cancels in the ratio. At NNLO, a small flavor-mixing
contribution enters through the sea-like non-singlet splitting function
$P_{\rm NS}^{s}$, inducing only a weak scale dependence in $R_n$.
Our results in Table~\ref{tab:Mellin} illustrate this stability:
$R_{1}=1.33$ and $R_{2}=1.61$ at both $\mu^{2}=16$ and $27$~GeV$^{2}$.

The ratio $R_n$ therefore largely isolates the SU(3)-breaking pattern
encoded in the LFWF at the hadronic scale $\mu_{0}$ and provides a useful
quantity for comparing determinations quoted at different resolution
scales, with only a negligible residual scale dependence at the present
level of accuracy. This is particularly useful for confronting our
results with recent phenomenological extractions of the kaon PDFs,
which are given at scales different from ours.

\begin{table}[htbp]
\centering
\begin{tabular}{lccc}
\hline\hline
 & $\mu^{2}$\,[GeV$^{2}$] & $R_{1}$ & $R_{2}$ \\
\hline
This work
& 16 & 1.33 & 1.61 \\
& 27 & 1.33 & 1.61 \\
Statistical model~\cite{Bourrely:2023yzi}
& 10 & 1.26 & 1.42 \\
JAM global fit~\cite{Barry:2025wjx}
& 4 & -- & 1.54(8) \\
DSE approach~\cite{Chen:2016sno}
& 27 & 1.29 & 1.55 \\
$\chi$CQM~\cite{Watanabe:2018qju}
& 27 & 1.04 & 1.06 \\
BLFQ-NJL~\cite{Lan:2019rba}
& 27 & 1.13 & 1.22 \\
lattice QCD~\cite{Alexandrou:2021mmi}
& 27 & 1.29 & 1.46 \\
\hline\hline
\end{tabular}
\caption{Flavor ratios
$R_{n}=\langle x^{n}\rangle^{s/K}_{\rm val}/
\langle x^{n}\rangle^{u/K}_{\rm val}$
of the kaon valence-quark distributions.
The ratios are only weakly scale dependent under NNLO evolution;
the quoted scales are shown for reference.}
\label{tab:ratios}
\end{table}

\begin{table}[t]
\centering
\begin{tabular}{lcccccc}
\hline\hline
$\mu^{2}$\,[GeV$^{2}$] & & valence $q$ & valence $\bar q$ &
valence & gluon & sea \\
\hline
$\mu_{0}^{2}$
 & $\pi^{+}$ & 0.500 & 0.500 & 1.000 & 0 & 0 \\
 & $K^{+}$   & 0.429 & 0.571 & 1.000 & 0 & 0 \\
\hline
$4$
 & $\pi^{+}$ & 0.281 & 0.281 & 0.562 & 0.352 & 0.086 \\
 & $K^{+}$   & 0.241 & 0.321 & 0.562 & 0.352 & 0.086 \\
\hline
$16$
 & $\pi^{+}$ & 0.247 & 0.247 & 0.495 & 0.392 & 0.113 \\
 & $K^{+}$   & 0.213 & 0.282 & 0.495 & 0.392 & 0.113 \\
\hline
$27$
 & $\pi^{+}$ & 0.238 & 0.238 & 0.476 & 0.400 & 0.124 \\
 & $K^{+}$   & 0.204 & 0.271 & 0.476 & 0.400 & 0.124 \\
\hline\hline
\end{tabular}
\caption{Momentum fractions $\langle x\rangle$ of the valence, gluon, and sea
components of the $\pi^{+}$ and $K^{+}$ at the model scale $\mu_{0}=0.6$ GeV and after
NNLO evolution to $\mu^{2}=4$, $16$, and $27~\mathrm{GeV}^{2}$. The quark and
antiquark valence contributions are shown separately, with their sum listed as
``valence.'' The sea contribution is defined as
$\langle x\rangle_{\rm sea}=\langle x\rangle_{\Sigma}
-\langle x\rangle_{\rm val}$. The $\mu_{0}$ entries denote the valence-only
model input.}
\label{tab:budget}
\end{table}

As seen in Table~\ref{tab:ratios}, our value $R_{2}\simeq 1.61$ is
consistent with $R_{2}=1.54(8)$ inferred from the moments of
Ref.~\cite{Barry:2025wjx}, and is close to the DSE result of
Ref.~\cite{Chen:2016sno} and to the lattice determination of
Ref.~\cite{Alexandrou:2021mmi}. The statistical-model extraction of
Ref.~\cite{Bourrely:2023yzi} yields somewhat weaker breaking,
\footnote{The moments needed for $R_{n}$ are not tabulated in
Ref.~\cite{Bourrely:2023yzi}, which quotes only the momentum fractions
$\langle x\rangle$ at $Q^{2}=10$~GeV$^{2}$. We have therefore evaluated
them from the statistical-model parametrization of that work, using the
best-fit parameters given at the input scale $Q_{0}^{2}=1$~GeV$^{2}$.
Forming the valence combinations $q_{v}=q-\bar q$ cancels the diffractive
terms identically, and since the parametrization is expressed as
$x\,q(x)$, the moments follow from
$\langle x^{n}\rangle_{q_{v}}
=\int_{0}^{1}dx\,x^{n-1}[x\,q_{v}(x)]$.
We quote $R_{n}$ without uncertainties, since the correlations among the
fitted parameters are not available.}
while the $\chi$CQM of Ref.~\cite{Watanabe:2018qju} and the BLFQ-NJL
calculation of Ref.~\cite{Lan:2019rba} predict considerably weaker
flavor asymmetry. A feature shared by the approaches with sizable
SU(3) breaking is that $R_{n}$ increases with $n$, 
reflecting the harder $s$-quark distribution relative to the $u$-quark distribution, a trend that our
fBT-LFQM reproduces.

We finally turn to the gluon and sea distributions generated by the evolution.
Table~\ref{tab:budget} lists the first moments of the valence, gluon and sea
distributions of the two mesons at the model scale and after evolution to
$\mu^{2}=4$, $16$ and $27~\mathrm{GeV}^{2}$. Here
$\langle x\rangle_{\rm val}$ denotes the sum of the two valence entries,
$\langle x\rangle^{u-\bar u}+\langle x\rangle^{\bar d-d}$ for the
$\pi^{+}$ and
$\langle x\rangle^{u-\bar u}+\langle x\rangle^{\bar s-s}$ for the
$K^{+}$. The pion and kaon differ in how this valence momentum is shared
between their two constituent flavors.

At $\mu_{0}$, the input consists of the two valence distributions alone, so
that $g_{H}(x,\mu_{0})=0$ and, by the momentum sum rule of
Eq.~\eqref{eq:Nq},
$\sum_{q}\langle x\rangle^{q}(\mu_{0})=1$ for $H=\pi,K$. Under DGLAP
evolution, the first moments of the quark singlet and gluon form a closed
$2\times2$ system whose solution depends only on the initial pair
$(\langle x\rangle_{\Sigma},\langle x\rangle_{g})=(1,0)$, on
$\alpha_{s}$ and on the number of active flavors, but not on the shape of the
input. The two total-valence combinations defined above likewise start from
unity and evolve with the same flavor-independent non-singlet kernel.
Consequently, $\langle x\rangle_{\rm val}$, $\langle x\rangle_{g}$ and
$\langle x\rangle_{\rm sea}
=\langle x\rangle_{\Sigma}-\langle x\rangle_{\rm val}$
remain identical for the two mesons at every scale; heavy-quark thresholds are
flavor blind and do not alter this conclusion.

The SU(3)-breaking pattern nevertheless remains visible in the
flavor-resolved valence moments and their ratios $R_{n}$ in
Table~\ref{tab:ratios}. In the gluon sector, the pion--kaon difference first
appears in moments with $n\ge2$, where the shapes of the singlet inputs enter.
Even there, the effect is small because the softer $u$ and harder $\bar s$
distributions largely compensate in the kaon singlet input
$f_{1}^{u/K}+f_{1}^{\bar s/K}$. Consequently, the ratio
$g^{K}(x)/g^{\pi}(x)$ stays within about $1\%$ of unity for $x\lesssim0.2$
and decreases to $\simeq0.80$ at $x=0.8$, deep in the large-$x$ tail where
$xg(x)$ has already fallen below $10^{-3}$ of its peak value. Finally, the
equality of the gluon momentum fractions is a consequence of the common
valence-only boundary condition rather than a prediction that the physical
pion and kaon carry the same gluon momentum. It would be lifted by
meson-dependent nonperturbative input, such as different effective input
scales or different gluon inputs at $\mu_{0}$, neither of which is included
in the present framework.

\begin{figure}[t]
\centering
\includegraphics[width=\columnwidth]{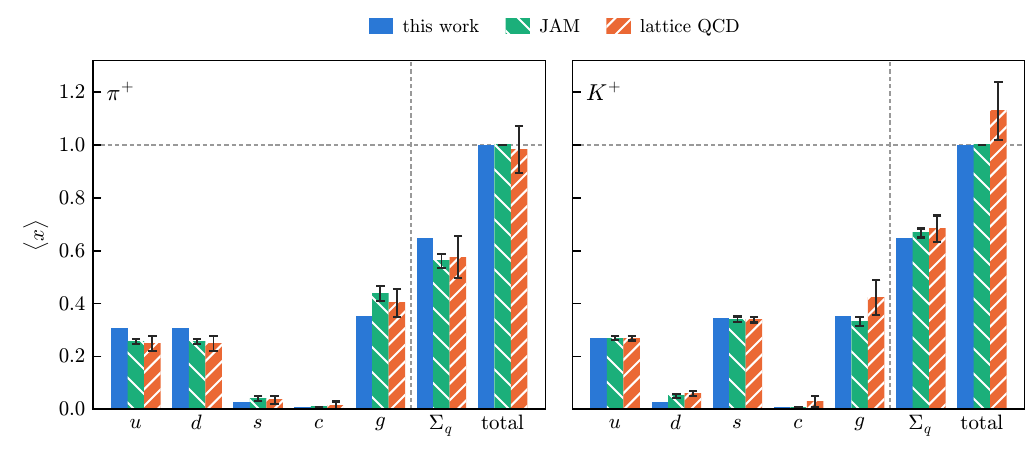}
\caption{Momentum fractions $\langle x\rangle^{q+\bar q}$ of the individual
quark flavors, the gluon fraction $\langle x\rangle^{g}$, the quark singlet
$\langle x\rangle^{\Sigma}$, and the total $\langle x\rangle^{\Sigma}
+\langle x\rangle^{g}$, for the $\pi^{+}$ (left) and the $K^{+}$ (right) at
$\mu=2~\mathrm{GeV}$, compared with the global QCD analysis of
Ref.~\cite{Barry:2025wjx} and with the lattice-QCD results of
Ref.~\cite{ExtendedTwistedMass:2024kjf}.
Our momentum sum and that of Ref.~\cite{Barry:2025wjx} are unity by
construction, whereas the lattice sum is $0.984(89)$ for the pion and
$1.13(11)$ for the kaon.}
\label{fig:lattice}
\end{figure}

Figure~\ref{fig:lattice} compares the flavor decomposition at
$\mu=2~\mathrm{GeV}$ with the lattice-QCD determination of
Ref.~\cite{ExtendedTwistedMass:2024kjf}, which is the only first-principles
momentum decomposition available for both mesons, and with the global QCD
analysis of Ref.~\cite{Barry:2025wjx}, which fits the same lattice moments
together with the Drell-Yan and leading-neutron data. For the kaon the
flavor-resolved quark momentum fractions agree well with the lattice values,
so that the SU(3)-breaking pattern of the valence sector survives the
evolution. For the pion our light-quark fraction is larger than the lattice
one, and in both mesons our gluon fraction is lower. These two differences are
not independent: both follow from the relatively large momentum retained in the
valence sector, $\langle x\rangle^{\pi}_{\rm val}=0.562$ at
$\mu=2~\mathrm{GeV}$, since the momentum sum rule fixes only the combined
remainder $\langle x\rangle_{g}+\langle x\rangle_{\rm sea}
=1-\langle x\rangle_{\rm val}$, whose partition is then determined by the
evolution and by the choice of $\mu_{0}$ and $\alpha_{s}(\mu_{0})$. The lattice
results quoted here instead constrain the flavor-resolved $q+\bar q$ moments
and do not provide the same valence--sea separation. The same reasoning applies
to the sea: the purely sea $d+\bar d$ channel of the kaon and the charm
fractions come out smaller than on the lattice, as expected when all sea
momentum is generated by evolution from a valence-only input rather than
carried by an independent nonperturbative component.

In the gluon sector the three determinations do not share a common pion--kaon
ordering. Our two mesons are degenerate by construction, whereas
Ref.~\cite{Barry:2025wjx} favors a smaller gluon fraction in the kaon and the
lattice values are compatible with either ordering, their momentum sum for the
kaon being $1.13(11)$ rather than unity. The two statements answer different
questions rather than conflict: in the global fit the pion and kaon gluon
normalizations are independent parameters at the input scale, so the difference
is already present in the nonperturbative input, while our common valence-only
boundary condition cannot generate one under evolution. This is not a matter of
tuning, since changing $\mu_{0}$ or $\alpha_{s}(\mu_{0})$ moves the two mesons
by the same amount. A meson-dependent nonperturbative input, as noted above,
is what the present framework would need in order to accommodate such a
difference.
\section{Summary}
\label{sec:summary}

We have presented a self--consistent fBT-LFQM for the kaon, built on the BT
construction, and applied it to the EM form factor, the scalar form factor,
and the full set of unpolarized $T$--even TMDs and their collinear PDFs. We
also analyzed the pion's scalar form factor and the associated twist-3 TMD
and PDF---topics not covered in our previous work~\cite{Choi:2024ptc}. The
defining feature of the fBT-LFQM is the uniform use of the invariant mass
$M_0$ in both the hadronic matrix elements and the Lorentz prefactors,
which maintains consistency with the on-shell constituent kinematics used
to construct the BT wave function and yields BT-consistent and current-component--independent
results for the integrated form factors in the kinematics considered here.

We emphasize the scope of the present framework. 
The LFQM used here is a valence–only effective model restricted to the $q\bar q$ sector at the initial hadronic scale before the QCD evolution takes place. 
The role of the fBT prescription is to enforce internal self–consistency and covariance of the projection procedure once 
the BT-based LFQM framework is adopted. 
It should therefore not be interpreted as incorporating the full QCD evolution dynamics of higher–twist physics,
which would require explicit higher Fock components and gluonic degrees of freedom beyond the scope of the present approach. 
Accordingly, the consistent treatment of the effective LF zero–mode corrections discussed here pertains to their 
observable-level incorporation within the BT-based valence LFQM, rather than to a complete description of dynamical 
higher–Fock–state effects in QCD.

The fBT-LFQM produces a single, component–independent kaon EM form factor $F_{K^+}(Q^2)$ from the 
$\Gamma=\gamma^+$, $\gamma^\perp$, and $\gamma^-$ projections, 
and it satisfies charge normalization at~$Q^2=0$. By contrast, in the pBT-LFQM the $\gamma^+$ and $\gamma^\perp$ extractions agree,
but the $\gamma^-$ channel requires the effective LF zero-mode correction missing from the pBT extraction to be included explicitly.
In the fBT-LFQM, implementing $M\to M_0$ at the integrand level 
incorporates this effective correction, at the level of the observable, through the BT-consistent Lorentz prefactor within the valence framework.

We also analyzed the scalar channel through the scalar form factor $F_S(Q^2)$
and the associated twist-3 TMD $e^q(x,\bm k_\perp)$ and PDF $e^q(x)$.
A central result of the present work is the explicit demonstration, within the
solvable manifestly covariant off-shell BS model discussed in
Appendix~\ref{app:scalarZM}, that the scalar form factor receives a
nonvanishing explicit LF zero-mode contribution.
In the on-shell BT-based LFQM, the fBT--pBT difference in the scalar channel
signals the
effective LF zero-mode correction missing from the pBT extraction at the observable level. In the fBT
prescription, this correction is incorporated through the BT-consistent
Lorentz prefactor via the replacement
$M\to M_0(x,\bm k_\perp)$, without implying an operator-level identification
with the explicit BS-model LF zero-mode term.

Numerically, we find that the pBT-LFQM yields substantially larger results for the scalar
form factor and the corresponding twist-3 PDF $e^q(x)$, whereas the fBT-LFQM
yields a consistent extraction in which the effective scalar LF zero-mode correction missing from the pBT extraction is
incorporated through the Lorentz prefactor. This behavior closely parallels
the well-known LF zero-mode problem in the $\gamma^-$ projection of the
electromagnetic current. The scalar channel therefore provides a nontrivial cross-channel
internal-consistency check of the fBT treatment of zero-mode-sensitive
observables within the BT-based valence LFQM.

We then investigated the T-even TMDs and PDFs obtained from the forward limits of the EM and scalar form factors.
At the model scale, the twist–2 TMD $f_1^q(x,\bm k_\perp)$
exhibits exact $\mathrm{SO}(2)$ symmetry, and its $x$-integrated transverse profile
$f_1^q(\bm k_\perp)$ is exactly Gaussian, with an RMS width fixed by the corresponding LFWF parameter $\beta_H$ ($H=\pi,K$),
corresponding to an exact Gaussianity ratio $R_G=1$. The higher–twist TMDs $f^{\perp q}$, $e^q$, and $f_4^q$ display systematic twist and flavor dependencies, 
including broader widths for the $s$ quark in $K^+$. 
Although the BT kinematic mass entering $f_4^q$ contains an explicit angular dependence at the integrand level through $|{\bm k}_\perp|^2\cos^2\theta$, 
the final TMD shown here is obtained
after azimuthal averaging in the $|{\bm q}_\perp|\to0$ limit and is therefore azimuthally symmetric. 
The competing $k_\perp$ dependences of the numerator and the BT
kinematic mass in the denominator make $f_4^q$ the least Gaussian
of the unpolarized $T$-even TMDs, as measured by $R_G$, while the
relative ordering of the $e^q$ and $f_4^q$ widths is flavor dependent.
In the fBT-LFQM, the twist–4 PDF $f_4^q(x)$ satisfies its forward–limit sum rule, whereas the pBT-LFQM violates it unless 
the effective LF zero–mode correction defined as the fBT--pBT difference is explicitly included. 
Likewise, the twist-3 PDF $e^q(x)$ exhibits sizable differences between
the two formulations, which are interpreted as the effective scalar LF zero-mode correction missing from the pBT extraction and incorporated through the BT replacement $M\to M_0$ in the scalar Lorentz prefactor.

For the collinear NNLO DGLAP evolution we adopt
$\mu_0=0.6~\mathrm{GeV}$ with a valence-only boundary condition and
$\alpha_s(\mu_0)=1$, which yields a pion valence-momentum fraction
$\langle x\rangle^{\pi}_{\rm val}=0.562$ at $\mu=2~\mathrm{GeV}$ and is applied
without change to the kaon. Because the two mesons then share the same initial
singlet--gluon momentum pair, their total valence, gluon, and sea momentum
fractions remain identical at every scale; SU(3) flavor-symmetry breaking
survives instead in the flavor-resolved valence moments and in the $x$
dependence of the evolved distributions. At $\mu=2~\mathrm{GeV}$ the
flavor-resolved kaon quark momentum fractions agree well with the lattice-QCD
determination of Ref.~\cite{ExtendedTwistedMass:2024kjf}, while our gluon
fraction lies below the lattice central value for both mesons and is lower for the pion but
slightly higher for the kaon with respect to
the global QCD analysis of Ref.~\cite{Barry:2025wjx}. A pion--kaon difference in
$\langle x\rangle_{g}$ would require meson-dependent nonperturbative input,
which deserves further investigation beyond the present valence-only framework.

Overall, the fBT-LFQM establishes a self-consistent framework
for the extraction of EM and scalar form factors, TMDs, and PDFs.
For the integrated form factors in the kinematics considered
here, the extraction is BT-consistent and current-component independent,
with the effective LF zero-mode corrections missing from the pBT
extractions consistently incorporated at the observable level in both the
$\gamma^-$ and scalar channels.
It provides a unified link between nonperturbative quark structure at the
model scale and perturbative QCD evolution at higher scales.
Future extensions include polarized TMDs, and in particular the
leading-twist $T$-odd Boer--Mulders function $h_1^{\perp q}$, whose
description requires the gauge-link (initial-/final-state interaction)
effects that are absent in a model without explicit gluon degrees of freedom,
generalized three--point correlators, and direct comparisons with precision
meson data from upcoming EIC experiments~\cite{AbdulKhalek:2021gbh}, as well
as with global TMD and PDF extractions~\cite{Bacchetta:2024qre,Barry:2025wjx}.

\section*{Acknowledgment}
We thank P. C. Barry for providing the pion and kaon valence distributions of
Ref.~\cite{Barry:2025wjx} at $\mu^{2}=27~\mathrm{GeV}^{2}$, together with their $95\%$
credible-interval bands, which are shown in Fig.~\ref{fig:evolution}.
Y. Choi and H.-M. Choi were supported by the National Research Foundation of Korea (NRF) grant funded by the Korea government
under Grant No. RS-2025-02634319 (Y. Choi) and RS-2026-25475296 (H.-M. Choi).
A.J.A. is supported by JAEA Postdoctoral Fellowship Program and was partly supported by the RCNP Collaboration Research Network program under project number COREnet 057, as well as by the PUTI Q1 Grant from the University of Indonesia under contract No. NKB-441/UN2.RST/HKP.05.00/2025.
C.-R. Ji was supported in part by the U.S. Department of Energy (Grant No. DE-FG02-03ER41260).
The National Energy Research Scientific Computing Center (NERSC) supported by the Office of Science of the U.S. Department of Energy
under Contract No. DE-AC02-05CH11231 is also acknowledged.

\appendix
\section{LF zero mode for scalar form factor in BS model}
\label{app:scalarZM}
\renewcommand{\theequation}{A\arabic{equation}}
\subsection{Manifestly covariant calculation}
\label{app:scalarZM1}
We examine the explicit LF zero-mode contribution to the scalar form factor,
defined by
$\Braket{{\cal J}_S}=\langle P'|\,\bar q(0)\mathbf{1}q(0)\,|P\rangle=2M\,F_S(Q^2)$,
within a solvable covariant BS model. By comparing the manifestly covariant result
with the corresponding LF calculation in the Drell-Yan-West frame $(q^+=0)$, we
identify the origin of the explicit LF zero-mode contribution.
In the solvable covariant BS model in $(3+1)$-dimensional fermion field theory~\cite{BCJ01},
the matrix element is given by
 \be\label{eq:31}
 \Braket{{\cal J}_S} = i N_c
 \int\frac{d^4 p_1}{(2\pi)^4} \frac{H_0 H'_0} {N_{\Lambda}
N_{1} N_{\bar q} N'_{1} N'_{\Lambda}} {\cal S},
 \ee
where $N_c$ is the color factor. The trace is 
\begin{align}\label{eq:s1}
{\cal S}
&=
\mathrm{Tr}\bigl[\gamma_5(\slashed{p}_1+m_1)(\slashed{p}'_1 + m_1)\gamma_5(-\slashed{k}+m_2)\bigr]
\nonumber\\
&=
4\left[ (m_1 - m_2) k\cdot (P + P') + k^2 (m_2 - 2 m_1) + m_2(P\cdot P') + m^2_1 m_2 \right].
\end{align}
Here $p_1=P-k$ and $p'_1=P'-k$ are the momenta of the active quark with mass $m_1$
and $k$ is the momentum of the spectator antiquark with mass $m_2$.
The denominator factors  are $N^{(\prime)}_1= p^{(\prime) 2}_{1} - m^2_1 + i\ep$ and
$N_{\bar q}= k^2 - m^2_2 + i\ep$.
Throughout this appendix the symbol ${\bm k}_\perp$ appearing in the LF wave
functions retains the meaning fixed in Eq.~\eqref{eq:15}, i.e., it is the relative
transverse momentum of the struck quark, ${\bm k}_\perp = {\bm p}_{1\perp}$ for
${\bm P}_\perp=0$, so that the transverse component of the spectator four-momentum
$k$ is $-{\bm k}_\perp$; correspondingly
${\bm k}'_\perp={\bm k}_\perp+(1-x){\bm q}_\perp$ with $q=P'-P$.
We take the vertex functions as  $H_0=H_0(p^2_1, k^2)=g/N_\Lambda$ and $H'_0=H'_0(p^{\prime 2}_1, k^2)=g/N'_\Lambda$  with
$N^{(\prime)}_{\Lambda}=p_1^{(\prime)2}-\Lambda^2+i\epsilon$.
The normalization constant $g$ is fixed by the charge-normalization condition for the pseudoscalar mesons at $q^2=0$.
To facilitate the loop integration, we decompose the five-denominator structure in Eq.~(\ref{eq:31}) as
\begin{align}\label{eq:34}
\frac{1}{N_{\Lambda}N_1N_{\bar q} N'_1 N'_{\Lambda}}
&=
\frac{1}{(\Lambda^2-m_1^2)^2}
\frac{1}{N_{\bar q}}
\left(\frac{1}{N_{\Lambda}}-\frac{1}{N_1}\right)
\left(\frac{1}{N'_{\Lambda}}-\frac{1}{N'_1}\right).
\end{align}
This reduces the original five-propagator integral to the Pauli--Villars combination
of four three-denominator integrals.
For a generic term $1/(N_aN_{\bar q}N'_b)$ with $a,b=m_1$ or $\Lambda$, Feynman
parametrization gives
\begin{align}
\frac{1}{N_aN_{\bar q}N'_b}
&=
2\int_0^1 dx\int_0^{1-x}dy\,
\frac{1}{
\left[N_{\bar q}+x(N_a-N_{\bar q})+y(N'_b-N_{\bar q})\right]^3},
\end{align}
where
\begin{align}
N_{\bar q}+x(N_a-N_{\bar q})+y(N'_b-N_{\bar q})
=
(k-xP-yP')^2 + C_{ab}+i\epsilon,
\end{align}
and
\begin{align}
C_{ab}
&=
(1-x-y)(xM^2+yM'^2)
+xyq^2
-xa^2-yb^2
-(1-x-y)m_2^2 .
\end{align}
After shifting the loop momentum by $\ell=k-xP-yP'$, the denominator becomes a
function of $\ell^2$ only. Therefore, all terms in the numerator that are odd
under $\ell\to-\ell$ vanish upon symmetric loop integration. Keeping only the
even terms, the trace can be reduced to
$S(\ell+K)=4(\alpha+\beta\ell^2)$, with
\begin{align}
\al &= \left[ (m_1 - m_2) K\cdot (P + P') +  K^2 (m_2 - 2 m_1) + m_2 (P\cdot P') + m^2_1 m_2 \right],
\nonumber\\
\beta &= (m_2 - 2 m_1),
\end{align}
where $K=xP + yP'$. Each three-denominator integral is evaluated using dimensional regularization.
Although the individual terms contain logarithmic divergences, these divergences
cancel in the Pauli--Villars combination generated by Eq.~\eqref{eq:34}. The finite
covariant matrix element is then
\begin{align}\label{eq:JS2}
\Braket{{\cal J}_S}
&=
\frac{N_cg^2}{4\pi^2(\Lambda^2-m_1^2)^2}
\int_0^1dx\int_0^{1-x}dy
\Biggl[
2\beta
\ln\!\left(
\frac{C_{\Lambda\Lambda}C_{m_1m_1}}
{C_{\Lambda m_1}C_{m_1\Lambda}}
\right)
-\alpha{\cal C}
\Biggr],
\end{align}
where
\begin{align}
{\cal C}
=
\frac{1}{C_{\Lambda\Lambda}}
-\frac{1}{C_{\Lambda m_1}}
-\frac{1}{C_{m_1\Lambda}}
+\frac{1}{C_{m_1m_1}} .
\end{align}
The Lorentz-invariant scalar form factor is finally obtained as
\be\label{eq:FSCov}
F^{\rm Cov}_S(q^2) = \frac{\Braket{{\cal J}_S}}{2M}.
\ee

\subsection{Light-front calculation}
\label{app:scalarZM2}
We now evaluate Eq.~\eqref{eq:31} in the $q^+=0$ frame and compare
the result with the manifestly covariant calculation. In the LF calculation, the
trace in Eq.~\eqref{eq:s1} is decomposed into the on-shell propagating part and the
instantaneous part by using $\slashed{p}+m = (\slashed{p}_{\rm on}+m) +\frac{1}{2}\gamma^+(p^- -p^-_{\rm on})$ so that
${\cal S} ={\cal S}_{\rm on} + {\cal S}_{\rm inst}$. Explicitly,
\begin{align}\label{eq:BStrace}
{\cal S}_{\rm on}
&= 4\left[ m_1 (k_{\rm on} \cdot p_{1\rm on} + k_{\rm on} \cdot p'_{1\rm on})
+ m_2 (p_{1\rm on} \cdot p'_{1\rm on} + m_1^2) \right], \nonumber\\
{\cal S}_{\rm inst}
&= 2\delta_1^- (m_1 k^+ + m_2 p_1^{\prime +})
+ 2\delta_1^{\prime -} (m_1 k^+ + m_2 p_1^+)
+ 2\delta_k^- m_1 (p_1 + p'_1)^+,
\end{align} 
where $\delta_1^{(\prime)-} = p_1^{(\prime)-}-p_{1\rm on}^{(\prime)-}$ and $\delta_k^- = k^- - k^-_{\rm on}$.

In the valence region $0<k^+<P^+$, the spectator antiquark pole ($k^- = k_{\rm on}^-$) 
lies in the lower half of the complex $k^-$ plane.  
Taking this pole gives
\bea\label{eq:319}
\Braket{{\cal J}_S}_{\rm val} &=&
\frac{N_c}{16\pi^3}\int^1_0 \frac{dx}{(1-x)}\int d^2{\bm k}_\perp
\chi' (x, {\bm k}'_\perp) \chi(x,{\bm k}_\perp) 
 {\cal S}_{\rm val},
  \eea
where ${\cal S}_{\rm val} = {\cal S}_{\rm on} + {\cal S}_{\rm inst}(\delta_k^- = 0)$ and
\be
{\cal S}_{\rm on} 
=4 \frac{ \left[ 2 m_1 ({\cal A}^2 + {\bm k}_\perp\cdot{\bm k}'_\perp ) + (1-x){\cal A}{\bm q}_\perp^2 \right]}{2 x(1-x)},
\ee
and
\be
{\cal S}_{\rm inst}(\delta_k^- = 0) = 2 {\cal A} (M^2 + M'^2 - M_0^2 - M_0^{\prime 2}).
\ee
The LF vertex functions are given by
 \be\label{eq:321}
 \chi(x,{\bm k}_\perp) = \frac{g}{ x^2 (M^2 - M^2_0)(M^2 - M^2_{\Lambda})},
 \qquad
\chi' (x, {\bm k}'_\perp)= \frac{g}{ x^2 (M'^2 - M'^2_0)(M'^2 - M'^2_{\Lambda})}.
\ee
where $M_{\Lambda}^{(\prime)}= M_0^{(\prime)}(m_1\to\Lambda)$.

Therefore, the LF valence contribution to the scalar form factor is
 \bea\label{eq:325}
[F_{S}(Q^2)]_{\rm BS}^{\rm val} &=& \frac{N_c}{16\pi^3}\int^1_0 \frac{dx}{(1-x)}
 \int d^2{\bm k}_\perp \chi'(x, {\bm k}'_\perp) \chi (x, {\bm k}_\perp)
   \frac{{\cal S}_{\rm val}}{2M}.
  \eea
In the nonvalence region $P^+<k^+<P'^+$, 
the relevant poles are
$p^{\prime -}_1=p^{\prime -}_{1\rm on}(m_1)$ from the outgoing struck-quark propagator and
$p^{\prime -}_1=p^{\prime -}_{1\rm on}(\Lambda)$ from the smeared vertex. Both poles lie in the
upper half of the complex $k^-$ plane. Since $q=P'-P$, one has
$P'^+/P^+=1+q^+/P^+\equiv1+\alpha$ with $\alpha\equiv q^+/P^+>0$, so that the
nonvalence region corresponds to $-\alpha<x<0$. Thus the explicit LF zero-mode contribution in
the $q^+\to0$ limit is obtained from the endpoint region $x\to0$.

A key feature of the zero mode is that the shrinking nonvalence interval can still
give a finite contribution, $\lim_{\alpha\to0}\int_0^\alpha dx\,(\cdots)\neq 0$,
when singular LF-energy factors compensate the vanishing integration domain.

With $x=-\alpha z$ and $0<z<1$, one has $p_1^+=xP^+=-\alpha zP^+$ and $p_1^{\prime +}=(x+\alpha)P^+=\alpha(1-z)P^+$.
Thus $p_1^+$ and $p_1^{\prime +}$ vanish linearly with $\alpha$, while $k^+$ remains
finite, with $k^+=(1-x)P^+\to P^+$ in the endpoint limit $x\to 0$. After taking the $p_1^{\prime -}$ pole, 
\[
p_1^{\prime -} =p^{\prime -}_{1\rm on}\sim\frac1{\al},\qquad
k^- =P'^- -p_1^{\prime -} \sim -\frac1{\al},\qquad
p_1^{-}=P^- -P'^- +p_1^{\prime -} \sim\frac1{\al}.
\]
Hence the endpoint singularity is controlled by the LF energy factors
$p_1^-$, $p_1^{\prime -}$, and $-k^-$.

The zero-mode is generated only by uncompensated singular LF-energy factors.
From the on-shell part, the relevant singular terms are 
\[
p_{1\rm on}\!\cdot k_{\rm on}
\supset \frac12 p^-_{1\rm on}k^+,
\qquad
p'_{1\rm on}\!\cdot k_{\rm on}
\supset \frac12 p^{\prime -}_{1\rm on}k^+ ,
\]
which give
\[
{\cal S}_{\rm on}\Big|_{\rm sing.}
=
2m_1k^+(p^-_{1\rm on}+p^{\prime -}_{1\rm on}).
\]
For the instantaneous part, after taking the $p_1^{\prime -}$ pole, $\delta_1^{\prime -}=0$, so only
the $\delta_1^{-}$ term contributes:
\[
{\cal S}_{\rm inst}\Big|_{\rm sing.}
=
2m_1k^+ (p_1^{-}-p^{-}_{1\rm on}) .
\]
Thus the $p^{-}_{1\rm on}$ terms cancel, and
\[
{\cal S}_{\rm nv}\Big|_{\rm sing.}
=
2m_1k^+(p^-_1+p_1^{\prime -}).
\]
Since $p_1^{\prime -}=p_1^-+{\cal O}(1)$ in the endpoint region, and
$k^+\to P^+$ as $x\to 0$, we obtain
\be\label{eq:330}
{\cal S}_{\rm nv}\big|_{\rm sing.}\ \xrightarrow[x\to0]{}\ 4m_1\,P^+\,p_1^-\sim\frac{1}{\al}.
\ee
The zero-mode source is therefore the singular LF-energy combination
$P^+p_1^-$, and this zero-mode operator is common to all four terms generated by
the decomposition in Eq.~\eqref{eq:34}.

Using Eq.~\eqref{eq:34} to avoid double $p^-_1$ poles, the explicit LF zero-mode contribution
from the LF energy $P^+p^-_1$ is
\be\label{eq:j4}
 I[P^+ p^-_1]_{\rm Z.M.}
=
iN_c \lim_{\al\to 0}
\int_{\rm nv} \frac{d^4 p_1}{(2\pi)^4}
\frac{H_0 H'_0}{N_1 N_{\bar q} N'_1}\,P^+ p_1^-
=
\frac{g^2 N_c}{16\pi^2(\Lambda^2-m_1^2)^2}
\int^1_0 dz
\ln\!\left(
\frac{B_{\Lambda m_1}B_{m_1\Lambda}}
{B_{\Lambda\Lambda}B_{m_1m_1}}
\right),
\ee
where $B_{ab}=z(1-z){\bm q}_\perp^2+(1-z)a^2+zb^2$.
Here $x=-\alpha z$ maps the shrinking nonvalence region onto $0<z<1$, and the
singular factor $P^+p_1^-\sim1/\alpha$ compensates the vanishing integration
domain $\int_{-\alpha}^{0}dx$.
The logarithmic structure results from the transverse-momentum integration; the
mass-independent divergent pieces cancel in the Pauli--Villars combination.

Therefore, the explicit LF zero-mode contribution to the scalar form factor is 
\be\label{eq:FSzero}
[F_S(Q^2)]_{\rm BS}^{\rm Z.M.} = \frac{4 m_1}{2M} I[P^+ p^-_1]_{\rm Z.M.}.
\ee

\subsubsection{Effective inclusion of the zero-mode in the valence region }
\label{app:scalarZM3}
In the solvable covariant BS model, the scalar-current matrix element contains
off-mass-shell contributions, which are responsible for the LF zero mode. As shown
in our previous work~\cite{JC}, the non-wave-function vertex in the $q^+>0$ frame
can be related to the ordinary valence wave function through the iteration of the
irreducible kernel in the bound-state equation. In the limit $q^+\to0$, this
nonvalence contribution becomes the explicit LF zero-mode contribution in the $q^+=0$ frame.
Thus, the zero mode can be represented by an effective operator acting within the
valence sector.
This effective treatment can be related to the LF covariant method of
Jaus~\cite{Jaus99}, where the orientation of the LF plane is specified by
$\omega\cdot x=0$~\cite{Kar76,Carbonell98}. The choice
$\omega=(\omega^+,\omega^-,{\bm\omega}_\perp)=(0,2,{\bm0}_\perp)$ gives the usual
LF plane $\omega\cdot x=x^+=0$. Although physical observables must be independent of
$\omega$, the valence contribution alone may not be sufficient to ensure this
covariance when zero-mode effects are present. The Jaus prescription identifies the
spurious $\omega$-dependent part and replaces the missing explicit LF zero-mode contribution by
an effective operator acting within the valence sector.
Below, we summarize how the Jaus prescription identifies the effective valence
operator corresponding to the explicit LF zero-mode contribution obtained directly from the
nonvalence endpoint region.
In the $\omega$-dependent LF covariant approach of
Refs.~\cite{Carbonell98,Jaus99},\rd{\footnote{Reference~\cite{Jaus99} adopts the
opposite convention for the momentum transfer, $q_{\rm J}=P_{\rm i}-P_{\rm f}=-q$,
and denotes $P_{\rm i}+P_{\rm f}$ by $P$, which is our $\bar P$. Under $q\to -q$
one has $q^2$ and $A^{(1)}_1$ unchanged, $q\cdot\bar P\to-q\cdot\bar P$,
${\bm q}_\perp\to-{\bm q}_\perp$, and $A^{(1)}_2\to-A^{(1)}_2$, while
$C^{(1)}_1$ and $Z_2$ are invariant as scalars although the signs of
$q\cdot\bar P$ and ${\bm k}_\perp\cdot{\bm q}_\perp$ in their explicit forms are
reversed. Equations~\eqref{eq:j1}--\eqref{eq:j3} are written throughout in our
convention $q=P'-P$.}} the four-vector $p_1^\mu$ is decomposed as
\be\label{eq:j1}
p^\mu_1
=
{\bar P}^\mu A^{(1)}_1
+q^\mu A^{(1)}_2
+\frac{\omega^\mu}{\omega\cdot{\bar P}}C^{(1)}_1,
\ee
where the coefficients are given by
 \bea\label{eq:j2}
A^{(1)}_1&=& \frac{\omega\cdot p_1}{\omega\cdot {\bar P}}=\frac{x}{2},
\nonumber\\
A^{(1)}_2&=& \frac{1}{q^2}\biggl( p_1\cdot q
- (q\cdot {\bar P})\frac{\omega\cdot p_1}{\omega\cdot {\bar P}}\biggr)
=-\frac{x}{2} - \frac{{\bm k}_\perp\cdot{\bm q}_\perp}{q^2},
\nonumber\\
C^{(1)}_1 &=& p_1\cdot {\bar P} - {\bar P}^2 A^{(1)}_1 - q\cdot {\bar P} A^{(1)}_2 = Z_2 - N_{\bar q},
 \eea
where 
 \be\label{eq:j3}
Z_2 = x(M^2 - M^2_0) + m^2_1 - m^2_2 + (1-2x)M^2
- [q^2 - q\cdot {\bar P}]\frac{{\bm k}_\perp\cdot{\bm q}_\perp}{q^2}.
 \ee
The coefficient $C^{(1)}_1$, which multiplies the LF orientation vector
$\omega^\mu$, contains the LF energy $p_1^-$ through $p_1\cdot{\bar P}$ and is
therefore the zero-mode-sensitive part of the decomposition.
Indeed, since ${\bar P}^+=2P^+$ in the $q^+=0$ frame, the part of
$p_1\cdot{\bar P}$ proportional to the off-shell LF energy is precisely $P^+p_1^-$,
the combination identified in the direct endpoint analysis of Eq.~\eqref{eq:330}.
Since $C^{(1)}_1=Z_2 - N_{\bar q}$, removing the spurious $\omega^\mu$ dependence amounts to
replacing the zero-mode associated with the spectator denominator by the valence operator
$Z_2$. 
Jaus's original prescription concerns the spectator denominator itself,
$N_{\bar q}\to Z_2$, i.e., $I[N_{\bar q}]_{\rm Z.M.}=I[Z_2]_{\rm val}$. In the endpoint
region, the singular part of the spectator denominator behaves as
$N_{\bar q}\big|_{\rm sing.}\simeq-P^+p_1^-$, and our direct nonvalence calculation in
Eq.~\eqref{eq:j4} accordingly gives $I[P^+p_1^-]_{\rm Z.M.}=-I[N_{\bar q}]_{\rm Z.M.}$.
This identity follows from the explicit endpoint integration performed here and is not
an additional prescription assumed from the outset. Combining the two relations, we
obtain
\be\label{eq:j4b}
I[P^+ p^-_1]_{\rm Z.M.}
=
-I[N_{\bar q}]_{\rm Z.M.}
=
-I[Z_2]_{\rm val}.
\ee
This relation implies that the explicit LF zero-mode contribution associated with the LF energy
combination $P^+p_1^-$ can be replaced by the effective operator $-Z_2$
acting within the valence region. The corresponding valence contribution generated by
$Z_2$ is~\cite{Jaus99}
\be\label{eq:j5}
I[Z_2]_{\rm BS}^{\rm val}
=
\frac{N_c}{16\pi^3}
\int_0^1\frac{dx}{1-x}
\int d^2{\bm k}_\perp\,
\chi'(x,{\bm k}'_\perp)\chi(x,{\bm k}_\perp)\,Z_2 .
\ee
Thus, the explicit LF zero-mode contribution associated with $P^+p^-_1$ can be
represented entirely by the effective valence operator $-Z_2$.
The rank-one covariant decomposition and the original replacement rule
$N_{\bar q}\to Z_2$ are detailed in Ref.~\cite{Jaus99}, while the identification of the
zero-mode source as $P^+p_1^-$ follows from the direct endpoint calculation presented
above.

In summary, the explicit LF zero-mode contribution to the scalar form factor
$F_S(Q^2)$ can be expressed in terms of the LF valence wave functions as
 \be\label{eq:FSLFZM}
 [F_S(Q^2)]_{\rm BS}^{\rm Z.M.} = \frac{N_c}{16\pi^3}\int^1_0 \frac{dx}{(1-x)}
 \int d^2{\bm k}_\perp \chi'(x, {\bm k}'_\perp) \chi (x, {\bm k}_\perp)
  \frac{\hat{{\cal S}}_{\rm Z.M.}}{2M},
  \ee
where $\hat{{\cal S}}_{\rm Z.M.} = - 4 m_1 Z_2$ is the effective zero-mode operator corresponding to the
zero mode associated with $4 m_1 P^+ p_1^-$.
We confirm numerically that this result is identical to the direct zero-mode result in Eq.~\eqref{eq:FSzero}.
Therefore, the full result is given by
 \be\label{eq:FSBSfull}
 [F_S(Q^2)]_{\rm BS}^{\rm full} = \frac{N_c}{16\pi^3}\int^1_0 \frac{dx}{(1-x)}
 \int d^2{\bm k}_\perp \chi'(x, {\bm k}'_\perp) \chi (x, {\bm k}_\perp)
  \frac{{\cal S}_{\rm val} + \hat{{\cal S}}_{\rm Z.M.}}{2M}.
  \ee
   \begin{figure}[t]
 	\centering
 	\includegraphics[width=0.6\textwidth]{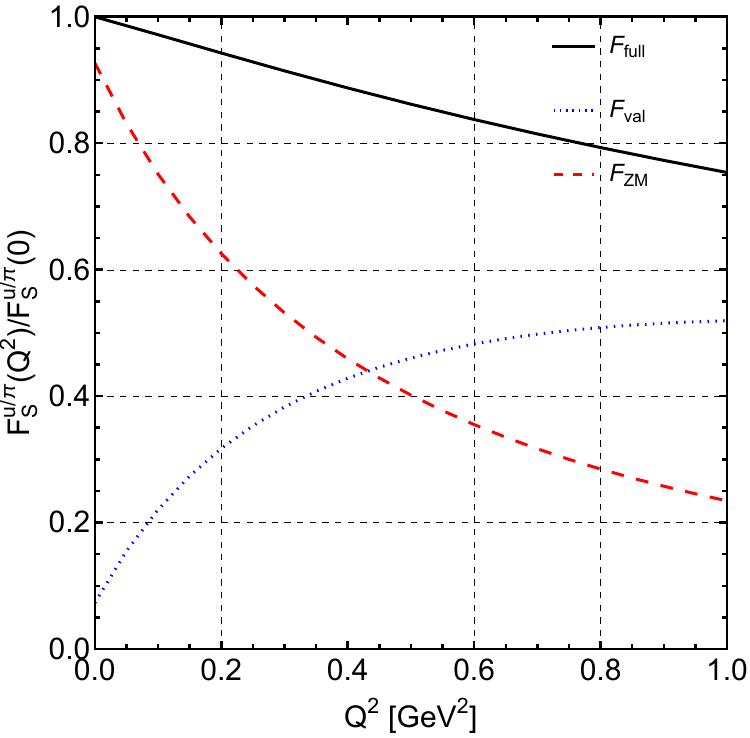}
 	\caption{The $u$-quark contribution to the scaled scalar form factor
$F_S^{u/\pi}(Q^2)/F_S^{u/\pi}(0)$ of the $\pi^+$ in the exactly solvable BS model.
The full result is compared with the LF valence contribution and the explicit LF zero-mode
contribution.\label{fig:FSpiBS}}
 \end{figure}
Figure~\ref{fig:FSpiBS} shows the $u$-quark contribution to the scaled pion scalar
form factor $F_S^{u/\pi}(Q^2)/F_S^{u/\pi}(0)$ obtained in the exactly solvable BS
model, using $m_1=m_2=0.3~{\rm GeV}$ and $\Lambda=0.6~{\rm GeV}$. The manifestly
covariant result in Eq.~\eqref{eq:FSCov} is reproduced by the sum of the LF valence
contribution in Eq.~\eqref{eq:325} and the explicit LF zero-mode contribution in
Eq.~\eqref{eq:FSLFZM}. We have also verified that Eq.~\eqref{eq:FSLFZM}, obtained
from the effective zero-mode operator, is identical to the direct zero-mode result
in Eq.~\eqref{eq:FSzero}. This demonstrates explicitly that the scalar form factor
receives a nonvanishing explicit LF zero-mode contribution in the covariant BS model.
Figure~\ref{fig:FSpiBS} also shows that the explicit BS-model LF zero-mode contribution is largest at $Q^2=0$ and 
decreases as $Q^2$ increases, while the valence contribution exhibits the opposite trend. 
Omitting this contribution therefore yields an incomplete scalar form factor within the BS model. 
This result establishes the zero-mode sensitivity of the scalar projection, but it does not by itself 
determine how the missing contribution must be represented in the dynamically distinct BT-based LFQM.

\subsubsection{Matching the BS valence result onto the BT-based LFQM}
\label{app:BS2fBT}

The analysis of Secs.~\ref{app:scalarZM1}--\ref{app:scalarZM3} establishes, within the
covariant BS model itself, that the scalar form factor receives a nonvanishing
explicit LF zero-mode contribution and identifies its endpoint origin. We now specify how
the valence part of that BS amplitude is carried over to the BT-based LFQM.

Before doing so it is useful to state the scope of what follows. The covariant
BS model and the BT-based LFQM are dynamically distinct frameworks. While the former
is formulated in terms of off-mass-shell quark and antiquark propagation and
contains, in addition to the on-shell propagating part, an instantaneous
contribution and a genuine LF zero mode whose form depends on the vertex
regularization, the latter is 
based on the BT construction of relativistic bound-states consistently including the interactions among the constituents in the specific formulation to fulfill the full Poincar\'e symmetry of the interacting particle systems. 
This is precisely why the matching is set up between the valence
contributions of the two frameworks. Because the two models differ in their
dynamical content, the zero-mode operator of the BS model is not an object that
can be transported into the Gaussian LFQM: it is tied to the off-shell
propagation and to the vertex regularization of the covariant model, neither of
which has a counterpart on the on-shell valence side. Importing it directly does
not restore covariance in the BT-based LFQM~\cite{CJ14,hhk8-g2bj}. What can be done
is to start from the BS valence contribution alone and
show how the fBT-LFQM result is obtained from it once the BT construction is
imposed at the integrand level. The prescription below therefore maps the BS
valence integrand onto the fBT-LFQM one, and we give an explicit account of what
it does to each structure of that integrand.

\paragraph{The matching prescription.}
When the LF calculation is free from treacherous contributions such as
instantaneous terms and zero modes, the covariant BS result is related to the
pBT-LFQM by the ordinary matching condition
\begin{equation}
  \sqrt{2N_c}\,\frac{\chi^{(\prime)}(x,\boldsymbol{k}^{(\prime)}_\perp)}{1-x}
  \;\longrightarrow\;
  \frac{\phi^{(\prime)}(x,\boldsymbol{k}^{(\prime)}_\perp)}
       {\sqrt{\mathcal{A}^2+\boldsymbol{k}^{(\prime)2}_\perp}} ,
  \label{eq:vertexlink}
\end{equation}
a well-known example being the pseudoscalar-meson electromagnetic form factor
obtained from the plus component of the current~\cite{Choi:2024ptc}. For the
scalar current, and likewise for the minus component of the vector current,
this replacement alone is not sufficient: the BS valence
integrand contains an instantaneous piece, and the full covariant amplitude in
addition receives the explicit LF zero-mode contribution identified in Sec.~\ref{app:scalarZM3}.

The prescription we use consists of Eq.~\eqref{eq:vertexlink} supplemented by
the on-mass-shell LF energy condition at the meson--quark vertex,
\begin{equation}
P^-_{\rm BT}=p_1^-+p_2^-
=\frac{M_0^2+\boldsymbol{P}_\perp^2}{P^+}
\qquad\Longleftrightarrow\qquad
M^{(\prime)}
\longrightarrow
M_0^{(\prime)}(x,\boldsymbol{k}^{(\prime)}_\perp),
\label{eq:onshellcond}
\end{equation}
imposed \emph{uniformly at the integrand level}, that is, on the spin
structure and on the Lorentz prefactor alike. We stress that Eq.~\eqref{eq:onshellcond} is
not an adjustable redefinition of the integrand introduced to reproduce a
desired answer: it is the defining condition of the BT construction to build the LFWF, and the
same condition was shown in Ref.~\cite{CJ14,hhk8-g2bj} to be what renders the
meson decay constants independent of the current component, the polarization
vector and the reference frame.

In particular, directly importing the BS-model zero-mode operator into the
Gaussian LFQM does not provide the appropriate correspondence between the two
frameworks. Instead, the fBT-LFQM result can be obtained directly from the BS-model valence
expression, without explicitly calculating or importing the BS-model zero-mode
contribution, by applying Eqs.~\eqref{eq:vertexlink} and \eqref{eq:onshellcond}
in the following form:
\bea\label{eq:matching2}
&&(1)\quad \text{vertex function:}\qquad
\sqrt{2N_c}\,
\frac{\chi^{(\prime)}(x,{\bm k}_\perp^{(\prime)})}{1-x}
\ \longrightarrow\
\frac{\phi^{(\prime)}(x,{\bm k}_\perp^{(\prime)})}
{\sqrt{{\cal A}^2+{\bm k}_\perp^{(\prime)2}}},
\nonumber\\[2mm]
&&(2)\quad \text{spin factor:}\qquad
{\cal S}_{\rm val}={\cal S}_{\rm on}+{\cal S}_{\rm inst}
\ \xrightarrow{\,M^{(\prime)}\to M_0^{(\prime)}\,}\
{\cal S}_{\rm on},
\nonumber\\[2mm]
&&(3)\quad \text{Lorentz factor:}\qquad
{\cal P}^{[\Gamma]}_{\rm BS}
\ \longrightarrow\
{\cal P}^{[\Gamma]}_{\rm fBT},
\eea
where ${\cal S}_{\rm val}$ denotes the Dirac trace of the BS valence
contribution, Eq.~\eqref{eq:BStrace}, to be distinguished from
the spin-orbit factor $S^{[\Gamma]}$ of the LFQM defined in
Eq.~\eqref{eq:17}.

The first replacement gives the standard correspondence between the BS vertex
function and the Gaussian LFQM wave function. It also fixes, at the same time,
the correspondence between the two spin structures. The two are in fact closely
related objects: they have the same numerator and differ only in the
normalization they carry. To see this, note that the covariant form of the
spin-orbit wave function, Eq.~\eqref{eq:2.5}, may be written as
\be\label{eq:Rcov}
{\cal R}^{00}_{\lambda_1\lambda_2}
=\frac{\bar u_{\lambda_1}(p_1)\gamma_5 v_{\lambda_2}(p_2)}{\sqrt{2}}\,
 \frac{\sqrt{x(1-x)}}{\sqrt{{\cal A}^2+{\bm k}_\perp^2}},
\ee
where we used $M_0^2-(m_1-m_2)^2=({\cal A}^2+{\bm k}_\perp^2)/[x(1-x)]$.
Inserting Eq.~\eqref{eq:Rcov} into the definition~\eqref{eq:17} and performing
the helicity sum with the on-shell spectator spin sum
$\sum_{\lambda_2} v_{\lambda_2}(p_2)\bar v_{\lambda_2}(p_2)=\slashed{p}_2-m_2$,
the numerator collapses to the Dirac trace
\be\label{eq:tracegeneric}
{\cal S}^{[\Gamma]}_{\rm on}
=\mathrm{Tr}\bigl[\gamma_5(\slashed{p}_{1\rm on}+m_1)\,\Gamma\,
 (\slashed{p}^{\prime}_{1\rm on}+m_1)\gamma_5(-\slashed{k}_{\rm on}+m_q)\bigr],
\ee
that is, to the on-shell part of Eq.~\eqref{eq:BStrace} with $k=p_2$ and
$m_q=m_2$. The remaining factors are exactly the Melosh and spinor
normalizations, so that
\be\label{eq:spindict}
\frac{1-x}{2}\,
\frac{{\cal S}^{[\Gamma]}_{\rm on}}
{\sqrt{{\cal A}^2+{\bm k}_\perp^{2}}\,\sqrt{{\cal A}^2+{\bm k}_\perp^{\prime 2}}}
= P^+\, S^{[\Gamma]} .
\ee
Nothing in this derivation refers to the specific form of $\Gamma$: the meson
vertices $\gamma_5$, the on-shell spectator spin sum and the normalization
factors are common to all projections, and $\Gamma$ is simply carried through
the trace unchanged. Equation~\eqref{eq:spindict} therefore holds for all Dirac projections
considered here, namely for the scalar current $\Gamma=1$ as well as for
all components of the vector current $\Gamma=\gamma^\mu$. Replacement (1) is thus not
merely a substitution of wave functions but also the generic dictionary between
the BS trace ${\cal S}^{[\Gamma]}_{\rm on}$ and the LFQM spin-orbit factor
$S^{[\Gamma]}$.

The second and third replacements are not independent, both being consequences
of Eq.~\eqref{eq:onshellcond}. In the BS-model valence expression, the
instantaneous contribution is proportional to a linear combination of the
off-shellness factors $M^2-M_0^2$ and $M^{\prime 2}-M_0^{\prime 2}$. It
therefore vanishes automatically when the BT replacement
$M^{(\prime)}\to M_0^{(\prime)}$ is implemented consistently at the integrand
level, reducing the valence spin factor to its on-shell part,
${\cal S}_{\rm val}\to{\cal S}_{\rm on}$.

The Lorentz prefactor of the BS model is built from the physical meson masses
and therefore coincides with the one used in the pBT-LFQM,
\be\label{eq:PBSpBT}
{\cal P}^{[\Gamma]}_{\rm BS}={\cal P}^{[\Gamma]}_{\rm pBT},
\ee
the two differing only in the wave function to which they are attached. It
should be noted, however, that ${\cal P}^{[\Gamma]}_{\rm fBT}$ in (3) is not
obtained from ${\cal P}^{[\Gamma]}_{\rm BS}$ by the substitution
$M^{(\prime)}\to M^{(\prime)}_0$ alone. The BT construction has to be applied to
the covariant structure itself,
${\cal P}^{[\gamma^\mu]}=\bar{P}^\mu-q^\mu(\bar{P}\cdot q)/q^2$ of
Eq.~\eqref{eq:formfactor}, whose second term vanishes identically for the elastic
kinematics of the BS model and of the pBT-LFQM, where
$\bar{P}\cdot q=M^{\prime 2}-M^2=0$, but survives at the integrand level once
$\bar{P}\cdot q\to M^2_-=M_0^{\prime 2}-M_0^2$, as emphasized in
Sec.~\ref{sec:EMFF}. For $\Gamma=\gamma^+$ this additional covariant term is absent and the
prefactor is unchanged, while for $\Gamma=1$ the correspondence reduces
to the simple replacement $M\to M_0$. For the transverse and minus
components, however, the full covariant structure must be carried along. With this
simultaneous implementation in the vertex, spin, and Lorentz structures, the
BS-model valence expression is mapped onto the corresponding fBT-LFQM result.

In summary, applying the correspondence of Eq.~\eqref{eq:matching2} to the
BS valence contribution of Eq.~\eqref{eq:325} gives
\bea\label{eq:FSfBT}
[F_S(Q^2)]_{\rm fBT}
&=&
\frac{1}{16\pi^3}
\int_0^1 \frac{1-x}{2}\,dx
\int d^2{\bm k}_\perp\,
\frac{\phi^{\prime}(x,{\bm k}_\perp^{\prime})}
{\sqrt{{\cal A}^2+{\bm k}_\perp^{\prime 2}}}
\frac{\phi(x,{\bm k}_\perp)}
{\sqrt{{\cal A}^2+{\bm k}_\perp^2}}
\frac{{\cal S}_{\rm on}}{2M_0},
\eea
which, by Eq.~\eqref{eq:spindict}, is identical to the direct fBT-LFQM result of
Eq.~\eqref{eq:19}. The essential point is that no explicit BS-model LF zero-mode contribution has to
be evaluated or imported on the way from Eq.~\eqref{eq:325} to
Eq.~\eqref{eq:FSfBT}: the valence contribution of the BS model, together with
the on-mass-shell condition imposed at the integrand level, already determines
the fBT-LFQM result.

\paragraph{Integrand kernels.}
It is convenient to display this correspondence directly at the level of the
integrand. For a given Dirac projection $\Gamma$, we write the BS valence
amplitude and the corresponding fBT-LFQM result as
\begin{align}
  \big[F^{[\Gamma]}\big]^{\rm val}_{\rm BS}
  &=
  \frac{N_c}{16\pi^3}
  \int_0^1\frac{dx}{1-x}
  \int d^2\boldsymbol{k}_\perp\,
  \chi'(x,\boldsymbol{k}'_\perp)
  \chi(x,\boldsymbol{k}_\perp)\,
  \mathcal{K}^{[\Gamma]}_{\rm BS},
  \label{eq:FBSop}\\[4pt]
  F^{[\Gamma]}_{\rm fBT}
  &=
  \int [d^3\boldsymbol{p}_1]\,
  \phi'(x,\boldsymbol{k}'_\perp)
  \phi(x,\boldsymbol{k}_\perp)\,
  \mathcal{K}^{[\Gamma]}_{\rm fBT},
  \label{eq:FfBTop}
\end{align}
where the corresponding integrand kernels are defined by
\begin{equation}
  \mathcal{K}^{[\Gamma]}_{\rm BS}
  \equiv
  \frac{
  {\cal S}^{[\Gamma]}_{\rm on}
  +{\cal S}^{[\Gamma]}_{\rm inst}}
  {\mathcal{P}^{[\Gamma]}_{\rm BS}},
  \qquad
  \mathcal{K}^{[\Gamma]}_{\rm fBT}
  \equiv
  \frac{S^{[\Gamma]}}
  {\mathcal{P}^{[\Gamma]}_{\rm fBT}}.
  \label{eq:opdef}
\end{equation}
Here, ${\cal S}^{[\Gamma]}_{\rm on}$ and $S^{[\Gamma]}$ are not the same
objects: they are related through the spin-structure correspondence in
Eq.~\eqref{eq:spindict}. Likewise, the Lorentz prefactors are related by
replacement (3) of Eq.~\eqref{eq:matching2}, together with
Eq.~\eqref{eq:PBSpBT}. Thus, Eqs.~\eqref{eq:FBSop}--\eqref{eq:opdef}
should be understood as an integrand-level correspondence between the BS
valence expression and the fBT-LFQM result, rather than as an
operator-level equivalence between the two dynamical frameworks.

Table~\ref{tab:BSfBTop} lists the BS and fBT integrand kernels for all four
Dirac projections considered here. For $\Gamma=\gamma^+$, the instantaneous
structure is absent and the BS and fBT Lorentz prefactors coincide,
$\mathcal{P}^{[\gamma^+]}_{\rm BS}
=\mathcal{P}^{[\gamma^+]}_{\rm fBT}=2P^+$.
Consequently, the ordinary vertex correspondence in
Eq.~\eqref{eq:vertexlink}, together with the spin-structure relation in
Eq.~\eqref{eq:spindict}, is sufficient to map the BS valence integrand onto
the fBT-LFQM one.

\begin{table}[t]
\centering
\renewcommand{\arraystretch}{2.2}
\setlength{\tabcolsep}{8pt}
\begin{tabular}{c c c c c}
\hline\hline
$\Gamma$ &
$\mathcal{P}^{[\Gamma]}_{\rm BS}=\mathcal{P}^{[\Gamma]}_{\rm pBT}$ &
$\mathcal{P}^{[\Gamma]}_{\rm fBT}$ &
$\mathcal{K}^{[\Gamma]}_{\rm BS}$ &
$\mathcal{K}^{[\Gamma]}_{\rm fBT}$ \\
\hline
$\gamma^+$ &
$2P^+$ & $2P^+$ &
$\dfrac{{\cal S}^{[\gamma^+]}_{\rm on}}{2P^+}$ &
$\dfrac{S^{[\gamma^+]}}{2P^+}$ \\
$\boldsymbol{\gamma}^j_\perp$ &
$\boldsymbol{q}^j_\perp$ &
$\boldsymbol{q}^j_\perp\Big(1+\dfrac{M^2_-}{\boldsymbol{q}^2_\perp}\Big)$ &
$\dfrac{{\cal S}^{[\gamma_\perp]}_{\rm on}
        +{\cal S}^{[\gamma_\perp]}_{\rm inst}}
       {\mathcal{P}^{[\gamma_\perp]}_{\rm BS}}$ &
$\dfrac{S^{[\gamma_\perp]}}{\mathcal{P}^{[\gamma_\perp]}_{\rm fBT}}$ \\
$\gamma^-$ &
$\dfrac{2M^2+\boldsymbol{q}^2_\perp}{P^+}$ &
$\dfrac{2M_0^{\prime 2}\boldsymbol{q}^2_\perp+\boldsymbol{q}^4_\perp+(M^2_-)^2}
       {\boldsymbol{q}^2_\perp P^+}$ &
$\dfrac{{\cal S}^{[\gamma^-]}_{\rm on}
        +{\cal S}^{[\gamma^-]}_{\rm inst}}
       {\mathcal{P}^{[\gamma^-]}_{\rm BS}}$ &
$\dfrac{S^{[\gamma^-]}}{\mathcal{P}^{[\gamma^-]}_{\rm fBT}}$ \\
$1$ &
$2M$ & $2M_0$ &
$\dfrac{{\cal S}^{[1]}_{\rm on}+{\cal S}^{[1]}_{\rm inst}}{2M}$ &
$\dfrac{S^{[1]}}{2M_0}$ \\
\hline\hline
\end{tabular}
\caption{Integrand operators of Eq.~\eqref{eq:opdef} for the BS-model valence
contribution and for the fBT-LFQM, related by the matching prescription of
Eq.~\eqref{eq:matching2}. Note that ${\cal S}^{[\Gamma]}_{\rm on}$ and
$S^{[\Gamma]}$ are different objects, related by Eq.~\eqref{eq:spindict}, and
that $M^2_-=M_0^{\prime 2}-M_0^2$. The entries $S^{[\Gamma]}$ and
$\mathcal{P}^{[\Gamma]}_{\rm pBT(fBT)}$ are those of
Table~\ref{tab:helicity}.}
\label{tab:BSfBTop}
\end{table}

For the other projections, the BT on-mass-shell condition must be implemented
consistently at the integrand level. Whenever an instantaneous contribution
is present in the BS valence trace, it is eliminated by
$M^{(\prime)}\to M_0^{(\prime)}$, while the same BT construction must be
applied simultaneously to the associated Lorentz structure. In particular,
for $\Gamma=1$ the Lorentz correspondence reduces to the replacement
$2M\to2M_0$, whereas for the transverse and minus components of the vector current,
the replacement $M^{(\prime)}\to M_0^{(\prime)}$ must be made in the
full covariant Lorentz decomposition before using
$\bar P\!\cdot q=0$ of the physical elastic kinematics.

Among the four projections, $\Gamma=\gamma^-$ and $\Gamma=1$ are
zero-mode sensitive: in the covariant BS model, the full amplitude for these
channels contains a nonvanishing contribution beyond the valence part.
The present matching does not transport that explicit BS-model LF zero-mode contribution into
the Gaussian LFQM. Rather, it shows how the BS \emph{valence} integrand is
mapped onto the corresponding fBT-LFQM integrand once the BT construction is
imposed consistently.

This pattern parallels Table~I of Ref.~\cite{hhk8-g2bj}, where the
$\gamma^{(+,\perp)}\gamma_5$ entries provide the cases that match without
additional zero-mode-sensitive contributions. The table also makes explicit the status of
the pBT extraction. By Eq.~\eqref{eq:PBSpBT}, the pBT-LFQM retains the same
Lorentz prefactor as the covariant BS model, built from the physical meson
mass, while evaluating the current matrix element in the on-shell valence
basis. For the zero-mode-sensitive projections, this mixed implementation is
precisely what leads to the inconsistency that is removed in the fBT-LFQM by
applying the BT construction uniformly to both the spin and Lorentz
structures.

\section{Derivation of Eq.~(\ref{eq:LFforward})}
\setcounter{equation}{0}
\renewcommand{\theequation}{B\arabic{equation}}
\label{app:forward}
Starting from the TMD correlator at fixed LF time $z^+=0$,
\begin{equation}\label{eq:AppPhi}
\Phi_q^{[\Gamma]}(x,\bm k_\perp)
=
\frac{1}{2}\int[\mathrm{d}^3 z]\,
e^{\,i p_1\cdot z}
\langle P|\,\bar q(0)\Gamma\mathcal W(0,z)q(z)\,|P\rangle
\Big|_{z^+=0},
\end{equation}
where $[\mathrm{d}^3 z]=\mathrm{d}z^-\,\mathrm{d}^2\bm z_\perp/[2(2\pi)^3]$.
Using $p_1^+=xP^+$ and $p_1\cdot z=\frac12 xP^+z^- -\bm k_\perp\cdot\bm z_\perp$,
we integrate over $x$ and $\bm k_\perp$:
Integrating Eq.~\eqref{eq:AppPhi} over $x$ and $\bm k_\perp$ yields
\begin{align}
\int\!\mathrm{d}x & \int\!\mathrm{d}^2\bm k_\perp\;\Phi_q^{[\Gamma]}(x,\bm k_\perp)\nonumber\\
&=\frac{1}{2}\int\![\mathrm{d}^3 z]\;
\int\!\mathrm{d}x\,e^{\,i(\tfrac12 P^+)x\,z^-}
\int\!\mathrm{d}^2\bm k_\perp\,e^{-i\bm k_\perp\cdot\bm z_\perp}
\langle P|\,\bar q(0)\,\Gamma\,\mathcal W(0,z)\,q(z)\,|P\rangle \nonumber\\[2pt]
&=\frac{1}{2}\,\frac{1}{2(2\pi)^3}\,\frac{4\pi}{P^+}\,(2\pi)^2
\int\!\mathrm{d}z^-\,\mathrm{d}^2\bm z_\perp\;
\delta(z^-)\,\delta^{(2)}(\bm z_\perp)\,
\langle P|\,\bar q(0)\,\Gamma\,\mathcal W(0,z)\,q(z)\,|P\rangle .
\end{align}
The delta functions set $z\to 0$, so that $\mathcal W(0,0)=\mathbf 1$.
Therefore,
\begin{equation}
\int\!\mathrm{d}x\int\!\mathrm{d}^2\bm k_\perp\,\Phi_q^{[\Gamma]}(x,\bm k_\perp)
=
\frac{1}{2P^+}\,\langle P|\,\bar q(0)\Gamma q(0)\,|P\rangle,
\end{equation}
which reproduces Eq.~\eqref{eq:LFforward}.
In deriving Eqs.~(B2) and (B3), the $x$ integration is understood to extend
over the full real axis. Since the correlator vanishes outside its physical
support, this is equivalent to integration over the full support of the
distribution. In the valence LFQM implementation, the active-quark and
active-antiquark sectors are represented separately, each with a positive
momentum fraction $0<x<1$.
 
\newpage
\bibliographystyle{JHEP}
\bibliography{references}

\end{document}